\documentclass[a4paper,11pt]{article}
\pdfoutput=1 % if your are submitting a pdflatex (i.e. if you have
\usepackage{jheppub} % for details on the use of the package, please
\usepackage{booktabs}
\usepackage{color}
\usepackage{graphicx}% Include figure files
\usepackage{dcolumn}% Align table columns on decimal point
\usepackage{bm}% bold math
\usepackage{graphicx}
\usepackage[T1]{fontenc} % if needed
\usepackage{color}

\title{\boldmath Relativistic Impulse Approximation in the Atomic Ionization Process induced by Millicharged Particles}

\author[a,b]{Chen-Kai Qiao, \note{The order of authors is arranged according to the contributions, rather than using the conventional alphabetical order.}}%\note{Corresponding author.}}
\author[b]{Shin-Ted Lin, }%\note{Corresponding author.}}
\author[c]{Hsin-Chang Chi,}
\author[b]{Hai-Tao Jia}
\affiliation[a]{College of Science, Chongqing University of Technology, Hongguang Avenue, Banan, Chongqing, 400054}
\affiliation[b]{College of Physics, Sichuan University, Yihuan Road, Chengdu, Sichuan, 610065}
\affiliation[c]{Department of Physics, National Dong Hwa University, Da Hsueh Road, Shoufeng, Hualien, 97401}
\emailAdd{chenkaiqiao@126.com; chenkaiqiao@cqut.edu.cn}
\emailAdd{\\ \ \ \ \ \ \ \ stlin@scu.edu.cn}
\emailAdd{\\ \ \ \ \ \ \ \ hsinchang@mail.ndhu.edu.tw}
\emailAdd{\\ \ \ \ \ \ \ \ jiahaitao@stu.scu.edu.cn}
\abstract
{
The millicharged particle has become an attractive topic to probe physics beyond the Standard Model. In direct detection experiments, the parameter space of millicharged particles can be constrained from the atomic ionization process. In this work, we develop the relativistic impulse approximation (RIA) approach, which can duel with atomic many-body effects effectively, in the atomic ionization process induced by millicharged particles. The formulation of RIA in the atomic ionization induced by millicharged particles is derived, and the numerical calculations are obtained and compared with those from free electron approximation and equivalent photon approximation. Concretely, the atomic ionizations induced by mllicharged dark matter particles and millicharged neutrinos in high-purity germanium (HPGe) and liquid xenon (LXe) detectors are carefully studied in this work. The differential cross sections, reaction event rates in HPGe and LXe detectors, and detecting sensitivities on dark matter particle and neutrino millicharge in next-generation HPGe and LXe based experiments are estimated and calculated to give a comprehensive study. Our results suggested that the next-generation experiments would improve 2-3 orders of magnitude on dark matter particle millicharge $\delta_{\chi}$ than the current best experimental bounds in direct detection experiments. Furthermore, the next-generation experiments would also improve 2-3 times on neutrino millicharge $\delta_{\nu}$ than the current experimental bounds. % This would be a wonderful news for theoretical and experimental studies of millicharged particles.

\

\noindent{Key Words: \\
Beyond Standard Model, Cosmology of Theories beyond the SM \\
Neutrino Physics, Solar and Atmospheric Neutrinos}
}

\begin{document}
\maketitle
\flushbottom

\section{Introduction \label{sec:1}}

Charge quantization is one of the most profound and fascinating open problems in physics. After exploring for several decades, the validity of charge quantization is still not revealed. All the observed elementary particles in Standard Model have quantized electric charge, but the underling nature is still a mystery and can't be solved in the context of Standard Model. The electric charge quantization can be predicted in many theories beyond the Standard Model, i.e., grand unifications \cite{Glashow}, magnetic monopoles \cite{Dirac,Deans}, and extra dimensions \cite{Klein,Gross}. However, no clear evidences have ever been provided as a confirmation for such theories.

Recently, a lot of studies propose a category of new particles, whose electric charge is tiny and non-quantized. These particles are named as ``millicharged particles'' \cite{Dobroliubov} \footnote{In some literatures, the ``millicharged particle'' is also called as ``minicharged particles, i.e., in reference \cite{Vinyoles}.} and has stimulated a number of theoretical and experimental investigations \cite{Vinyoles,Davidson1991}. The experimental studies on millicharged particles can be carried out through reactor experiments \cite{Singh2019}, positronium decays \cite{Badertscher}, solar and celestial bodies observations \cite{Vinyoles}, cosmic microwave background (CMB) \cite{Dolgov}, big bang nucleosynthesis (BBN), supernovas \cite{Vogel}, 21-cm line observations \cite{LiuHongwan}, accelerator and collider experiments \cite{LiuZuowei2019,Liang2020,Jaeckel}. These experiments strictly constrain the mass of millicharged particles, as well as its electric charge over the past few years.

The millicharged particle has tiny electric charge, it could have electromagnetic interactions with target atoms or molecules. For instance, it can cause the following atomic ionization process
\begin{equation}
\chi + A \rightarrow \chi + A^{+} + e^{-} \label{Atomic Ionization Millicharged Particle}
\end{equation}
Therefore, in the direct detection experiments, the millicharged particles can be caught and detected from the signals produced by atomic ionization processes. In future super-terranean or underground experiments, i.e., the China Dark matter EXperiment (CDEX) experiments located at China Jin-Ping underground Laboratory (CJPL) \cite{CDEX}, through searching the ionization, scintillation or heat signals in real detectors produced by atomic ionization processes, we can detect these unknown millicharged particles and give a constrain on their parameter space.

Previous studies on atomic ionization process induced by millicharged particles are often proceeded based on the free electron approximation (FEA) \cite{Singh2019}. In the FEA, the calculation is the simplest and atomic many-body effects are neglected. The FEA approach works well in the high-energy transfer region, where atomic binding effects are negligible and the atomic electrons are approximately free. However, in the low-energy transfer region, the atomic many-body effects (including atomic binding effects, electron shielding, electron correlation effects and other many-body interactions) become dominant, thus the FEA breaks down and new approaches inspired by many-body physics are needed. It is worth noting that, the low-energy transfer region plays a crucial role in the experimental search for millicharged particles through atomic ionization processes, because the differential cross sections and reaction event rates in detectors have a dominant enhancement in this region. Particularly, detailed studies shown that, the FEA results may underestimate the differential cross section by more than 1 order in the low-energy transfer region, compared with the results from many-body physics \cite{Chen2014}. Therefore, in this case, the FEA approach may produce large errors and it must be corrected through considering atomic many-body effects.

Apart from the FEA, there are other approaches used in the studies of atomic ionizations induced by millicharged particles. For instance, the equivalent photon approximation (EPA) is frequently used \cite{Singh2019}. In the EPA approach, the atomic many-body effects can be partly considered. In this formulation, contributions coming from virtual photons in the electromagnetic interaction are equivalent to contributions from real photons. However, there is a fatal weakness: the EPA is valid only when energy transfer is extremely small (in the $T\rightarrow0$ limit). Therefore, a precise method which could deal with atomic many-body effects in the entire energy region is especially needed.

Recently, a new approach---the relativistic random-phase approximation---is applied to the studies of millicharged particles to treat the atomic many-body effects \cite{Chen2014,Chen2015,Chen2017,Hsieh2019}. However, this method is complicated in numerical calculations, especially when the incident particle energy is extremely high \cite{private}. A relatively simple approach that could contain atomic many-body effects in the entire energy region would be helpful.

Inspired by the previous researches in many-body physics, in this work, we develop the relativistic impulse approximation (RIA) in the atomic ionization process induced by millicharged particles. The original ideas and framework of RIA approach are developed in previous years to handle a number of electromagnetic interactions in atomic physics, such as atomic Compton scattering \cite{Eisenberger1970,Ribberfors1975,Ribberfors1982}, electron impact \cite{Bell1989} and other atomic processes \cite{Brandt1983,Lee1990}. The atomic many-body effects can be treated effectively in the RIA approach. With the advantages of simplicity and flexibility, the RIA formulation has been widely applied to atomic and molecular physics \cite{Toth1996,Pratt,Pratt2010}, condensed matter physics \cite{Kubo,Cooper}, nuclear and elementary particle physics \cite{Brusa,Salvat,Ramanathan}. In particular, in the conventional Monte Carlo simulation program Geant4 \cite{Geant4}, which are extensively used in nuclear and elementary particle physics experiments, many processes are treated using the formulation of RIA \cite{Geant4b,Livermore,Monash}.

In the present work, we develop the RIA approach for the atomic ionization process induced by millicharged particles. The formulation of RIA is derived, and the numerical calculations are obtained and compared with those from FEA and EPA approaches. These comparisons should give us information of the influences brought by atomic many-body effects, especially in the low-energy transfer regions. Our RIA approach developed in this work is quite general, irrelevant to the material composition of detectors and the underling nature of millicharged particles.

In this work, we consider two categories of millicharged particles: millicharged dark matter particles and millicharged neutrinos. On the one hand, dark matter problem is one of the most important topics in elementary particle physics, astrophysics, astronomy and cosmology. Currently, accumulating evidences in special rotation curves of spiral galaxies \cite{Zwicky,Rubin,Corbelli}, gravitational lensing \cite{Clowe}, large scale structure formation \cite{Blumenthal,Davis}, cosmic microwave background and baryon acoustic oscillations \cite{WMAP,Planck} have indicated that there are large amount of non-luminous dark matter in our universe \cite{Undagoitia,Bertone}. Therefore, direct detection of dark matter particle becomes an extremely significant and urgent work \cite{Undagoitia}. Traditionally, the most promising candidates for these unknown dark matter particles are weakly interacting massive particles (WIMPs), which only interact with ordinary matter through weak interaction beyond the Standard Model \cite{Undagoitia,Bertone,Feng2010}. Other candidates, such as axion, sterile neutrino, mirror dark matter and so on \cite{Wilczek1983,Perez2020,Boyarsky2019,Foot2018}, are also actively studied in the dark matter detections. The millicharged particles, due to its ultra-tiny electromagnetic interaction, can successfully generate the dark matter relic abundance and give consistent results with astrophysical observations, which makes it become a candidate of dark matter particles \cite{Liu2012,Petraki2014,Foot2016}. On the other hand, neutrino physics also becomes a promising field in elementary particle physics, astrophysics, astronomy and cosmology, for it can reveal many aspects of physics beyond the Standard Model, e.g., baryon non-conservation \cite{Vergados1986}, matter-antimatter asymmetry \cite{Dine2004,Buchmuller}, neutrino oscillation \cite{Super-Kamiokande,King2013}, seesaw mechanism \cite{Mohapatra2006}, and their Dirac or Majorana nature of fermions \cite{Rodejohann,Bilenky}. Interestingly, there is an overlap, recent studies suggested that neutrino may become millicharged particles and they may have tiny electromagnetic interactions \cite{Giunti2008,Giunti2015}.

In the numerical calculations, we choose Ge and Xe elements as detector materials to study the atomic ionization processes induced by millicharged particles. The Ge and Xe elements are ideal materials for experimental detection of charged or neutral particles. With sufficient low threshold, large effective volume, high efficiency and ultra low background, high-purity germanium (HPGe) and liquid xenon (LXe) detectors are most generally used in particle physics experiments, especially for dark matter direct detections and neutrinoless double beta decay experiments \cite{Undagoitia,Rodejohann,CDMS,CDEX,GERDA,PandaX,XENON,XENON2,EXO,KamLAND-Zen}. Concretely, in the present work, we study the ionization of Ge and Xe atoms by millicharged particles, calculating the differential cross section and the reaction event rate of these ionization processes in real HPGe and LXe detectors. The low-energy transfer and near threshold regions, where atomic many-body effects could have great impacts, are especially considered. The estimation of detecting sensitivities for dark matter particle and neutrino millicharge in next-generation HPGe and LXe based experiments is also provided according to the reaction event rate the and experimental background level.

Furthermore, in the actual \emph{ab initio} calculations, the influence coming from relativistic effects of atomic electrons is also a critical point in dealing with atomic ionization process induced by millicharged particles. For the deep inner-shell atomic electrons, their motion around atomic nucleus could be very rapid with relativistic corrections at the order of $Z\alpha_{\text{em}}$ \footnote{In a semiclassical point of view, the velocity of deep inner-shell atomic electrons can be approximate as $v_{e}/c \sim p_{e}/mc \sim Z\alpha_{\text{em}}$. This relation can be easily derived from the Bohr model of Hydrogen-like ions in the old-fashioned quantum theory.}, where $\alpha_{\text{em}} \approx 1/137$ is the fine structure constant for electromagnetic interactions \cite{Friedrich2006,Amusia2012}. Particularly, for Xe atomic, this value could reach $Z\alpha_{\text{em}} \approx 0.4$, suggesting that electron relativistic effects would play a significant role the same as electron many-body effects. Some previous studies have shown that the relativistic effects have non-negligible contributions to the scattering of atomic electrons with photon, ions, neutrinos and dark matter particles \cite{Chen2014,Chen2015,Chen2017,Toth1996,Pratt,Pratt2010,Roberts2016a}. Recently, B. M. Roberts et al. concluded that the electron relativistic effect could give a large enhancement on cross sections as well as reaction event rates in the ionization process of atomic electrons induced by weakly interacting massive particles (WIMPs) \cite{Roberts2016a,Roberts2016b,Roberts2019}. These studies strongly imply that in the atomic ionization process induced by millicharged particles, which we are interested, the relativistic effects of atomic electrons may also become a non-negligible issue.

In atomic and molecular physics, the relativistic effect and atomic many-body effects can be treated efficiently in the Dirac-Fock theory, which is the relativistic extension of the self-consistent Hartree-Fock method \cite{Grant1961,Desclaux1971,Desclaux,Grant,Zanna}. In this theory, the ground state wavefunctions are obtained by solving the fully relativistic many-body Dirac-Fock equation for atomic systems. The Dirac-Fock theory, since it was developed in the 1970s, has been widely applied to a number of atomic and molecular processes and has been confirmed by spectroscopic observations and scattering experiments in the past few decades. Therefore, in order to incorporate relativistic effects and many-body effects in the calculation of ground state wavefunctions and electron momentum distributions for atomic systems, we adopt the fully relativistic Dirac-Fock theory in this work.

This paper is organised as follows: section \ref{sec:2} gives an introduction of the millicharged particle; section \ref{sec:3} briefly describes the general ideals for RIA approach; section \ref{sec:4} is devoted to theoretical derivation of RIA approach in the atomic ionization process induced by millicharged particles; numerical results and discussions are given in section \ref{sec:5} and section \ref{sec:6} for millicharged dark matter particle as well as millicharged neutrino; and conclusions and future perspectives are summarized in section \ref{sec:7}. Furthermore, in the appendices, we give descriptions on free electron approximation (FEA), equivalent photon approximation (EPA) and the Dirac-Fock theory.

\section{Millicharged Particles \label{sec:2}}

This section gives a brief introduction to the millicharged particles. The mechanism giving rise to the millichaged particle and the current experimental bounds for millicharged particles are mainly discussed.

The millicharged particle can be obtained from theories beyond the Standard Model \cite{Jaeckel,Vogel}. In particular, we will describe two mechanisms that could give rise to millicharged particles in this section.

First, millicharged particles can be generated in the extension of Standard Model by introducing an additional unbroken local $U(1)_{h}$ gauge group to the Standard Model gauge group \cite{Feldman2007,Holdom1986,Kors2004,Vogel}. All Standard Model particles are singlets under the new gauge group $U(1)_{h}$. We also add a massive hidden fermion $\chi$ charged under the new gauge group $U(1)_{h}$ only. Therefore, together with the Abelian gauge group $U(1)_{Y}$ in the Standard Model, there are two Abelian gauge groups: $U(1)_{Y}$ and $U(1)_{h}$. The two gauge fields associated with gauge groups $U(1)_{Y}$ and $U(1)_{h}$ can couple to each other through the kinetic mixing. The Lagrangian for this model is \footnote{In this work, we have made the speed of light explicitly, rather than taking the natural unit $c=1$. In the \emph{ab initio} calculations in atomic or molecular physics, the atomic unit is frequently adopted, and the speed of light takes the value $c \approx 137.036$ in this unit.}:
\begin{eqnarray}
\mathcal{L} & = & \mathcal{L}_{0} + \mathcal{L}_{1} \nonumber
\\
            & = & -\frac{1}{4}F_{\mu\nu}F^{\mu\nu}-\frac{1}{4}V_{\mu\nu}V^{\mu\nu}-\frac{\kappa}{2}F_{\mu\nu}V^{\mu\nu}
                  +J_{\mu}^{B}B^{\mu}+J_{\mu}^{C}C^{\mu} \nonumber
\\
            &   & +\bar{f}(ic\hbar\gamma^{\mu}\partial_{\mu}-m_{f}c^{2})f+\bar{\chi}(ic\hbar\gamma^{\mu}\partial_{\mu}-m_{\chi}c^{2})\chi
\label{millicharge interaction lagrangian0}
\end{eqnarray}
Here, $\gamma^{\mu}$ is the conventional Dirac--$\gamma$ matrices, $B^{\mu}$ is the gauge field of $U(1)_{Y}$ group, $C^{\mu}$ is the gauge field of $U(1)_{h}$ group, $f$ is Standard Model fermion and $\chi$ is the hidden fermion charged under new gauge group $U(1)_{h}$. The $F_{\mu\nu}=\partial_{\mu}B_{\nu}-\partial_{\nu}B_{\mu}$ and $V_{\mu\nu}=\partial_{\mu}C_{\nu}-\partial_{\nu}C_{\mu}$ are the field strength for $B^{\mu}$ and $C^{\mu}$, respectively. Moreover, $J_{\mu}^{B}=g_{S}\bar{f}\gamma_{\mu}f$ is the current associating with the $U(1)_{Y}$ gauge field $B^{\mu}$, and $J_{\mu}^{C}=g_{h}\bar{\chi}\gamma_{\mu}\chi$ is the current associating with the additional $U(1)_{h}$ gauge field $C_{\mu}$. The $\kappa$ denotes the kinetic mixing parameter between two gauge fields $B^{\mu}$ and $C^{\mu}$.

To make the physical picture clearer, we introduce the following two gauge fields $A^{\mu}$ and $\tilde{A}^{\mu}$ as the combination of gauge field $B^{\mu}$ and $C^{\mu}$:
\begin{subequations}
\begin{eqnarray}
B^{\mu} & = & \frac{1}{\sqrt{1-\kappa^{2}}}A^{\mu}
\\
C^{\mu} & = & -\frac{\kappa}{\sqrt{1-\kappa^{2}}}A^{\mu}+\tilde{A}^{\mu}
\end{eqnarray}
\end{subequations}
After the definition and re-coupling of $A^{\mu}$ and $\tilde{A}^{\mu}$, the Lagrange density can be rewritten as \cite{Feldman2007}:
\begin{eqnarray}
\mathcal{L}_{1} & = & J_{\mu}^{B}B^{\mu}+J_{\mu}^{C}C^{\mu}
                      +\bar{\chi}(ic\hbar\gamma^{\mu}\partial_{\mu}-m_{\chi}c^{2})\chi \nonumber
\\
                & = & \bigg[
                        \frac{1}{\sqrt{1-\kappa^{2}}}J_{\mu}^{B}-\frac{\kappa}{\sqrt{1-\kappa^{2}}}J_{\mu}^{C}
                      \bigg]A^{\mu}
                      +J_{\mu}^{C}\tilde{A}^{\mu}
                      +\bar{\chi}(ic\hbar\gamma^{\mu}\partial_{\mu}-m_{\chi}c^{2})\chi \nonumber
\\
                & = & J_{\mu}^{A}A^{\mu}+\tilde{J}_{\mu}^{\tilde{A}}\tilde{A}^{\mu}
                      +\bar{\chi}(ic\hbar\gamma^{\mu}\partial_{\mu}-m_{\chi}c^{2})\chi
\label{millicharge interaction lagrangian}
\end{eqnarray}
From this rearrangement, it can be clearly manifested that, the bosonic field $A^{\mu}$ is coupled with currents $J_{\mu}^{B}$ and $J_{\mu}^{C}$, while the bosonic field $\tilde{A}^{\mu}$ is coupled with $J_{\mu}^{C}$ only. In this picture, our universe can be divided into two parts: the Standard Model sector and the ``hidden sector''. Accordingly, $J_{\mu}^{B}=g_{S}\bar{f}\gamma_{\mu}f$ is the current in the Standard Model sector, while $J_{\mu}^{C}=g_{h}\bar{\chi}\gamma_{\mu}\chi$ can be viewed as ``current'' in the hidden sector, with $A^{\mu}$ and $\tilde{A}^{\mu}$ to be the ordinary photon and ``dark photon'' in the Standard Model sector and hidden sector \footnote{For simplicity, we omit the electro-weak mixing in this section. More complicated cases should include the electro-mixing as well as the Higgs mechanism in the Standard Model.}. The $g_{h}$ is the coupling between the ``dark photon'' $\tilde{A}^{\mu}$ and the hidden sector fermion $\chi$, and $g_{S}$ is the coupling between photon $A^{\mu}$ and Standard Model fermion $f$. There is one important point should be noted, based on Eq. (\ref{millicharge interaction lagrangian}), the ``current'' $J_{\mu}^{C}=g_{h}\bar{\chi}\gamma_{\mu}\chi$ in the hidden sector not only couples with the dark photon $\tilde{A}^{\mu}$, but also couples with photon $A^{\mu}$. Therefore, in this picture, a fermion $\chi$ living in the hidden sector not only acts as a charged particle in the hidden sector, but also behaves likes a charged particle in the Standard Model sector. Its electric charge can be determined through the coupling between ``current'' $J_{\mu}^{C}$ and photon $A^{\mu}$:
\begin{eqnarray}
& &
-\frac{\kappa}{\sqrt{1-\kappa^{2}}}J_{\mu}^{C}=-\frac{\kappa}{\sqrt{1-\kappa^{2}}}g_{h}\bar{\chi}\gamma_{\mu}\chi \nonumber
\\
& \Rightarrow &
q_{\chi}\equiv\delta_{\chi} e=-\frac{\kappa}{\sqrt{1-\kappa^{2}}}g_{h} \nonumber
\\
& \Rightarrow &
\delta_{\chi}=-\frac{\kappa}{\sqrt{1-\kappa^{2}}}\frac{g_{h}}{e}
\end{eqnarray}
Assuming the kinetic mixing is extremely small, namely $\kappa \ll 1$, then the electric charge of hidden sector fermion $\chi$ is tiny, which makes it to be a millicharged particle. In this case, the electric charge of the millicharged particle $\chi$ can be further simplified as $q_{\chi}=\delta_{\chi} e \approx -\kappa g_{h}$.

\begin{figure}
\centering
\includegraphics[scale=0.385]{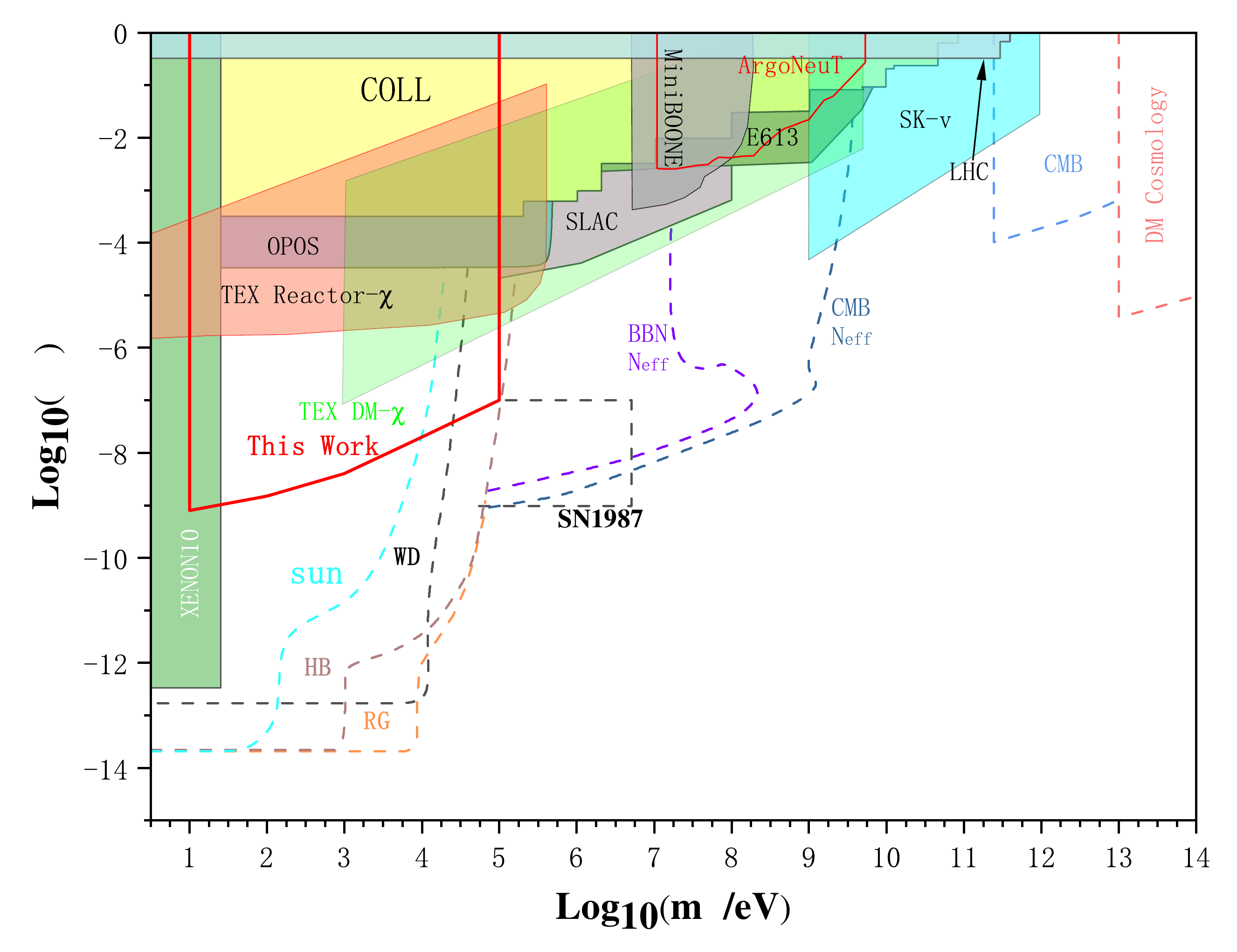}
\includegraphics[scale=0.385]{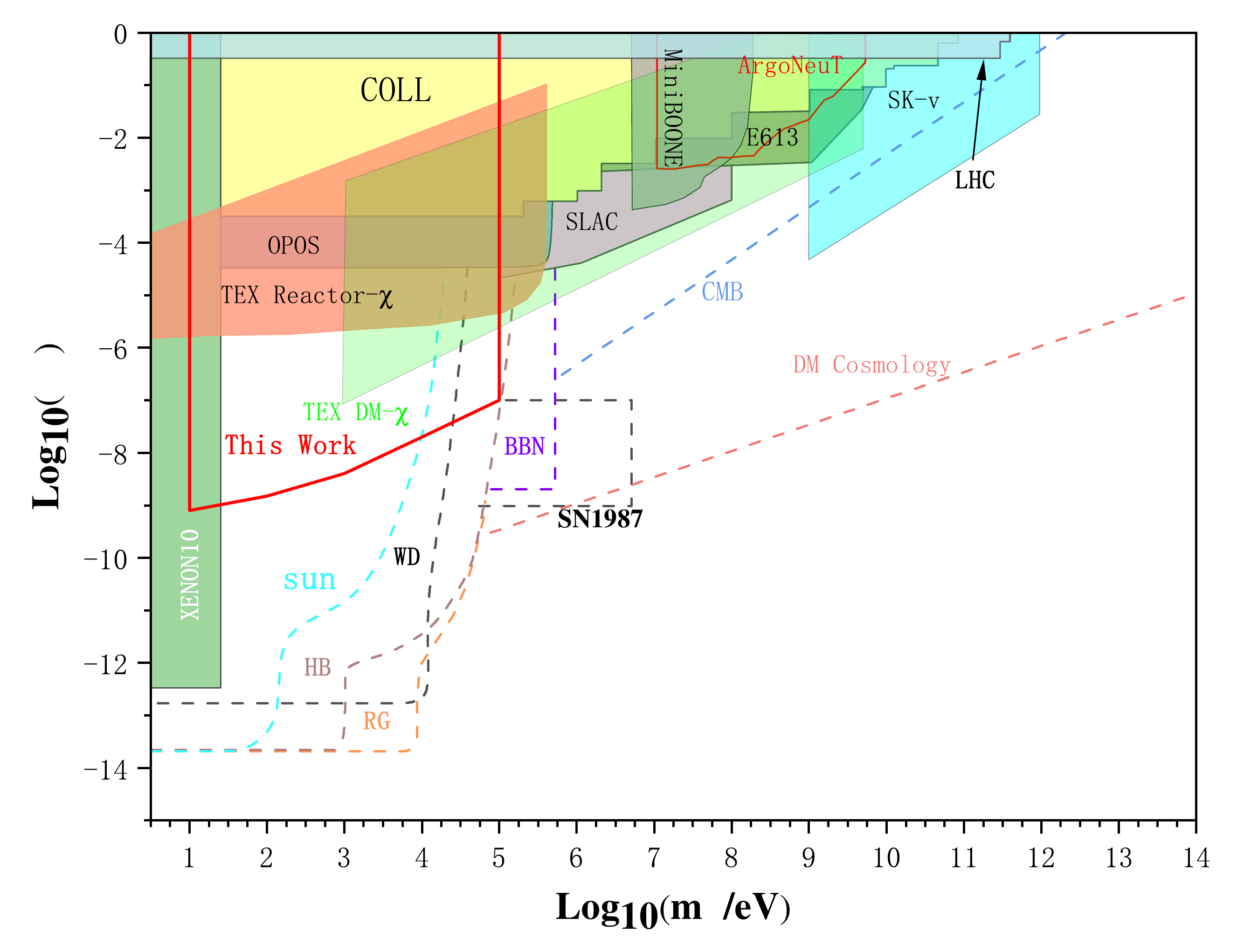}
\caption{The current experimental bounds for millicharged particles. In this figure, the horizontal axis labels the mass of millicharged particles $m_{\chi}$, and the vertical axis labels the millicharge $\delta_{\chi}$. Both the horizontal axis and vertical axis are plotted employing logarithmic coordinates. The upper and lower panels correspond to two different mechanisms: upper panel shows the exclusion region for models with an additional $U(1)_{h}$ gauge group; lower panel shows the exclusion region for models with right-handed massive fermions, which are $SU(2)_{L}$ singlets in the Standard Model gauge group. This figure present measurements and observations from astrophysics, cosmology, and particle physics experiments. The cosmological and astrophysical bounds from the sun (SUN) \cite{Vinyoles}, horizontal branch stars (HB) \cite{Vogel}, red-giant (RG) \cite{Davidson2000,Vogel}, white dwarf (WD) \cite{Davidson2000,Vogel}, supernova (SN1987A) \cite{Mohapatra}, cosmic microwave background (CMB) \cite{Vogel,Dubovsky,Dolgov} and big bang nucleosynthesis (BBN) \cite{Vogel} are denoted as dashed lines. Direct detections from the underground experiments (XENON \cite{AnHP2013} and Super-Kamiokande \cite{HuPK}), reactor experiments (TEXONO) \cite{Gninenko,Singh2019}, positronium decays (OPOS) \cite{Badertscher} are displayed as solid lines. The experimental constrains from accelerators and colliders (COLL \cite{Davidson1991,Davidson2000}, SLAC \cite{Prinz}, LHC \cite{Jaeckel}, E613 \cite{Soper2014}, MiniBooNE \cite{Magill}, ArgoNeuT \cite{Acciarri}) are also presented in this figure as comparisons. Furthermore, our estimations of detecting sensitivity for millicharged dark matter particles in next-generation LXe based experiments calculated using our RIA approach are also shown in this figure.}
\label{millicharge experiment}
\end{figure}

In the aforementioned models, the ``dark photon'' $\tilde{A}^{\mu}$ could acquire mass through the Higgs mechanism or St\"uckelberg mechanism \cite{Feldman2007,Kors2004}. Therefore, the parameter space of ``dark photon'' naturally consists of its mass $m_{\tilde{A}}$ and kinetic mixing $\kappa$. In recent years, the studies of ``dark photon'' have attracted considerable attention \cite{AnHP,Raggi2015}.

Furthermore, the millicharged particle can also be generated by other mechanisms. For instance, a class of models can be constructed by introducing right-handed massive fermion, which is a singlet under the $SU(2)_{L}$ Standard Model gauge group \cite{Vinyoles}. The Lagrangian for these models is:
\begin{eqnarray}
\mathcal{L} & = & \mathcal{L}_{0} + \mathcal{L}_{1} \nonumber
\\
            & = & -\frac{1}{4}F_{\mu\nu}F^{\mu\nu}-\bar{\chi}(ic\hbar\gamma^{\mu}\partial_{\mu}-m_{\chi}c^{2})\chi
                  +\delta_{\chi}e\bar{\chi}\gamma_{\mu}B^{\mu}\chi
\end{eqnarray}
where $\gamma^{\mu}$ is the conventional Dirac--$\gamma$ matrices, and $B^{\mu}$ is the gauge boson in the $U(1)_{Y}$ Standard Model gauge group. In these models, the right-handed massive fermion $\chi$ is the millicharged particle with mass to be $m_{\chi}$, and its electric charge is related to the millicharge $\delta_{\chi}$ via $q_{\chi}=\delta_{\chi} e$. Particularly, the neutrino millicharge, which will be discussed in section \ref{sec:6}, can be obtained in this way by introducing right-handed Dirac neutrinos \cite{Giunti2015}.

The parameter space of the millicharged particle is defined by $(m_{\chi},\delta_{\chi})$. Many experimental investigations have strongly constrained the parameter space of millicharged particles. Figure \ref{millicharge experiment} gives the current experimental bounds for millicharged particles. This figure presents measurements and observations from astrophysics, cosmology, and particle physics experiments. The cosmological and astrophysical bounds from the sun (SUN) \cite{Vinyoles}, horizontal branch stars (HB) \cite{Vogel}, red-giant (RG) \cite{Davidson2000,Vogel}, white dwarf (WD) \cite{Davidson2000,Vogel}, supernova (SN1987A) \cite{Mohapatra}, cosmic microwave background (CMB) \cite{Vogel,Dubovsky,Dolgov} and big bang nucleosynthesis (BBN) \cite{Vogel} are denoted as dashed lines. Direct detections from the underground experiments (XENON \cite{AnHP2013} and Super-Kamiokande \cite{HuPK}), reactor experiments (TEXONO) \cite{Gninenko,Singh2019}, positronium decays (OPOS) \cite{Badertscher} are displayed as solid lines. The experimental constrains from accelerators and colliders (COLL \cite{Davidson1991,Davidson2000}, SLAC \cite{Prinz}, LHC \cite{Jaeckel}, E613 \cite{Soper2014}, MiniBooNE \cite{Magill}, ArgoNeuT \cite{Acciarri}) are also presented in this figure as comparisons. Furthermore, this figure also gives our estimations of detecting sensitivity for millicharged dark matter particles in next-generation LXe based experiments calculated using our RIA approach developed in this work.

\section{General Pictures for the RIA approach \label{sec:3}}

In this section, we give an introduction of the RIA approach used in electromagnetic interactions in atomic physics. The general ideas, physical pictures, and theoretical formulation of the RIA approach are introduced in details. The development of the RIA approach in the atomic ionization process induced by millicharged particles is given in section \ref{sec:4}.

In the formulation of RIA, due to atomic binding effects, the atomic bound electrons in an atom have a momentum distribution, which can be determined through its ground state wavefunctions. In the scattering process, electrons with different momentum scattered with incident particle independently, the interference term between different momentum electrons is omitted for simplicity. With the advantages of simplicity and flexibility, the RIA formulation has been extensively used in many atomic physics processes, especially in the atomic Compton scattering \cite{Eisenberger1970,Eisenberger1974,Ribberfors1975,Ribberfors1975b,Ribberfors1982,Ribberfors1983,Qiao}, electron impact \cite{Bell1989}, and other atomic processes \cite{Brandt1983,Lee1990,Toth1996}.

In the following part, we will use the atomic Compton scattering
\begin{equation}
\gamma + A \rightarrow \gamma + A^{+} + e^{-}
\end{equation}
as an example to illustrate the general pictures and basic ideas for RIA formulation. In atomic Compton scattering, consider an incident photon with energy $\omega_{i}$ and momentum $\boldsymbol{k}_{i}$ scattering with an atomic bound electron with energy $E_{i}$ and momentum $\boldsymbol{p}_{i}$. After scattering, the energy and momentum of emitted photon are $\omega_{f}$ and $\boldsymbol{k}_{f}$, and energy and momentum of final state electron are $E_{f}$ and $\boldsymbol{p}_{f}$, respectively. Then the doubly-differential cross section (DDCS) of Compton scattering in RIA formulation is given by \cite{Ribberfors1975,Ribberfors1982}:
\begin{eqnarray}
\bigg(\frac{d^{2}\sigma}{d\omega_{f}d\Omega_{f}}\bigg)_{\text{RIA}}
& = & \frac{r_{0}^{2}m^{2}c^{4}}{2} \frac{\omega_{f}}{\omega_{i}} \iiint{d^{3}p_{i}\rho(\boldsymbol{p}_{i})
      \frac{X(K_{i},K_{f})}{E_{i}E_{f}}}
      \delta (E_{i}+\omega_{i}-E_{f}-\omega_{f})
\label{doubly differential cross section1}
\end{eqnarray}
where $r_{0}$ is the electron classical charge radius, $E_{i}=\sqrt{p_{i}^{2}c^{2}+m^{2}c^{4}}$ and $E_{f}=\sqrt{p_{f}^{2}c^{2}+m^{2}c^{4}}$ are the energies of initial and final state electrons, respectively. The functions $K_{i}$ and $K_{f}$ are defined as:
\begin{subequations}
\begin{eqnarray}
K_{i} & = & k_{i}^{\mu}\cdot p_{i\mu}=\frac{E_{i}\cdot \omega_{i}}{c^{2}}-\boldsymbol{p}_{i} \cdot \boldsymbol{k}_{i}
\\
K_{f} & = & k_{f}^{\mu}\cdot p_{i\mu}=\frac{E_{i}\cdot \omega_{f}}{c^{2}}-\boldsymbol{p}_{i} \cdot \boldsymbol{k}_{f}
        =   K_{i}-\frac{\omega_{i}\omega_{f}(1-\cos\theta)}{c^{2}}
\end{eqnarray}
\end{subequations}
The function $X(K_{i},K_{f})$ is proportional to the reaction probability of the free electron Compton scattering $\gamma + e \rightarrow \gamma + e$, which is the scattering between the incident photon $\gamma$ and electron momentum eigenstate $|\boldsymbol{p}_{i}\rangle$. It is defined as:
\begin{eqnarray}
X(K_{i},K_{f}) & = & \frac{K_{i}}{K_{f}}+\frac{K_{f}}{K_{i}}
                     +2m^{2}c^{2}
                     \bigg(
                       \frac{1}{K_{i}}-\frac{1}{K_{f}}
                     \bigg)
                     +m^{4}c^{4}
                     \bigg(
                       \frac{1}{K_{i}}-\frac{1}{K_{f}}
                     \bigg)^{2}
\label{function X}
\end{eqnarray}
Here, $\rho(\boldsymbol{p}_{i})$ denotes the momentum distribution of atomic electrons, which is calculated through ground state wavefunctions. From Eq. (\ref{doubly differential cross section1}), it is easy to see that in the RIA formulation electrons with different momentum eigenstate $|\boldsymbol{p}_{i}\rangle$ scattered with photon $\gamma$ independently, and the interference terms between different momentum eigenstates are omitted. In this approach, atomic many-body effects are mainly reflected in the momentum distribution of atomic bound electrons.

\begin{figure}
\centering
\includegraphics[scale=0.5]{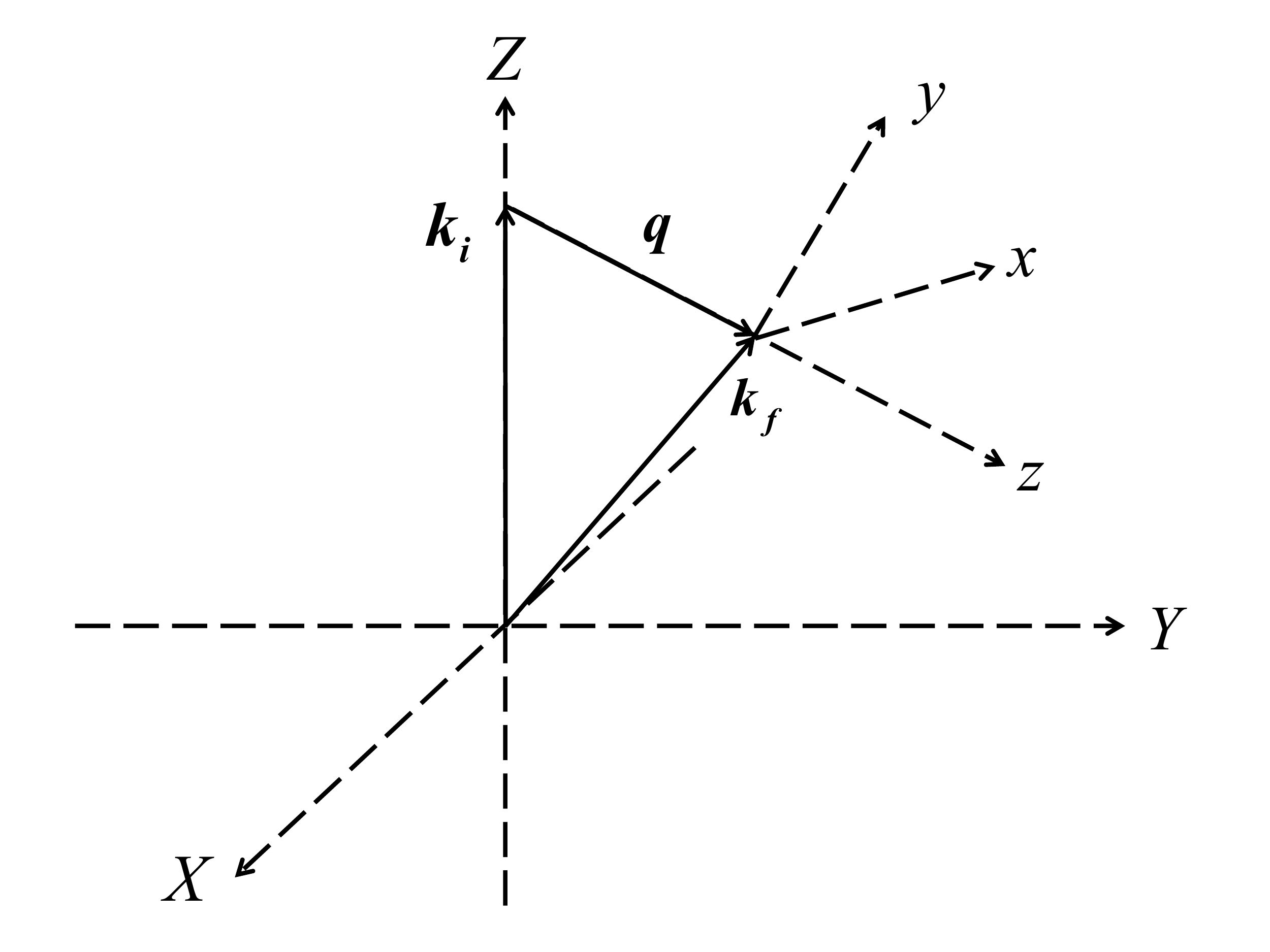}
\caption{Coordinate system $XYZ$ and $xyz$. Coordinate system $XYZ$ is chosen such that the $Z$ axis is along the direction of initial photon $\gamma$, and $X$ axis can be chosen as arbitrary direction perpendicular to the $Z$ axis. While the coordinate system $xyz$ is chosen such that the $z$  axis represents the momentum transfer direction. After Compton scattering, the momentum of the scattered photon is denoted as $\boldsymbol{k}_{f}$, and the momentum transfer vector $\boldsymbol{q}$ is defined as $\boldsymbol{q} \equiv \boldsymbol{k}_{f}-\boldsymbol{k}_{i}$. }
\label{coordinate0}
\end{figure}

In the previous studies, Roland Ribberfors et al. pointed out that the reaction probability function $X(K_{i},K_{f})$ in Eq. (\ref{doubly differential cross section1}) is a slow-varying function with respect to the integration variable $\boldsymbol{p}_{i}$. Therefore, it can be pulled out of the integration \cite{Ribberfors1975,Ribberfors1982,Brusa}. Successively, Ribberfors et al. made an approximation for function $X(K_{i},K_{f})$:
\begin{eqnarray}
X(K_{i},K_{f}) & \approx & \overline{X}(p_{z}) \nonumber
\\
               &    =    & \frac{K_{i}(p_{z})}{K_{f}(p_{z})}
                           +\frac{K_{f}(p_{z})}{K_{i}(p_{z})}
                           +2m^{2}c^{2}
                            \bigg(
                              \frac{1}{K_{i}(p_{z})}-\frac{1}{K_{f}(p_{z})}
                            \bigg)
                           +m^{4}c^{4}
                            \bigg(
                              \frac{1}{K_{i}(p_{z})}-\frac{1}{K_{f}(p_{z})}
                            \bigg)^{2} \nonumber
\\ \label{function X Pmin}
\end{eqnarray}
with $K_{i}(p_{z})$ and $K_{f}(p_{z})$ defined as:
\begin{subequations}
\begin{eqnarray}
K_{i}(p_{z}) & = & \frac{\omega_{i}E(p_{z})}{c^{2}}+\frac{\omega_{i}(\omega_{i}-\omega_{f}\cos\theta)p_{z}}{c^{2}q}
\\
K_{f}(p_{z}) & = & K_{i}(p_{z})-\frac{\omega_{i}\omega_{f}(1-\cos\theta)}{c^{2}}
\end{eqnarray}
\end{subequations}
In the above expressions, $q$ is the modulus of the momentum transfer vector $\boldsymbol{q} \equiv \boldsymbol{k}_{f}-\boldsymbol{k}_{i}$, and $p_{z}$ is the projection of the electron's initial momentum on the momentum transfer direction
\begin{equation}
p_{z} = \frac{\boldsymbol{p}\cdot\boldsymbol{q}}{q}
      = \frac{\omega_{i}\omega_{f}(1-\cos\theta)-E(p_{z})(\omega_{i}-\omega_{f})}{c^{2}q} \label{projection momentum}
\end{equation}
with energy $E(p_{z})$ defined by
\begin{equation}
E(p_{z}) = \sqrt{m_{e}^{2}c^{4}+p_{z}^{2}c^{2}}
\end{equation}
In the above calculations, the coordinate system $xyz$ is chosen such that the $z$ axis represents the momentum transfer direction in the Compton scattering process. The coordinate systems $xyz$ and $XYZ$ are defined and illustrated in figure \ref{coordinate0}. In many literatures \cite{Ribberfors1982,Ribberfors1983,Brusa}, a convenient approximation for $p_{z}$ component is proposed as follows:
\begin{equation}
p_{z} \approx \frac{\omega_{i}\omega_{f}(1-\cos\theta)-mc^{2}(\omega_{i}-\omega_{f})}{c^{2}q} \label{projection momentum2}
\end{equation}
This approximation works well for small $p_{z}$ values, however, it can cause notable discrepancies for large $p_{z}$ values.

Using the above assumptions, the DDCS of atomic Compton scattering process in the RIA formulation is given by:
\begin{equation}
\bigg(\frac{d^{2}\sigma}{d\omega_{f}d\Omega_{f}}\bigg)_{\text{RIA}}
=
\frac{r_{0}^{2}}{2}\frac{m}{q}
\frac{mc^{2}}{E(p_{z})}
\frac{\omega_{f}}{\omega_{i}}
\overline{X}(p_{z})
J(p_{z})
=
\overline{Y}^{\text{RIA}} \cdot J(p_{z}) \label{RIA}
\end{equation}
In this expression, the correction factor $J(p_{z})$ in the DDCS is called as the atomic Compton profile \cite{Biggs}
\begin{equation}
J(p_{z})\equiv \iint\rho(\boldsymbol{p}_{i})dp_{x}dp_{y} \label{Compton profile}
\end{equation}
with $\rho(\boldsymbol{p}_{i})$ to be the ground state electron momentum density of the atomic system. For most of the atomic systems, the momentum distribution is spherical symmetric, then the atomic Compton profile reduces to
\begin{equation}
J(p_{z})=2\pi\int\limits_{|p_{z}|}^{\infty}p_{i}\rho(p_{i})dp_{i}
\label{electron profile2}
\end{equation}
In this work, we only consider the spherical symmetric cases, and we use a fully relativistic Dirac-Fock theory to calculate the ground states of atomic systems and obtain their atomic Compton profiles.

The DDCS of atomic Compton scattering in the RIA formulation in Eq. (\ref{RIA}) can be further simplified. In previous studies, an alternative and simpler approximation of the reaction probability function $X(K_{i},K_{f})$ was made by taking the $p_{z}\rightarrow0$ limit of $\overline{X}(p_{z})$, which finally gives its FEA value (also called as the Klein-Nishina value) \cite{Ribberfors1982,Ribberfors1983}
\begin{equation}
X(K_{i},K_{f}) \approx X_{\text{KN}}=\frac{\omega_{i}}{\omega_{f}}+\frac{\omega_{f}}{\omega_{i}}-\sin^{2}\theta
\label{function X FEA}
\end{equation}
Therefore, the simplified results of DDCS for atomic Compton scattering in RIA formulation can be expressed as:
\begin{equation}
\bigg(\frac{d^{2}\sigma}{d\omega_{f}d\Omega_{f}}\bigg)_{\text{RIA}}
=
\frac{r_{0}^{2}}{2}
\frac{m}{q}\frac{\omega_{f}}{\omega_{i}}
X_{\text{KN}}
J(p_{z})
=
Y_{\text{KN}}^{\text{RIA}} \cdot J(p_{z}) \label{RIA simplified}
\end{equation}

From Eq. (\ref{RIA}) and Eq. (\ref{RIA simplified}), it is obvious that the DDCS of atomic Compton scattering in the RIA approach factorizes into two parts
\begin{equation}
\bigg(\frac{d^{2}\sigma}{d\omega_{f}d\Omega_{f}}\bigg)_{\text{RIA}}=Y^{\text{RIA}} \cdot J(p_{z})
\label{RIA factorization}
\end{equation}
The factor $Y^{\text{RIA}}$ is dependent on the kinematical and dynamical properties of atomic Compton scattering, and it is irrelevant to the electronic structure of target materials. The correction factor $J(p_{z})$, known as the Compton profile, is related to the momentum distributions of electrons in the atomic or molecular ground state. In the RIA approach, all the atomic many-body effects can be incorporated into atomic Compton profiles.

Given the DDCS in atomic Compton scattering, the differential cross section with respect to the energy transfer can be calculated through the integration
\begin{eqnarray}
\bigg(
  \frac{d\sigma}{dT}
\bigg)_{\text{RIA}}
& = &
\int d\Omega_{f}
\bigg(
  \frac{d^{2}\sigma}{d\omega_{f}d\Omega_{f}}
\bigg)_{\text{RIA}}
=
\int d\Omega_{f}
\frac{r_{0}^{2}}{2}\frac{m_{e}}{q}
\frac{m_{e}c^{2}}{E(p_{z})}
\frac{\omega_{f}}{\omega_{i}}
\overline{X}(p_{z})
J(p_{z})
\label{RIA spectrum}
\\
\bigg(
  \frac{d\sigma}{dT}
\bigg)_{\text{RIA}}
& = &
\int d\Omega_{f}
\bigg(
  \frac{d^{2}\sigma}{d\omega_{f}d\Omega_{f}}
\bigg)_{\text{RIA}}
=
\int d\Omega_{f}
\frac{r_{0}^{2}}{2}\frac{m_{e}}{q}
\frac{\omega_{f}}{\omega_{i}}
X_{\text{KN}}
J(p_{z})
\end{eqnarray}
with $T=\omega_{i}-\omega_{f}$ to be the energy transfer in Compton scattering. With the atomic many-body effects incorporated into atomic Compton profiles, the RIA formulation could overcome the shortcomings in the FEA formulation. Therefore, it is a practical approach to calculate the Compton scattering in the low-energy transfer region or near photoionization threshold region. With the advantages of simplicity and flexibility, the RIA formulation has been widely applied to atomic \cite{Ribberfors1982}, condensed matter \cite{Kubo,Cooper}, nuclear and elementary particle physics \cite{Brusa,Ramanathan}. In particular, in the Monte Carlo simulation program Geant4 \cite{Geant4}, which are extensively used in nuclear and elementary particle physics, several algorithms employ the RIA approach to treat the Compton scattering process \cite{Geant4b,Livermore,Monash}. Furthermore, atomic Compton profile $J(p_{z})$ can also reflect some important information in condensed matter physics and material science, i.e. the electronic structure \cite{Gillet,Sahariya}, electron momentum distribution \cite{Gillet,Aguiar}, electron correlation \cite{Kubo,Pisani}, band structure, and Fermi surface \cite{Wang,Rathor}.

\section{The RIA Approach for the Atomic Ionization induced by Millicharged Particles \label{sec:4}}

In this section, we develop the RIA approach to the atomic ionization process induced by millicharged particles
\[
\chi + A \rightarrow \chi + A^{+} + e^{-}
\]
The derivation of the RIA formulation for the atomic ionization process is presented in detail. The differential cross section of the atomic ionization process is focused and discussed. The general results of the doubly-differential cross section (DDCS) are given in subsection \ref{sec:4a}, and simplified results of DDCS are given in subsection \ref{sec:4b}. In subsection \ref{sec:4c}, we give comments on our newly developed RIA approach for atomic ionization process induced by millicharged particles. Finally, the explicit expressions for differential cross section with respect to energy transfer are presented in \ref{sec:4d}.

\subsection{General Result of the Doubly-Differential Cross Section (DDCS) \label{sec:4a}}

For the atomic ionizations induced by millicharged particles $\chi + A \rightarrow \chi + A^{+} + e^{-}$, consider the millicharged particle with energy $E_{\chi}$ and momentum $\boldsymbol{k}_{\chi}$. After the ionization, the energy and momentum of millicharged particle become $E_{\chi}'$ and $\boldsymbol{k}'_{\chi}$, respectively. Similar to the cases discussed in section \ref{sec:2}, in RIA formulation, the electron in an atom has a momentum distribution $\rho(\boldsymbol{p}_{i})$, which is calculated through the electron momentum wavefunction of atomic ground state. Further, atomic electrons with different momentum $\boldsymbol{p}_{i}$ scatter independently with millicharged particle $\chi$, and interactions between electrons with different momentum are omitted. After the ionization process, final state electron gets momentum $\boldsymbol{p}_{f}$. Therefore, in the RIA approach, the DDCS of the atomic ionization process $\chi + A \rightarrow \chi + A^{+} + e^{-}$ is calculated by summing over contributions from all possible momentum $\boldsymbol{p}_{i}$:
\begin{eqnarray}
\bigg(\frac{d^{2}\sigma}{dE_{\chi}'d\Omega_{\chi}'}\bigg)_{\text{RIA}}
& = &
\frac{r_{0}^{2}m_{e}^{2}c^{4}}{2} \frac{E_{\chi}'}{E_{\chi}}
\iiint{d^{3}p_{i}\rho(\boldsymbol{p}_{i})
\frac{X}{E_{i}E_{f}}}
\delta (E_{i}+E_{\chi}-E_{f}-E_{\chi}')
\label{millicharge doubly differential cross section}
\end{eqnarray}
where $\Omega_{\chi}'$ is the solid angle for scattered millicharged particles, $E_{i}=\sqrt{p_{i}^{2}c^{2}+m^{2}c^{4}}$ and $E_{f}=\sqrt{p_{f}^{2}c^{2}+m^{2}c^{4}}$ are energies of initial and final state electrons, respectively. The function $X$ is proportional to the reaction probability of the scattering between millicharged particle and electron momentum eigenstate, namely the scattering process $\chi + e^{-} \rightarrow \chi + e^{-}$ with electron momentum eigenstate $|\boldsymbol{p}_{i}\rangle$. In the atomic Compton scattering, the function $X$ is given by Eq. (\ref{function X}) in section \ref{sec:2}.
%\begin{eqnarray}
%X = X(K_{i},K_{f}) & = & \frac{K_{i}}{K_{f}}+\frac{K_{f}}{K_{i}}
%                         +2m_{e}^{2}c^{2}
%                          \bigg(
%                            \frac{1}{K_{i}}-\frac{1}{K_{f}}
%                          \bigg)
%                         +m_{e}^{4}c^{4}
%                          \bigg(
%                            \frac{1}{K_{i}}-\frac{1}{K_{f}}
%                          \bigg)^{2}
%\end{eqnarray}
In the atomic ionization process induced by millicharged particles, the probability function $X$ should be calculated through the scattering amplitude of $\chi + e^{-} \rightarrow \chi + e^{-}$. This process is very similar to the Rutherford scattering process $p^{+} + e^{-} \rightarrow p^{+} + e^{-}$. In analogy with the Rutherford scattering, the function $X$ can be written as \cite{Schwartz}:
\begin{equation}
X = \delta_{\chi}^{2}\frac{u^{2}+s^{2}+4t(m_{e}^{2}c^{4}+m_{\chi}^{2}c^{4})-2(m_{e}^{2}c^{4}+m_{\chi}^{2}c^{4})^{2}}{t^{2}}
\label{function X millicharge}
\end{equation}
The function $X$ in Eq. (\ref{function X millicharge}) is directly obtained from the probability of Rutherford scattering with the replacement: $p \rightarrow \chi$, $m_{p} \rightarrow m_{\chi}$ and $e \rightarrow q_{\chi}=\delta_{\chi}e$. Here, $s$, $t$, $u$ are Mandelstam variables defined as:
\begin{subequations}
\begin{eqnarray}
s \equiv (p_{i}+k_{\chi})^{2}c^{2} & = & m_{e}^{2}c^{4}+m_{\chi}^{2}c^{4}
                                         + 2 \big( E_{i}E_{\chi}-\boldsymbol{p}_{i}\cdot\boldsymbol{k}_{\chi} \cdot c^{2} \big)
\\
t \equiv (k_{i}-k'_{\chi})^{2}c^{2} & = & 2m_{\chi}^{2}c^{4}
                                          - 2 \big( E_{\chi}E_{\chi}'-\boldsymbol{k}_{\chi}\cdot\boldsymbol{k}'_{\chi} \cdot c^{2} \big) \nonumber
\\
                                    & = & 2m_{\chi}^{2}c^{4} - 2 \big( E_{\chi}E_{\chi}'-k_{\chi}k'_{\chi}c^{2}\cos\theta \big)
\\
u \equiv (p_{i}-k'_{\chi})^{2}c^{2} & = & m_{e}^{2}c^{4}+m_{\chi}^{2}c^{4}
                                          - 2 \big( E_{i}E_{\chi}'-\boldsymbol{p}_{i}\cdot\boldsymbol{k}'_{\chi}\cdot c^{2} \big)
\end{eqnarray}
\end{subequations}
According to the property of Mandelstam variables $s+t+u=2m_{e}^{2}c^{4}+2m_{\chi}^{2}c^{4}$, the variable $u$ can be simplified as:
\begin{equation}
u = m_{e}^{2}c^{4}-m_{\chi}^{2}c^{4}
    - 2 \big( E_{i}E_{\chi}-\boldsymbol{p}_{i}\cdot\boldsymbol{k}_{\chi} \cdot c^{2} \big)
    + 2 \big( E_{\chi}E_{\chi}'-k_{\chi}k'_{\chi}c^{2}\cos\theta \big)
\end{equation}

\begin{figure}
\centering
\includegraphics[scale=0.5]{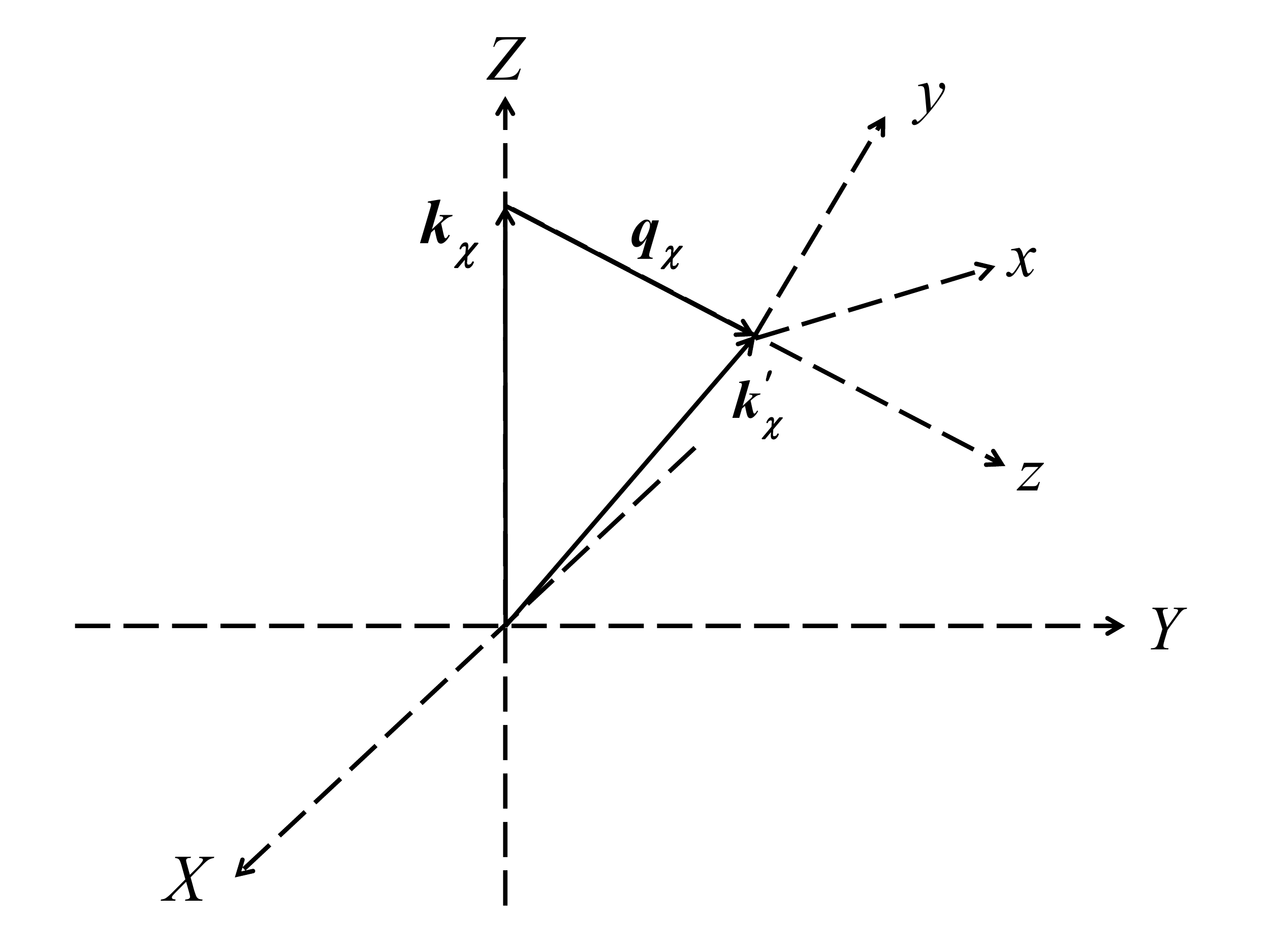}
\caption{In the atomic ionization process induced by millicharged particles, the coordinate systems $XYZ$ and $xyz$ are chosen similar to the cases in atomic Compton scattering (in figure \ref{coordinate0}). The coordinate system $XYZ$ is chosen such that the $Z$ axis is along the direction of initial momentum $\boldsymbol{k}_{\chi}$ for millicharged particle, and $X$ axis can be chosen as arbitrary direction perpendicular to the $Z$ axis. The momentum for the scattered millicharged particle is denoted as $\boldsymbol{k}'_{\chi}$, and the momentum transfer vector $\boldsymbol{q}_{\chi}$ in atomic ionization process is defined as $\boldsymbol{q}_{\chi} \equiv \boldsymbol{k}'_{\chi}-\boldsymbol{k}_{\chi}$. In this coordinate, the axis $z$ represents the momentum transfer direction.}
\label{coordinate}
\end{figure}

Choosing an appropriate coordinate system will benefit the numerical calculation. In this work, the coordinate system $xyz$ is chosen similar to the cases in atomic Compton scattering (in figure \ref{coordinate0}). In the $xyz$ system displayed in figure \ref{coordinate}, the $z$ axis represents the momentum transfer direction. After introducing such coordinate system, the momentum component $p_{z}$ is determined by energy and momentum conservations $p_{i}^{\mu}+k_{\chi}^{\mu}=p_{f}^{\mu}+(k'_{\chi})^{\mu}$. The explicit expression for $p_{z}$ is given by:
\begin{equation}
p_{z} = \frac{\boldsymbol{p}_{i}\cdot\boldsymbol{q}_{\chi}}{q_{\chi}}
      = \frac{E_{\chi}E_{\chi}'-E(p_{z})(E_{\chi}-E_{\chi}')}{c^{2}q_{\chi}}-\frac{k_{\chi}k'_{\chi}\cos\theta+m_{\chi}^{2}c^{2}}{q_{\chi}}
\end{equation}
with $q_{\chi}$ to be the modulus of the momentum transfer $\boldsymbol{q}_{\chi}=\boldsymbol{k}'_{\chi}-\boldsymbol{k}_{\chi}$ in the scattering process
\begin{equation}
q_{\chi} = \sqrt{k_{\chi}^{2}+(k'_{\chi})^{2}-2k_{\chi}k'_{\chi}\cos\theta}
\end{equation}
and $E(p_{z})=\sqrt{m_{e}^{2}c^{4}+p_{z}^{2}c^{2}}$. Assuming that millicharged particles are massive particles, then the initial and final state momentum $k_{\chi}$ and $k'_{\chi}$ can be calculated as:
\begin{equation}
k_{\chi}=\frac{\sqrt{E_{\chi}^{2}-m_{\chi}^{2}c^{4}}}{c}, \ \ \ \ \ k'_{\chi}=\frac{\sqrt{{E_{\chi}'}^{2}-m_{\chi}^{2}c^{4}}}{c}
\end{equation}
Similar to the cases in Compton scattering, from the energy and momentum conservations, it can be revealed that $p_{z}$ and $E(p_{z})$ are exactly the minimal energy and momentum of the initial state electrons activated in the ionization process, namely
\begin{equation}
p_{i}^{\text{min}}=|p_{z}|;\ \ E_{i}^{\text{min}}=E(p_{z}) \label{minimal momentum}
\end{equation}
Furthermore, the momentum component $p_{z}$ can be approximated as:
\begin{equation}
p_{z} \approx \frac{E_{\chi}E_{\chi}'-m_{e}c^{2}(E_{\chi}-E_{\chi}')}{c^{2}q_{\chi}}-\frac{k_{\chi}k'_{\chi}\cos\theta+m_{\chi}^{2}c^{2}}{q_{\chi}}
\label{apprximate pz}
\end{equation}
However, it should be noted that Eq. (\ref{apprximate pz}) only valid when $p_{z}$ is sufficiently small.

Similar to the cases in Compton scattering, the probability function $X$ in the integrand of Eq. (\ref{millicharge doubly differential cross section}) is can be averaged and pulled out of the integration as Roland Ribberfors et al. did in Compton scattering in reference \cite{Ribberfors1975,Ribberfors1982} (see Eq. (\ref{function X Pmin}) in section \ref{sec:2}).
Concretely, we can take the following approximation:
\begin{eqnarray}
X & \approx & \overline{X}(\overline{s}(p_{z}),\overline{t}(p_{z}),\overline{u}(p_{z})) \nonumber
\\
& = & \delta_{\chi}^{2}
      \frac{\overline{u}(p_{z})^{2}+\overline{s}(p_{z})^{2}+4\overline{t}(p_{z})(m_{e}^{2}c^{4}+m_{\chi}^{2}c^{4})-2(m_{e}^{2}c^{4}+m_{\chi}^{2}c^{4})^{2}}
           {\overline{t}(p_{z})^{2}}
\label{kernel function approximation}
\end{eqnarray}
And the corresponding values of Mandelstam variables can be expressed by:
\begin{subequations}
\begin{eqnarray}
\overline{s}(p_{z}) & = & m_{e}^{2}c^{4}+m_{\chi}^{2}c^{4}
                          + 2 \bigg[ E(p_{z})E_{\chi}+\frac{k_{\chi}(k_{\chi}-k'_{\chi}\cos\theta)p_{z}c^{2}}{q_{\chi}} \bigg]
\\
\overline{t}(p_{z}) & = & t = 2m_{\chi}^{2}c^{4} - 2 \big( E_{\chi}E_{\chi}'-k_{\chi}k'_{\chi}c^{2}\cos\theta \big)
\\
\overline{u}(p_{z}) & = & m_{e}^{2}c^{4}-m_{\chi}^{2}c^{4}
                          + 2 \big( E_{\chi}E_{\chi}'-k_{\chi}k'_{\chi}c^{2}\cos\theta \big)
                          - 2 \bigg[ E(p_{z})E_{\chi}+\frac{k_{\chi}(k_{\chi}-k'_{\chi}\cos\theta)p_{z}c^{2}}{q_{\chi}} \bigg] \nonumber
\\
\end{eqnarray}
\end{subequations}
Obviously, the above approximation of probability function $X$ in Eq. (\ref{kernel function approximation}) made from Eq. (\ref{function X millicharge}) indicates that the electron initial momentum $\boldsymbol{p}_{i}$ is specified only in the momentum transfer direction $z$, while momentum components in other directions $p_{x}$ and $p_{y}$ are omitted for simplicity.

Based on the above assumptions, we substitute the approximation (\ref{kernel function approximation}) into Eq. (\ref{millicharge doubly differential cross section}) and simplify the energy of electron as $E_{i} \approx E_{i}^{\text{min}}=E(p_{z})$. Finally, the DDCS of atomic ionization process induced by millicharged particles can be expressed as:
\begin{eqnarray}
\bigg(\frac{d^{2}\sigma}{dE_{\chi}'d\Omega_{\chi}'}\bigg)_{\text{RIA}}
& = & \frac{r_{0}^{2}}{2}\frac{m_{e}}{q_{\chi}}
      \frac{m_{e}c^{2}}{E(p_{z})}
      \frac{E_{\chi}'}{E_{\chi}}
      \overline{X}(\overline{s}(p_{z}),\overline{t}(p_{z}),\overline{u}(p_{z}))
      J(p_{z}) \nonumber
\\
& = & \overline{Y}^{\text{RIA}} \cdot J(p_{z})
\label{RIA millicharge}
\end{eqnarray}

\subsection{Simplified Result of the Doubly-Differential Cross Section (DDCS) \label{sec:4b}}

In this subsection, we provide a simpler version in the calculation of DDCS in atomic ionization process induced by millicharged particles. Simpler results of DDCS can be achieved by making more simplified approximation for probability function $X$ in the calculation of Eq. (\ref{millicharge doubly differential cross section}). For instance, similar to the cases in atomic Compton scattering, an alternative and simpler approximation of function $X$ in Eq. (\ref{kernel function approximation}) can be made by taking the $p_{z}\rightarrow0$ limit of $\overline{X}(\overline{s}(p_{z}),\overline{t}(p_{z}),\overline{u}(p_{z}))$, which finally gives:
\begin{eqnarray}
X & \approx & X_{\text{sim}}
       =      \delta_{\chi}^{2}
              \frac{u_{\text{sim}}^{2}+s_{\text{sim}}^{2}+4t_{\text{sim}}(m_{e}^{2}c^{4}+m_{\chi}^{2}c^{4})-2(m_{e}^{2}c^{4}+m_{\chi}^{2}c^{4})^{2}}
                   {t_{\text{sim}}^{2}}
\label{kernel function approximation2}
\end{eqnarray}
Correspondingly, the 3 Mandelstam variables $s$, $t$, $u$ can be further simplified as:
\begin{subequations}
\begin{eqnarray}
s_{\text{sim}} & = & m_{e}^{2}c^{4}+m_{\chi}^{2}c^{4}+ 2m_{e}c^{2}E_{\chi}
\\
t_{\text{sim}} & = & t = 2m_{\chi}^{2}c^{4} - 2 \big( E_{\chi}E_{\chi}'-k_{\chi}k'_{\chi}c^{2}\cos\theta \big)
\\
u_{\text{sim}} & = & m_{e}^{2}c^{4}-m_{\chi}^{2}c^{4}-2m_{e}c^{2}E_{\chi}
                          + 2 \big( E_{\chi}E_{\chi}'-k_{\chi}k'_{\chi}c^{2}\cos\theta \big)
\end{eqnarray}
\end{subequations}
Using the approximation (\ref{kernel function approximation2}), the DDCS of atomic ionization process induced by millicharged particles can be further simplified as:
\begin{equation}
\bigg(\frac{d^{2}\sigma}{dE_{\chi}'d\Omega_{\chi}'}\bigg)_{\text{RIA}}
=\frac{r_{0}^{2}}{2}
 \frac{m_{e}}{q_{\chi}}\frac{E_{\chi}'}{E_{\chi}}
   X_{\text{sim}}
   J(p_{z})
=Y_{\text{sim}}^{\text{RIA}} \cdot J(p_{z}) \label{RIA simplified millicharge}
\end{equation}

\subsection{Some Comments \label{sec:4c}}

From the above results in Eqs. (\ref{RIA millicharge}) and (\ref{RIA simplified millicharge}), it is evident that the DDSC of the atomic ionization process induced by millicharged particles can be summarized as:
\begin{equation}
\bigg(\frac{d^{2}\sigma}{dE_{\chi}'d\Omega_{\chi}'}\bigg)_{\text{RIA}}
=Y^{\text{RIA}} \cdot J(p_{z}) \label{RIA factorization millicharge}
\end{equation}
This results of DDCS for atomic ionization process induced by millicharged particle in Eq. (\ref{RIA factorization millicharge}) is similar to the cases in atomic Compton scattering introduced in section \ref{sec:3} (the Eq. (\ref{RIA factorization})). From Eq. (\ref{RIA millicharge}), it is clearly that the DDCS of the atomic ionization process also factorizes into two parts: the factor $Y^{\text{RIA}}$ and the atomic Compton profile $J(p_{z})$. The kinematical and dynamical property of the atomic ionization process is incorporated in factor $Y^{\text{RIA}}$, irrespective of the material elements and electron structures in atomic systems. The correction from the atomic effects and electronic structures is mainly incorporated into the atomic Compton profile $J(p_{z})$ as in atomic Compton scattering.

It should be noted that the above results in Eqs. (\ref{millicharge doubly differential cross section}) -- (\ref{RIA simplified millicharge}) only correspond to the single electron systems. However, the detector atom is usually a multi-electron system and consists of electrons form different subshells. Summing over contributions from all subshell electrons, the DDCS of the multi-electron atomic system can be calculated as:
\begin{eqnarray}
\bigg(\frac{d^{2}\sigma}{dE_{\chi}'d\Omega_{\chi}'}\bigg)_{\text{RIA}}
& = &
\frac{r_{0}^{2}}{2}
\frac{E_{\chi}'}{E_{\chi}}
\frac{m_{e}}{q_{\chi}}
\frac{m_{e}c^{2}}{E(p_{z})}
\overline{X}(\overline{s}(p_{z}),\overline{t}(p_{z}),\overline{u}(p_{z})) \nonumber
\\
&   &
\times \sum_{njl}Z_{njl}J_{njl}(p_{z})\Theta(E_{\chi}-E'_{\chi}-E_{njl}^{B}) \nonumber
\\
& = &
\frac{r_{0}^{2}}{2}
\frac{E_{\chi}'}{E_{\chi}}
\frac{m_{e}}{q_{\chi}}
\bigg[ 1+ \bigg( \frac{p_{z}c}{m_{e}c^{2}} \bigg)^{2} \bigg]^{-1/2}
\overline{X}(\overline{s}(p_{z}),\overline{t}(p_{z}),\overline{u}(p_{z})) \nonumber
\\
&   &
\times \sum_{njl}Z_{njl}J_{njl}(p_{z})\Theta(E_{\chi}-E'_{\chi}-E_{njl}^{B})
\label{RIA millicharge full}
\\
\bigg(\frac{d^{2}\sigma}{dE_{\chi}'d\Omega_{\chi}'}\bigg)_{\text{RIA}}
& = &
\frac{r_{0}^{2}}{2}
\frac{m_{e}}{q_{\chi}}\frac{E_{\chi}'}{E_{\chi}}
X_{\text{sim}}
\times \sum_{njl}Z_{njl}J_{njl}(p_{z})\Theta(E_{\chi}-E'_{\chi}-E_{njl}^{B})
\label{RIA simplified millicharge full}
\end{eqnarray}
In the above expressions, $J_{njl}(p_{z})$ is the atomic Compton profile for subshell ($njl$)
\begin{equation}
J_{njl}(p_{z})\equiv\iint\rho_{njl}(\boldsymbol{p})dp_{x}dp_{y} \label{subshell Compton profile}
\end{equation}
The $E_{njl}^{B}$ is the atomic binding energy of subshell ($njl$), $Z_{njl}$ is the number of electron in subshell ($njl$) \footnote{In relativistic atomic theories, due to spin-orbit couplings, the electron state in a spherical symmetrical system is specified by quantum number $(njl)$ or $(n\kappa)$. Furthermore, if magnetic quantum number is taken into account, the quantum number of electron state becomes $(njlm_{j})$ or $(n\kappa m_{j})$. This is different with the quantum number ($nlm_{l}$) in the non-relativistic atomic theories. More details can be found in appendix \ref{appendix3} (see Eq. (\ref{Dirac orbital}) in appendix \ref{appendix3}).}, and $\Theta(E_{\chi}-E'_{\chi}-E_{njl}^{B})$ is the Heaviside step function
\begin{equation}
\Theta(x) \equiv \bigg\{
                   \begin{array}{cc}
                     1 & \ x \geq 0 \\
                     0 & \ x < 0
                   \end{array}
\label{Heaviside}
\end{equation}
When the energy transfer $T=E_{\chi}-E_{\chi}'$ is less than the subshell binding energy $E_{njl}^{B}$, electron in subshell $(njl)$ is inactive in the atomic ionization process induced by millicharged particles $\chi + A \rightarrow \chi + A^{+} + e^{-}$, and this subshell gives zero contributions in the DDCS.

\subsection{Differential Cross Section with respect to Energy Transfer \label{sec:4d}}

Similarly, given the DDCS of the atomic ionization process induced by millicharged particles, the differential cross section with respect to the energy transfer $T$ in this process can be obtained through the integration
\begin{eqnarray}
\bigg(
  \frac{d\sigma}{dT}
\bigg)_{\text{RIA}}
& = &
\int d\Omega_{\chi}'
\bigg(
  \frac{d^{2}\sigma}{dE_{\chi}'d\Omega_{\chi}'}
\bigg)_{\text{RIA}}
\label{RIA Energy Spectrum millicharge}
\end{eqnarray}
with $T=E_{\chi}-E'_{\chi}$ to be the energy transfer for atomic ionization process induced by millicharged particles. Put the DDCS in Eq (\ref{RIA millicharge full}) and Eq. (\ref{RIA simplified millicharge full}) into the integration, we finally get the explicit expressions for differential cross section with respect to energy transfer
\begin{eqnarray}
\bigg(
  \frac{d\sigma}{dT}
\bigg)_{\text{RIA}}
& = &
\int d\Omega_{\chi}'
\bigg(
  \frac{d^{2}\sigma}{dE_{\chi}'d\Omega_{\chi}'}
\bigg)_{\text{RIA}} \nonumber
\\
& = &
\int d\Omega_{\chi}'
\bigg\{
\frac{r_{0}^{2}}{2}
\frac{E_{\chi}'}{E_{\chi}}
\frac{m_{e}}{q_{\chi}}
\bigg[ 1+ \bigg( \frac{p_{z}c}{m_{e}c^{2}} \bigg)^{2} \bigg]^{-1/2}
\overline{X}(\overline{s}(p_{z}),\overline{t}(p_{z}),\overline{u}(p_{z})) \nonumber
\\
&   &
\times \sum_{njl}Z_{njl}J_{njl}(p_{z})\Theta(E_{\chi}-E'_{\chi}-E_{njl}^{B})
\bigg\}
\label{RIA millicharge full DCS}
\\
\bigg(
  \frac{d\sigma}{dT}
\bigg)_{\text{RIA}}
& = &
\int d\Omega_{\chi}'
\bigg(
  \frac{d^{2}\sigma}{dE_{\chi}'d\Omega_{\chi}'}
\bigg)_{\text{RIA}} \nonumber
\\
& = &
\int d\Omega_{\chi}'
\bigg\{
\frac{r_{0}^{2}}{2}
\frac{m_{e}}{q_{\chi}}\frac{E_{\chi}'}{E_{\chi}}
X_{\text{sim}}
\times \sum_{njl}Z_{njl}J_{njl}(p_{z})\Theta(E_{\chi}-E'_{\chi}-E_{njl}^{B})
\bigg\}
\label{RIA simplified millicharge full DCS}
\end{eqnarray}

To summarise, this section gives the theoretical derivation of our RIA approach in the atomic ionization process induced by millicharged particles. A promising feature is that our approach is quite general, depend neither on the underling nature or mechanism of millicharged particles, nor on the composition of detector materials. Therefore, it can be extensively applied to the studies of millicharged particles. In this work, we also develop a numerical program based on the above approach. The numerical calculations are presented in the next two sections for millicharged dark matter particles and millicharged neutrinos.

\section{Numerical Results and Discussions on Millicharged Dark Matter Particles \label{sec:5}}

This section is devoted to the numerical results of the atomic ionization process induced by millicharged dark matter particles. Based on our RIA approach derived in section \ref{sec:4}, a numerical program is developed utilizing the basic Fortran language. In subsection \ref{sec:5a}, the differential cross sections with respect to energy transfer are obtained for Ge and Xe atom, and results from our RIA approach are compared with those from FEA and EPA approaches. In subsection \ref{sec:5b}, the differential reaction event rates in HPGe and LXe detectors are given for typical experimental environments. Furthermore, in subsection \ref{sec:5c}, we give an estimation of the detecting sensitivities on dark matter particle millicharge $\delta_{\chi}$ in next-generation HPGe and LXe based experiments. These numerical results presented in this section can shed light on theoretical investigations as well as experimental explorations for millicharged dark matter particles.

\subsection{Differential Cross Section \label{sec:5a}}

In this subsection, we provide numerical calculations on differential cross section with respect to energy transfer for the atomic ionization process induced by millicharged dark matter particles.

\begin{figure}
\centering
\includegraphics[scale=0.6]{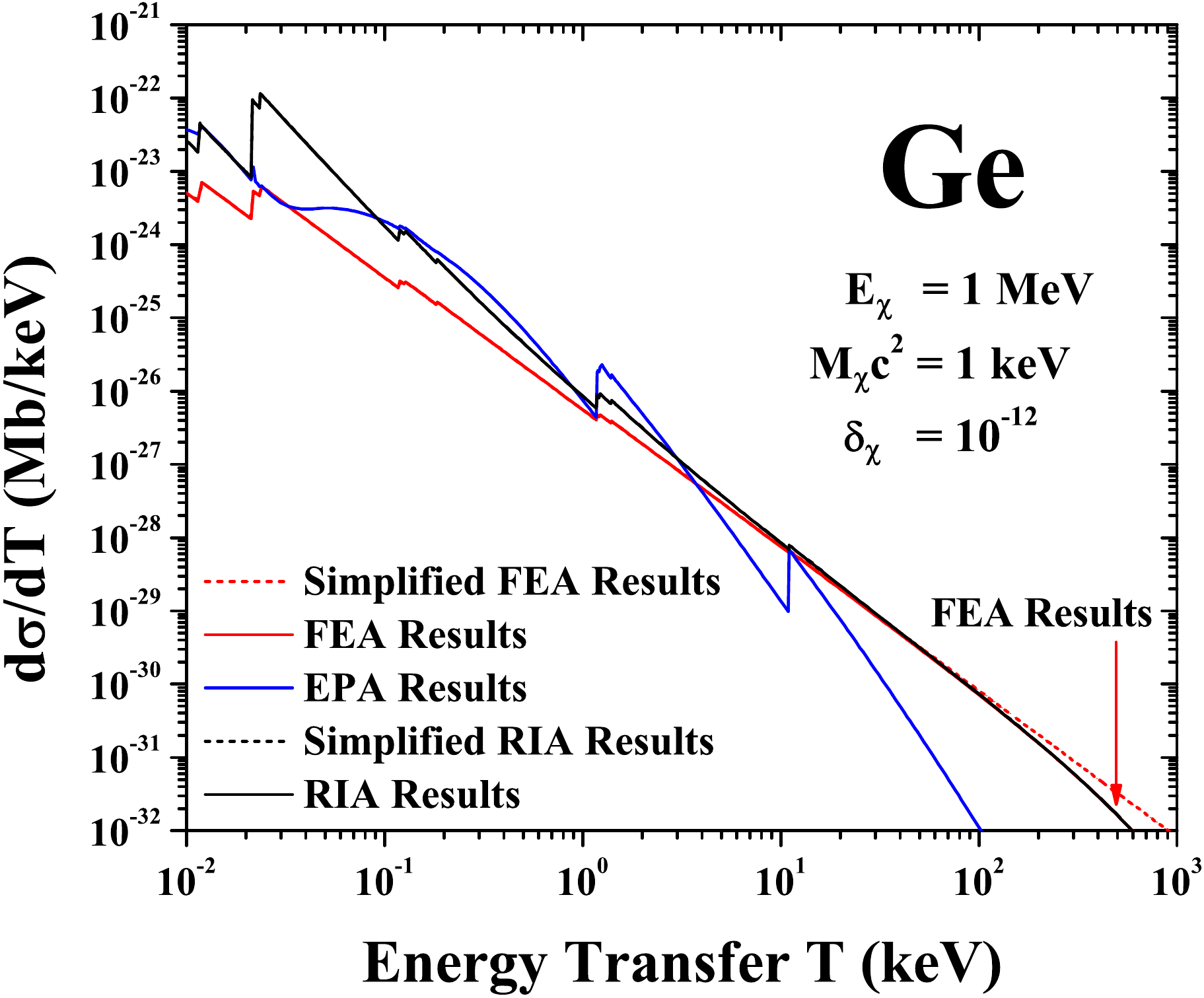}
\includegraphics[scale=0.6]{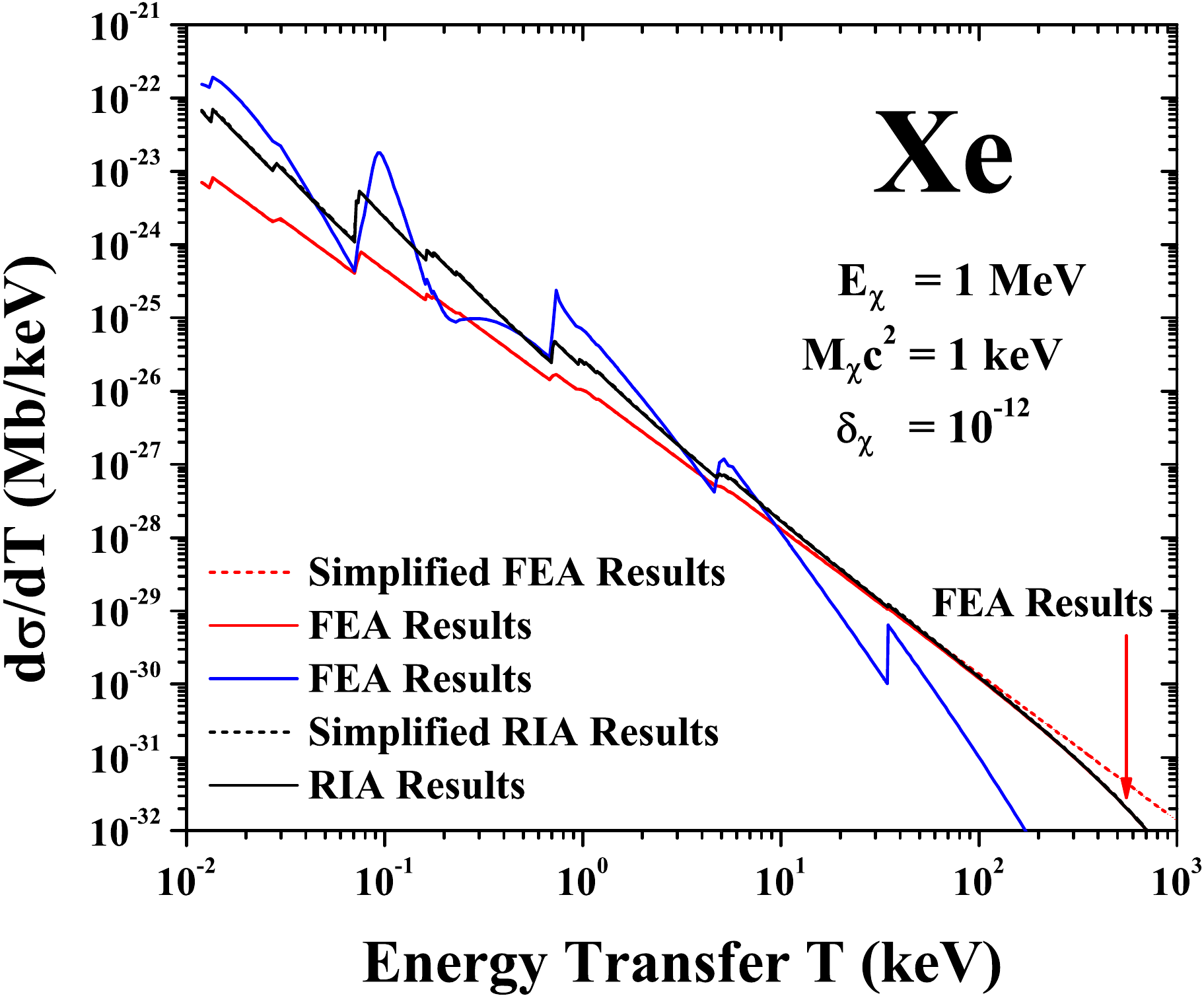}
\caption{Differential cross sections of atomic ionization process for Ge and Xe atoms induced by millicharged dark matter particles. The mass and initial energy of millicharged particle are chosen as $m_{\chi}c^{2}=1$ keV and $E_{\chi}=1$ MeV, and the dark matter particle millicharge is chosen to be $\delta_{\chi}=10^{-12}$. In this figure, we compare the numerical results on differential cross section $d\sigma/dT$ calculated in the FEA, EPA and RIA approaches. The red solid lines correspond to the FEA results calculated through Eq. (\ref{millicharge FEA}); red dashed lines represent the simplified FEA results calculated through Eq. (\ref{millicharge FEA2}); blue lines stand for the EPA results calculated from Eq. (\ref{millicharge EPA}); black solid lines show the RIA results calculated using Eq. (\ref{RIA millicharge full DCS}); and black dashed lines present the simplified RIA results calculated using Eq. (\ref{RIA simplified millicharge full DCS}). }
\label{Millicharge cross section figure}
\end{figure}

Figure \ref{Millicharge cross section figure} shows the differential cross section $d\sigma/dT$ for atomic ionization process induced by high-energy millicharged dark matter particles. In this figure, the mass and initial energy of millicharged particle are chosen as $m_{\chi}c^{2}=1$ KeV and $E_{\chi}=1$ MeV, respectively. The millicharge of dark matter particle is chosen to be $\delta_{\chi}=10^{-12}$ as a typical example. The numerical results come from FEA, EPA and RIA approaches are given in this figure for comparisons. Among these approaches, the FEA results calculated through Eq. (\ref{millicharge FEA}) neglect all the atomic many-body effects, and they could provide a approximate result only in the high-energy transfer region, in which the atomic electron is nearly free and atomic binding effects become very weak. However, FEA results fail to give a precise prediction in the low-energy transfer region because atomic many-body effects have a strong effect on the atomic ionization process. The simplified FEA results, which are calculated through Eq. (\ref{millicharge FEA2}), can be reduced from FEA results when $m_{\chi} \ll m_{e}$, $T \ll m_{e}c^{2}$, $T \ll E_{\chi}$ are satisfied. The EPA results calculated using Eq. (\ref{millicharge EPA}) could provide a more precise results than the FEA results in the ultra-low-energy transfer region by including atomic many-body effects partly. The EPA approach can be derived from quantum field theory when energy and momentum transfer are extremely small, namely in the $T \rightarrow 0$ limit, and it becomes invalid in high-energy region. The introduction of FEA and EPA approaches is give in the appendices. Our RIA approaches developed in this work could deal with atomic many-body effects in the entire region, regardless of the underlining nature of millicharged particles and the composition of detector materials.

From figure \ref{Millicharge cross section figure}, it can be clearly manifested that the differential cross sections $d\sigma/dT$ from FEA, EPA and RIA calculations all diminish as energy transfer $T$ increases. In the low-energy transfer region, both EPA and RIA results acquire larger cross sections than FEA results, indicating that the atomic many-body effects, including the atomic binding, electron shielding and electron correlation, could greatly enhance the atomic ionization process induced by millicharged particles and enlarge their differential cross sections. Particularly, the EPA results for Xe atom present large peak when energy transfer $T \sim 100$ eV. This is because in the EPA approach, the differential cross section $d\sigma/dT$ for atomic ionization process induced by millicharged particles is proportional to the photonabsorption cross section, as shown in appendix \ref{appendix2} (see Eq. (\ref{millicharge EPA})). For photoabsorption cross section, there is a giant resonance for $4d$ electrons of Xe atom in the 100 eV region \cite{Johnson1992,Andersen,Toffoli,Kumar2009,Qiao2019} \footnote{For Xe atom, there is also a peak in $T \sim 700$ eV region due to the resonance for $3d$ electrons of Xe atom in the photoabsorption cross section \cite{Amusia}. The peak of $3d$ electrons for Xe atom in $T \sim 700$ eV is relatively smaller than that of $4d$ electrons in $T \sim 100$ eV}. For our RIA results, in the low-energy transfer region, our RIA results get larger cross section than those from FEA results; while in the high-energy transfer region, the RIA results do not exhibit notable differences with respective to FEA results. The physical reason can be explained naturally: when energy transfer $T$ is much larger than the atomic binding energy  $E^{B}_{1s}$ for $1s$ electron (which is 11.1 keV for Ge atom and 34.5 keV for Xe atom), the atomic effects can be neglected and the atomic electron is approximately free. However, when energy transfer $T$ is sufficient low and is comparable to the atomic binding energy $E^{B}_{1s}$, atomic binding, electron shielding as well as electron correlation effects become dominant. In these cases, atomic electrons can no longer be treated as free electrons, which lead to large deviations between RIA and FEA results in the low-energy transfer region. Furthermore, figure \ref{Millicharge cross section figure} also indicates that our RIA results are approaching to the EPA results when energy transfer $T$ is extremely small, especially in the $T \rightarrow 0$ limit. It can be viewed as a demonstration for the validity and availability of our RIA approach developed in the present work.

Figure \ref{Millicharge cross section figure} also manifested that, for large incident particle energy $E_{\chi}=1$ MeV, the simplified RIA results calculated using Eq. (\ref{RIA simplified millicharge full DCS}) converge to the RIA results calculated using Eq. (\ref{RIA millicharge full DCS}) in the entire region of energy transfer $T$. Therefore, for high-energy millicharged dark matter particles, among the approximations of probability function $X$ in the integrand in Eq. (\ref{millicharge doubly differential cross section}), the more simplified approximation $X \approx X_{\text{sim}}$ is good enough, and it does not lead to notable deviations compared with the more accurate approximation $X \approx \overline{X}(\overline{s}(p_{z}),\overline{t}(p_{z}),\overline{u}(p_{z}))$.

\begin{figure*}
\centering
\includegraphics[scale=0.4]{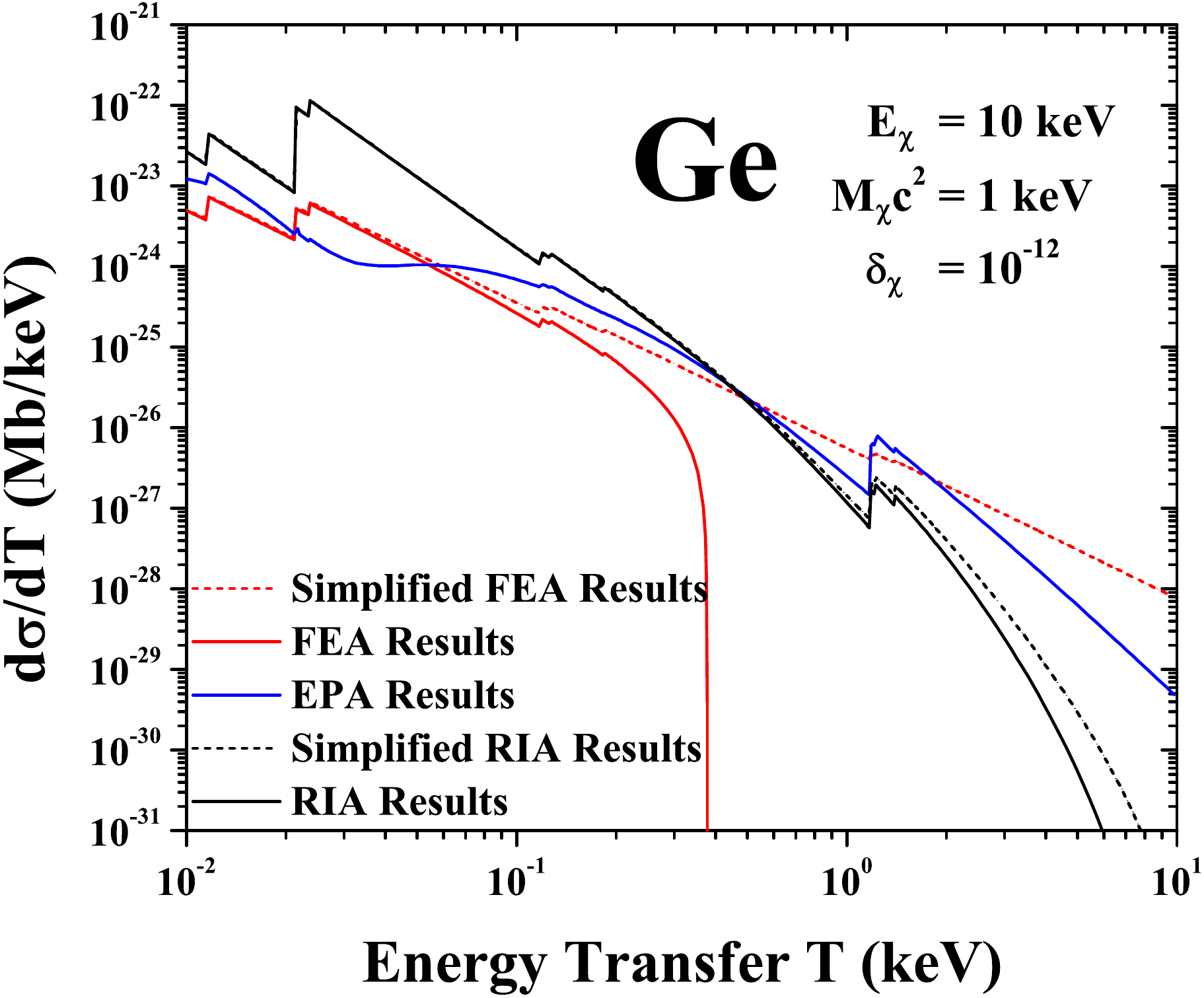}
\includegraphics[scale=0.4]{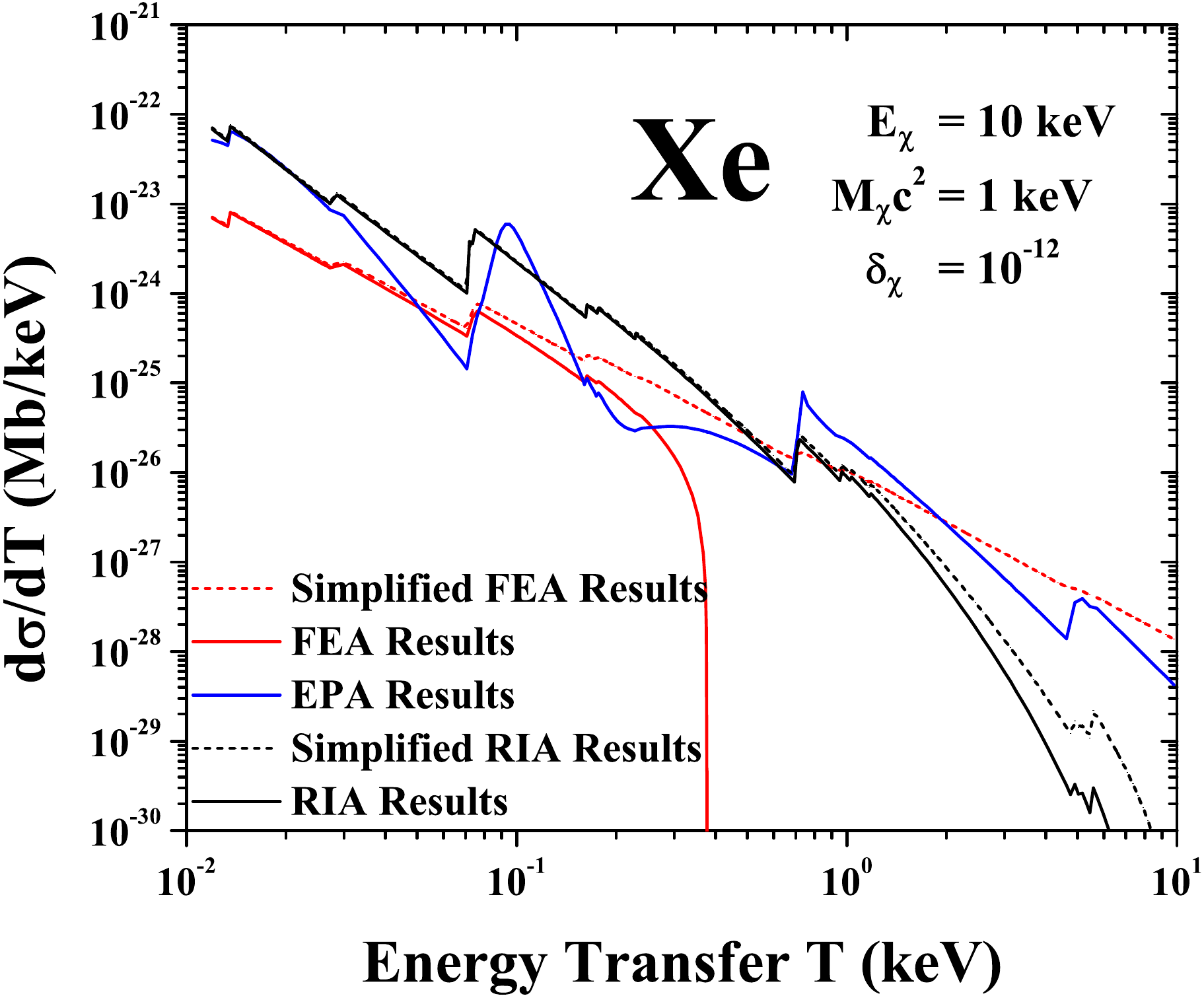}
\includegraphics[scale=0.4]{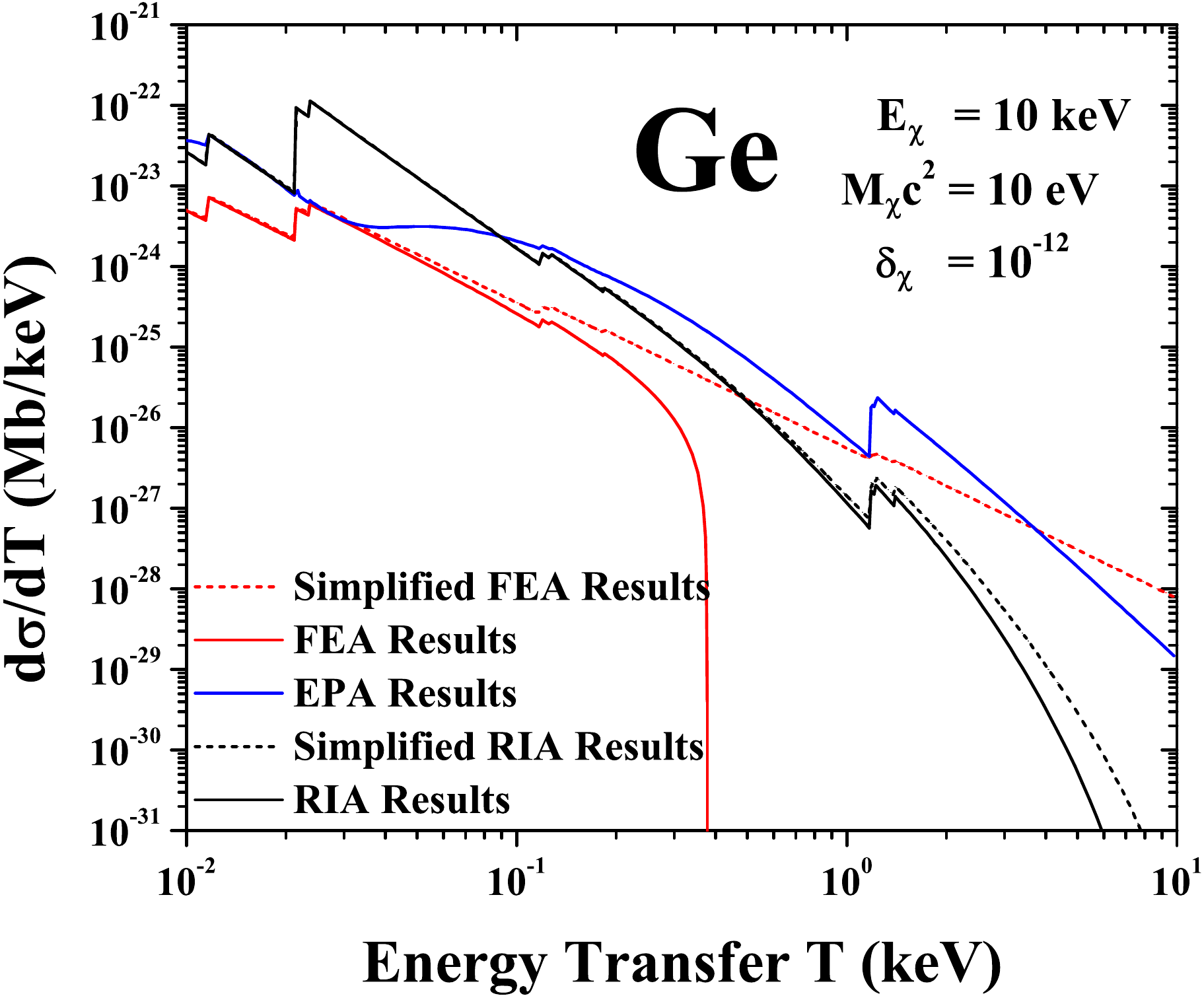}
\includegraphics[scale=0.4]{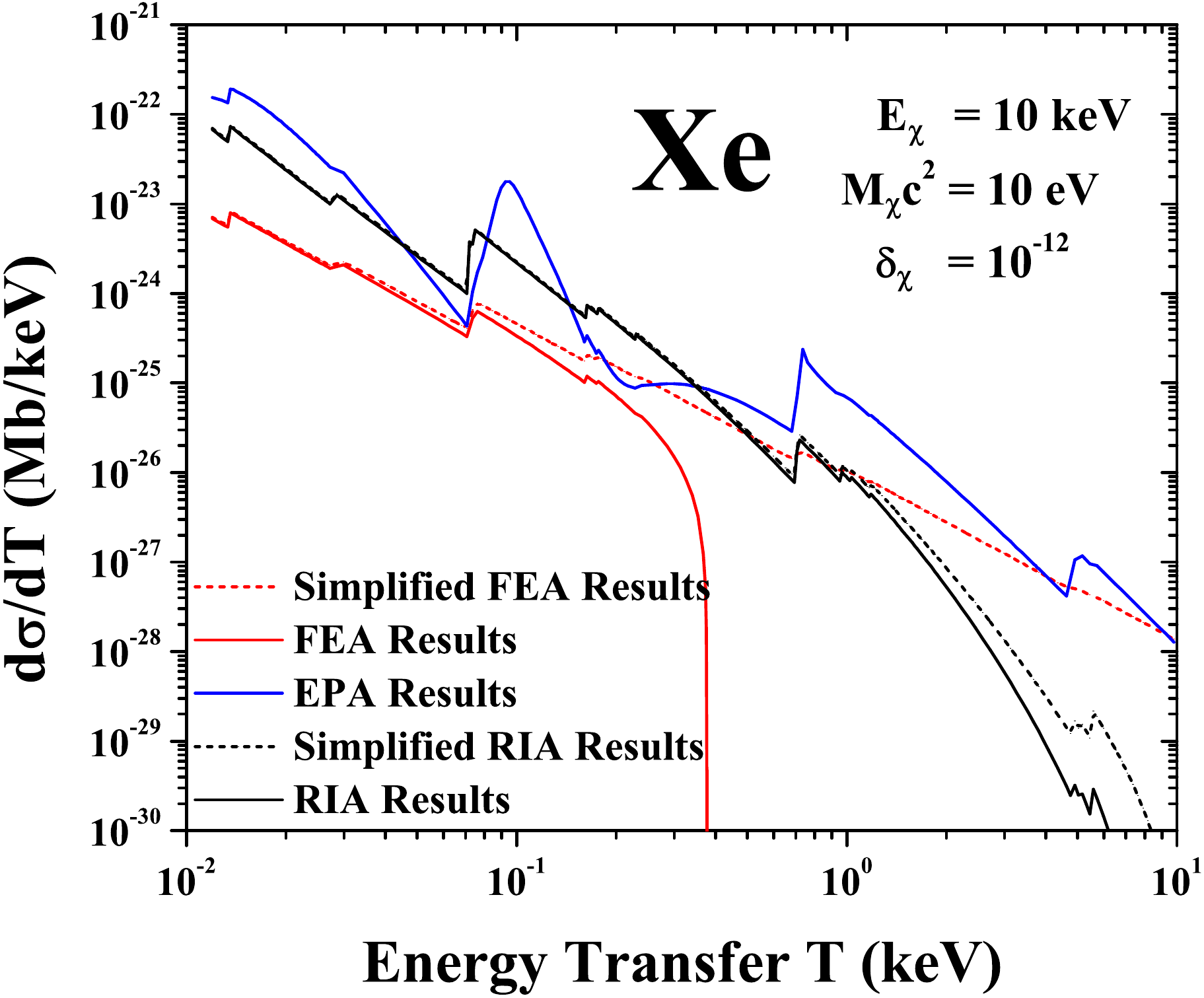}
\caption{Differential cross sections for atomic ionization process of Ge and Xe atoms induced by millicharged dark matter particles. In this figure, the mass and initial energy of millicharged particle is chosen as $m_{\chi}c^{2}=1$ keV, $E_{\chi}=10$ keV, and the dark matter particle millicharge is chosen to be $\delta_{\chi}=10^{-12}$. The upper and lower panels show the cases of different millicharged particle mass: the upper panel correspond to $m_{\chi}c^{2}=1$ keV, and the lower panel correspond to $m_{\chi}c^{2}=10$ eV. In this figure, we compare the numerical results on differential cross section $d\sigma/dT$ calculated in the FEA, EPA and RIA approaches. The red solid lines correspond to the FEA results calculated through Eq. (\ref{millicharge FEA}); red dashed lines represent the simplified FEA results calculated through Eq. (\ref{millicharge FEA2}); blue lines stand for the EPA results calculated from Eq. (\ref{millicharge EPA}); black solid lines show the RIA results calculated using Eq. (\ref{RIA millicharge full DCS}); and black dashed lines present the simplified RIA results calculated using Eq. (\ref{RIA simplified millicharge full DCS}). }
\label{Millicharge cross section figure0}
\end{figure*}

In figure \ref{Millicharge cross section figure0}, we given the differential cross sections for low-energy millicharged dark matter particles. The initial energy of millicharged dark matter particle is chosen as $E_{\chi}=10$ keV, while the dark matter particle millicharge is chosen to be $\delta_{\chi}=10^{-12}$ the same as in figure \ref{Millicharge cross section figure}. The upper and lower panels correspond to $m_{\chi}c^{2}=1$ keV and $m_{\chi}c^{2}=10$ eV, respectively. In these cases, the incoming particle energy $E_{\chi}$ is not large enough to make the atomic many-body effects negligible. Therefore, our RIA results does not converge to the FEA results in the entire region of $T$, but our RIA results still come close to the EPA results in the ultra-low-energy $T \rightarrow 0$ limit. From this figure, we can also observe that the EPA results acquire larger cross sections for smaller millicharged dark matter particle mass $m_{\chi}$. There is another notable point should be noted: large differences between the simplified FEA results and the FEA results emerge in energy range $T>0.1$ keV. In the appendix \ref{appendix1}, it would be clarified that the simplified FEA results on differential cross section calculated using Eq. (\ref{millicharge FEA2}) converge to the full FEA results calculated using Eq. (\ref{millicharge FEA}) only when $m_{\chi} \ll m_{e}$, $T \ll m_{e}c^{2}$, $T \ll E_{\chi}$ are satisfied. For low-energy millicharged dark matter particles, e.g. for incident particle energy $E_{\chi}=10$ keV in figure \ref{Millicharge cross section figure0}, the condition $T \ll E_{\chi}$ is not satisfied when $T>0.1$ keV. Therefore, only FEA results calculated through Eq. (\ref{millicharge FEA}) are reasonable in such cases. Compare with the FEA results, the simplified FEA results calculated through Eq. (\ref{millicharge FEA2}) overestimate the differential cross sections of the atomic ionization process for Ge and Xe atoms. Furthermore, for small incident particle energy $E_{\chi}=10$ keV, there are discrepancies between simplified RIA results and RIA results when energy transfer $T>1$ keV. For low-energy millicharged dark matter particles, the more simplified approximation $X \approx X_{\text{sim}}$ of probability function $X$ could bring about some deviations, and it is better to use the more accurate approximation $X \approx \overline{X}(\overline{s}(p_{z}),\overline{t}(p_{z}),\overline{u}(p_{z}))$ to evaluate the differential cross section $d\sigma/dT$.

Particularly, results in figure \ref{Millicharge cross section figure} and figure \ref{Millicharge cross section figure0} could reflect some sort of generality. In various approaches, i.e. FEA, EPA and RIA, the differential cross section $d\sigma/dT$ for the atomic ionization process induced by millicharged dark matter particles is proportional to $\delta_{\chi}^{2}$. For the same incoming energy, results correspond to other millicharge $\delta_{\chi}$ can be obtained from figure \ref{Millicharge cross section figure} and figure \ref{Millicharge cross section figure0} by proportional magnifying or shrinking the results by $\delta_{\chi}^{2}$ times.

\subsection{Reaction Event Rate in HPGe and LXe Detectors \label{sec:5b}}

In this subsection, we shall give the numerical calculations of differential reaction event rate for atomic ionization process induced by millicharged particles in typical super-terranean or underground experiments.

In a typical experimental environment, such as CDEX experiment located in CJPL as well as other super-terranean or underground experiments, the differential reaction event rate in detectors for atomic ionization process induced by millicharged particles can be expressed as:
\begin{equation}
\frac{dR}{dT} = \rho_{A} \int_{E_{\chi}^{\text{min}}}^{E_{\chi}^{\text{max}}} dE_{\chi} \frac{d\sigma}{dT} \frac{d\phi_{\chi}}{dE_{\chi}}
\label{millicharge count rate}
\end{equation}
where $\rho_{A}$ is the number density of detector atoms, $\phi_{\chi}$ is the total flux of millicharged dark matter particles, and $d\phi_{\chi}/dE_{\chi}$ is the flux spectrum at a given incoming energy $E_{\chi}$. In Eq. (\ref{millicharge count rate}), $E_{\chi}^{\text{min}}$ and $E_{\chi}^{\text{max}}$ are the maximal and minimal energy of the millicharged dark matter particles that could enter into the detectors. From the Eq. (\ref{millicharge count rate}), it can be clearly manifested that the energy spectrum of the ionization process $d\sigma/dT$ and the flux spectrum of the millicharged particles $d\phi_{\chi}/dE_{\chi}$ totally determine the differential reaction event rate $dR/dT$ in a typical experiment environment.

In this work, for simplicity, we assume that incoming millicharged dark matter particles all come from the cosmic rays \footnote{There are other sources could give rise to millicharged dark matter particle flux, e.g., using theoretical calculation and experimental measurements, reference \cite{Singh2019} also consider the millicharged particles come from nuclear reactors as well as earth atmosphere.}. Although the behaviour of dark matter particles in the cosmic rays is still an open question, several recent studies suggested that the millicharged dark matter particles could be accelerated analogous to Standard Model charged particles in cosmic rays \cite{HuPK,Singh2019} through the Fermi acceleration mechanism \cite{Blandford,Perkins2003,Gaisser1990}. Therefore, as a result, the flux spectrum of millicharged dark matter particle $d\phi_{\chi}/dE_{\chi}$ obeys a simple power low \cite{HuPK,Singh2019}:
\begin{equation}
\frac{d\phi_{\chi}}{dE_{\chi}} = 30\delta_{\chi}^{\alpha-1}
                                 \bigg(
                                   \frac{\text{GeV}}{m_{\chi}c^{2}}
                                 \bigg)
                                 \bigg(
                                   \frac{E_{\chi}}{\text{GeV}}
                                 \bigg)^{-\alpha}
                                 \text{cm}^{-2}\text{s}^{-1}\text{GeV}^{-1}\text{sr}^{-1}
\label{dark cosmic ray}
\end{equation}
where $\alpha$ is the power index with a fixed value $\alpha=2.7$ \cite{HuPK}, and $\text{sr}$ represents the steradian. A key point should be mentioned is that: Eq. (\ref{dark cosmic ray}) is satisfied under certain conditions, demanding that the millicharged dark matter particle should be ultra relativistic \cite{HuPK}. In this work, the minimal incoming energy of millicharged dark matter particle is chosen to be $E_{\chi}^{\text{min}}=10\ m_{\chi}c^{2}$ in the numerical calculations, the same as in reference \cite{Singh2019}.

\begin{figure*}
\centering
\includegraphics[scale=0.6]{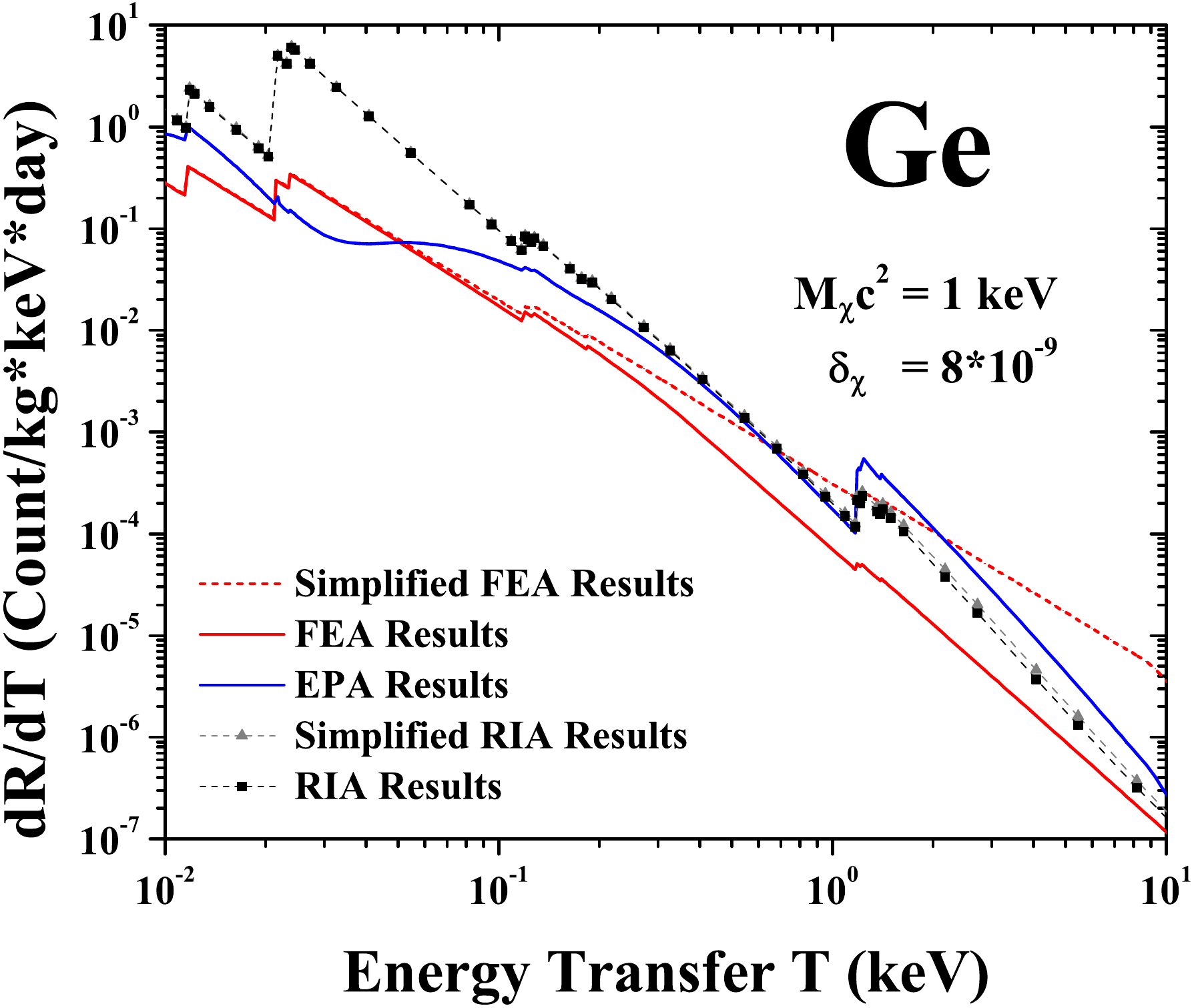}
\includegraphics[scale=0.6]{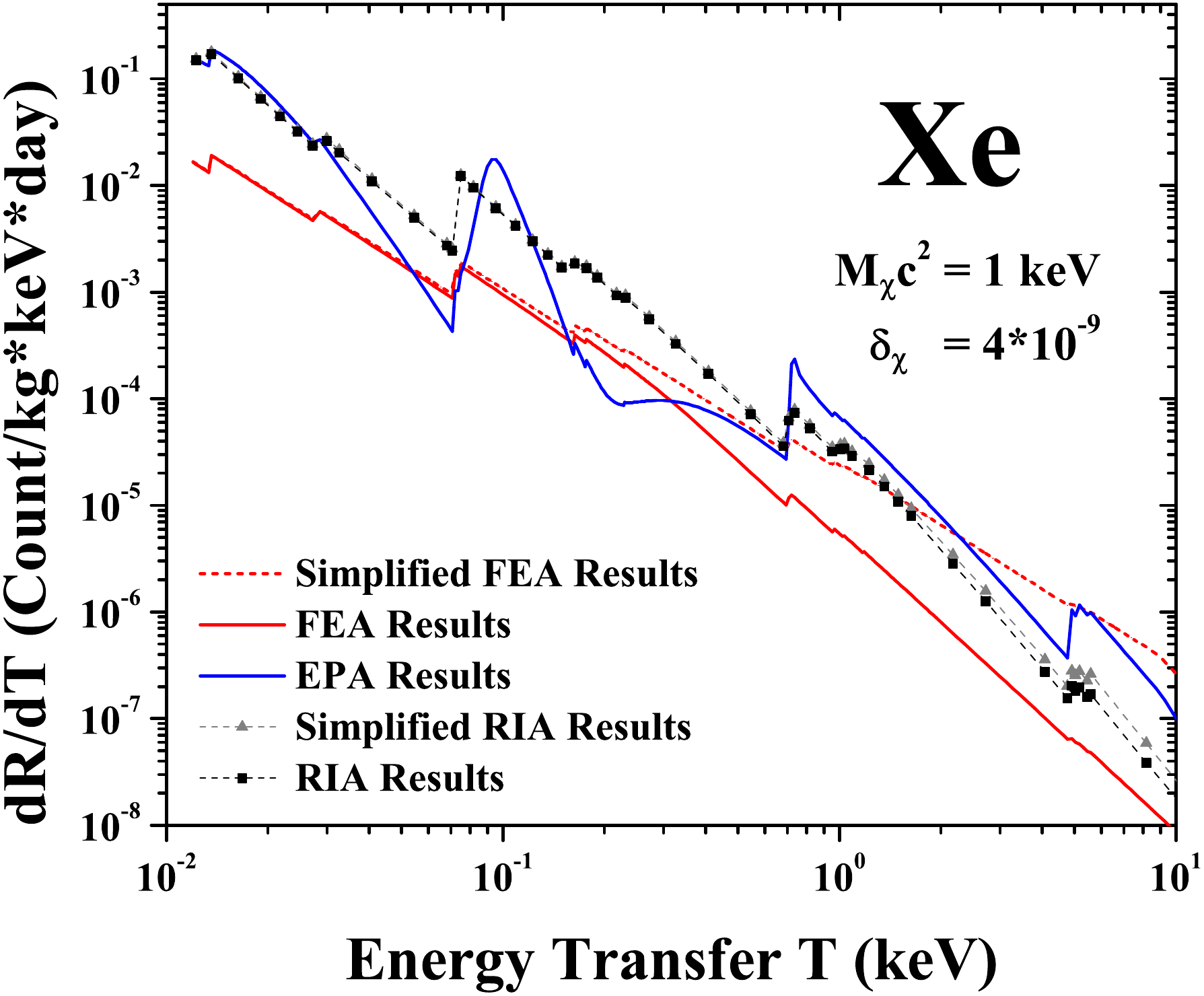}
\caption{The differential reaction event rates for atomic ionization process induced by millicharged particles. The mass of millicharged dark matter particle is set as $m_{\chi}c^{2}=1$ keV, and its millicharge is chosen to be $\delta_{\chi}=8\times10^{-9}$ for HPGe detector and $\delta_{\chi}=4\times10^{-9}$ for LXe detector, respectively. In this figure, the horizontal axis represents the energy transfer $T$, and the vertical axis represents the differential event rate $dR/dT$ in unit of cpkkd. The red solid lines correspond to the FEA results; red dashed lines represent the simplified FEA results; blue lines stand for the EPA results; black squares display the RIA results; gray triangles show the simplified RIA results.}
\label{Millicharge count rate figure}
\end{figure*}

In typical experimental environments, the differential reaction event rates for atomic ionization process induced by millicharged particles are given in the figure \ref{Millicharge count rate figure} for HPGe and LXe detectors. The numerical results obtained from FEA, EPA, and RIA approaches are displayed for comparisons. The FEA, EPA and RIA results on differential event rates $dR/dT$ are obtained by integrating the differential cross sections $d\sigma/dT$ through Eq. (\ref{millicharge count rate}). The differential cross sections $d\sigma/dT$ are calculated using FEA, EPA and RIA approaches as in subsection \ref{sec:5a}. In this figure, the horizontal axis represents the energy transfer $T$, and the vertical axis represents the differential event rate $dR/dT$ in unit of cpkkd \footnote{The unit ``cpkkd'', which is the abbreviation for ``counts per kilogram per keV per day'', stands for the number of reaction events in a real detector for 1 keV effective mass and 1 keV energy transfer interval.}. The mass of millicharged dark matter particle is set as $m_{\chi}c^{2}=1$ keV, and its millicharge is chosen to be $\delta_{\chi}=8\times10^{-9}$ for HPGe detector and $\delta_{\chi}=4\times10^{-9}$ for LXe detector, respectively.

From figure \ref{Millicharge count rate figure}, it is indicated that the reaction event rates in HPGe and LXe detectors decrease rapidly as energy transfer $T$ becomes higher. Therefore, the low-energy transfer region is dominant in the atomic ionization process induced by millicharged dark matter particles, and this region should be pay close attention to in the direct detection experiments for millicharged dark matter particles. For this reason, figure \ref{Millicharge count rate figure} only presents the event rates in $T<10$ keV region, and the $T>10$ keV region is omitted. Particularly, figure \ref{Millicharge count rate figure} shows that, for RIA results with millicharged dark matter particle mass $m_{\chi}c^{2}=1$ keV, the differential event rate in HPGe detector at energy range $T \sim 0.1$ keV is about $dR/dT \sim 0.1$ cpkkd, and the differential event rate in LXe detector at energy range $T \sim 0.5$ keV is roughly $dR/dT \sim 10^{-4}$ cpkkd.

The same as figure \ref{Millicharge cross section figure} in subsection \ref{sec:5a}, results in figure \ref{Millicharge count rate figure} can also reveal the influences come from atomic many-body effects that act on the atomic ionization process induced by millicharged particles. In low-energy transfer region, the EPA and RIA approaches obtain more reaction event rates in HPGe and LXe detectors, compared with those from FEA results. This exhibits a similar tendency with the differential cross sections presented in figure \ref{Millicharge cross section figure} in subsection \ref{sec:5a}. In the low-energy transfer region, the atomic binding, electron shielding as well as electron correlation effects can greatly enhance the atomic ionization processes induced by millicharged particles. This would bring desirable news to the next-generation direct detection experiments for millicharged dark matter particles. Furthermore, in the $T \rightarrow 0$ limit, in which range the EPA approach is derived, our RIA results do not appear large deviations from the EPA results, both for HPGe and LXe detectors. It can also be provided as an indication for the validity of our RIA approach developed in this work in the low-energy transfer region.

From figure \ref{Millicharge count rate figure}, there is another notable point: the reaction event rates in HPGe and LXe detectors obtained through FEA and simplified FEA results have large deviations when energy transfer $T>0.2$ keV. This is caused by the differential cross section in FEA approach. The simplified FEA results on differential cross section calculated using Eq. (\ref{millicharge FEA2}) converge to the FEA results calculated using Eq. (\ref{millicharge FEA}) only when $m_{\chi} \ll m_{e}$, $T \ll m_{e}c^{2}$, $T \ll E_{\chi}$ are satisfied. However, for the millicharged dark matter particle coming from cosmic rays, the flux spectrum represent a power low as in Eq. (\ref{dark cosmic ray}). When energy of incident particle is lower, the flux spectrum becomes larger. In typical super-terranean or underground experiments, there are large amount of low-energy millicharged dark matter particles entering into the HPGe and LXe detectors, which would destroy the condition $T \ll E_{\chi}$ and make the simplified FEA results inappropriate \footnote{The minimal energy of incident millicharged dark matter particle is chosen as $E_{\chi}^{\text{min}}=10\ m_{\chi}c^{2}$ in the numerical calculations. In figure \ref{Millicharge count rate figure}, the minimal energy corresponds to $E_{\chi}^{\text{min}}=10\ m_{\chi}c^{2}=10$ keV, and the condition $T \ll E_{\chi}$ is not satisfied with great accuracy when $T>0.2$ keV.}. When energy transfer $T$ becomes higher, the condition $T \ll E_{\chi}$ is harder to satisfy, and there are more differences between the simplified FEA results and FEA results. This is similar to the cases of differential cross sections discussed in subsection \ref{sec:5a}. Therefore, only FEA results calculated through Eqs. (\ref{millicharge FEA}) and (\ref{millicharge count rate}) are reasonable in this region. In such cases, compare with the FEA result, the simplified FEA results calculated through Eqs. (\ref{millicharge FEA2}) and (\ref{millicharge count rate}) overestimate the reaction event rates in HPGe and LXe detectors.

\subsection{Detecting Sensitivity on Dark Matter Particle Millicharge in Next-Generation HPGe and LXe Based Experiments \label{sec:5c}}

According to the calculations of reaction event rates in HPGe and LXe detectors, we can give an estimation of the detecting sensitivity for dark matter particle millicharge $\delta_{\chi}$ in next-generation HPGe and LXe based direct detection experiments. The estimation is carried out according to the following assumptions:
\begin{itemize}
\item Consider the dark matter particle with millicharge $\delta_{\chi}=\delta_{0}$, if the calculated reaction event rates in energy range above the experimental threshold surpass the experimental background, then the signals from atomic ionization process induced by millicharged particles can be catched and identified effectively. In this case, next-generation experiments have the ability to detect dark matter particles with millicharge $\delta_{\chi}=\delta_{0}$\footnote{Only when reaction event rate overwhelm the experimental background in energy range above the experimental threshold, signals produced from the atomic ionization process induced by millicharged particles could be catched and identified effectively. Otherwise, the energy transfer is too small to track the atomic ionization signals in detectors, or these atomic ionization signals may be overwhelmed by background signals and couldn't be identified and analyzed effectively.}.
\item On the other hand, if the calculated reaction event rates in energy range above the experimental threshold are less than the experimental background, then the atomic ionization signals would not be effectively identified and next-generation experiments could't set a constrain on dark matter particles with millicharge $\delta_{\chi}=\delta_{0}$.
\end{itemize}
In the numerical calculations, for any dark matter particle mass $m_{\chi}$, we calculate the differential event rates $dR/dT$ in HPGe and LXe detectors using FEA, EPA and RIA methods at a given dark matter particle millicharge $\delta_{\chi}$, then adjust the value of millicharge $\delta_{\chi}$ such that reaction event rates in HPGe and LXe detectors in energy region above the experimental threshold surpass the experimental backgrounds. Finally, we can obtain the lower limit of milliarge $\delta_{\chi}$ satisfying the above conditions. This is the estimation of detecting sensitivity on dark matter particle millicharge $\delta_{\chi}$ in the next-generation HPGe and LXe based direct detection experiments.

\begin{table*}
\centering
\caption{Estimation of detecting sensitivity on dark matter particle millicharge $\delta_{\chi}$ in the next-generation direct detection experiments. The results for HPGe and LXe based experiments in the FEA, EPA and RIA calculations are given in this table. In the HPGe based next-generation experiments, the energy threshold and background level have been assumed as 100 eV and 0.1 cpkkd, respectively. While in the LXe based next-generation experiments, the energy threshold and background level are assumed as 500 eV and $10^{-4}$ cpkkd, respectively.}
\label{table}
\vspace{5mm}
%\begin{ruledtabular}
%\begin{tabular}{lccccccccc}
\begin{tabular}{|ccccccccc|}
\hline
& \multicolumn{7}{c}{HPGe Based Experiments} &
\\
\hline
& $m_{\chi}c^{2}$ && \multicolumn{5}{c}{detecting sensitivity on dark matter particle millicharge\ $\delta_{\chi}$} &
\\
\hline
& keV  && FEA Results        && EPA Results         && RIA Results &
\\
\hline
& 0.01 && $3.2\times10^{-9}$ && $3.5\times10^{-10}$ && $1\times10^{-9}$ &
\\
& 0.1  && $6.0\times10^{-9}$ && $1.9\times10^{-9}$  && $2\times10^{-9}$  &
\\
& 1    && $1.4\times10^{-8}$ && $1.0\times10^{-8}$  && $8\times10^{-9}$  &
\\
& 10   && $7.0\times10^{-8}$ && $5.3\times10^{-8}$  && $4.5\times10^{-8}$  &
\\
& 100  && $3.7\times10^{-7}$ && $2.8\times10^{-7}$  && $2.5\times10^{-7}$  &
\\
\hline
& \multicolumn{7}{c}{LXe Based Experiments} &
\\
\hline
& $m_{\chi}c^{2}$ && \multicolumn{5}{c}{detecting sensitivity on dark matter particle millicharge\ $\delta_{\chi}$} &
\\
\hline
& keV  && FEA Results        && EPA Results         && RIA Results &
\\
\hline
& 0.01 && $1.7\times10^{-9}$ && $2.5\times10^{-10}$ && $8\times10^{-10}$ &
\\
& 0.1  && $3.1\times10^{-9}$ && $6.0\times10^{-10}$ && $1.5\times10^{-9}$  &
\\
& 1    && $5.9\times10^{-9}$ && $3.2\times10^{-9}$  && $4\times10^{-9}$  &
\\
& 10   && $2.5\times10^{-8}$ && $1.8\times10^{-8}$  && $2\times10^{-8}$  &
\\
& 100  && $1.3\times10^{-7}$ && $9.3\times10^{-8}$  && $1\times10^{-7}$  &
\\
\hline
\end{tabular}
%\end{ruledtabular}
\end{table*}

The estimation of detecting sensitivity on dark matter particle millicharge $\delta_{\chi}$ in the next-generation HPGe and LXe based experiments is shown in table \ref{table} for several dark matter particle mass $m_{\chi}$. The results from the FEA, EPA and RIA calculations are given in this table for comparisons. For the HPGe based next-generation experiments, the energy threshold and background level have been assumed as 100 eV and 0.1 cpkkd, respectively. For the LXe based next-generation experiments, the energy threshold and background level are assumed as 500 eV and $10^{-4}$ cpkkd, respectively. From this table, it can be clearly shown that, for several dark matter particle mass, the detecting sensitivities of millicharge $\delta_{\chi}$ calculated from RIA and EPA approaches are much larger than those calculated from the FEA approach. In subsection \ref{sec:5a} and subsection \ref{sec:5b}, we have learned that the atomic any-body effects can greatly enhance the atomic ionization process induced by millicharged dark matter particles in the low-energy transfer region, leading to the increase of differential cross sections $d\sigma/dT$ as well as differential reaction event rates $dR/dT$ in this region. Therefore, for the same experimental background, atomic many-body effects make it more easy to let reaction event rates surpass the experimental background, which eventually leads to a more strong constrain on dark matter particle millicharge $\delta_{\chi}$. This would be beneficial for direct detection of millicharged dark matter particles in next-generation experiments. These results shown that atomic many-body effects would play a significant role in the electromagnetic interactions of millicharged particles, and it may open an new window for the explorations of millicharge particles. Furthermore, with relatively lower experimental background, the next-generation LXe based experiments could set a lower bound on dark matter particle millicharge $\delta_{\chi}$, no matter which approach is employed in the numerical calculations. For HPGe based experiments, the EPA results get smaller dark matter millicharge $\delta_{\chi}$ than RIA results in the low-mass cases ($m_{\chi}c^{2} \leq 0.1$ keV). While for LXe based experiments, the EPA results get smaller millicharge $\delta_{\chi}$ than RIA results in all cases (10 eV $<m_{\chi}c^{2}<$ 100 keV) because of the giant resonance for $d$ electrons in the photoabsorption cross section of Xe atom \cite{Johnson1992,Andersen,Toffoli,Kumar2009,Qiao2019,Amusia}. Detailed numerical results giving rise to the detecting sensitivities in figure \ref{table} are presented in the appendix \ref{appendix4}.

The estimations of detecting sensitivity on dark matter particle millicharge $\delta_{\chi}$ in this work can contribute to the parameter space of millicharged dark matter particles. In the figure \ref{millicharge experiment}, we also present our estimations of detecting sensitivity on millicharge $\delta_{\chi}$ in RIA calculations for next-generation LXe based experiments. From figure \ref{millicharge experiment}, the indirect searches from astronomy and cosmology set stronger constrains on dark matter particle millicharge $\delta_{\chi}$. While the direct detection experiments and accelerator/collider experiments, i.e. XENON10, TEX, OPOS, COLL, SLAC, LHC in figure \ref{millicharge experiment}, set looser bounds for millicharge $\delta_{\chi}$. However, it is remarkable that the next-generation LXe based direct detection experiments would greatly increase the detecting sensitivity on dark matter particle millicharge $\delta_{\chi}$. In the range 10 eV $<m_{\chi}c^{2}<$ 100 keV, The current best experimental bound in direct detection experiments and accelerator/collider experiments is roughly $\delta_{\chi} \sim 10^{-5}$ to $\delta_{\chi} \sim 10^{-7}$, which is 2-3 order of magnitude larger than our estimation for next-generation LXe based experiments.

There is one point need to be mentioned: the calculations of reaction event rates in HPGe and LXe detectors as well as the calculations of detection sensitivities on dark matter particle millicharge $\delta_{\chi}$ in next-generation HPGe and LXe based experiments are just a leading order estimation. In our numerical calculations, we have made some simplified assumptions. The electromagnetic interactions between millicharged dark matter particles in cosmic rays and the charged particles in earth atmosphere, as well as the electromagnetic interactions between millicharged dark matter particles and atoms and molecules in environmental rocks, are not taken into considerations. These interactions may lead to an upper bound in the parameter space of millicharge dark matter particles, as inidicated in reference \cite{Singh2019}. If the millicharge of dark matter particle is much too large, then the electromagnetic interactions between millicharged dark matters in cosmic rays and the charged particles in earth atmosphere would be too strong, which lead to tremendous attenuation of dark matter particle flux in the atmosphere. As a result, it will prevent millicharged dark matter particles entering into HPGe and LXe detectors in super-terranean or underground experiments.

\section{Numerical Results and Discussions on Millicharged Neutrinos \label{sec:6}}

In section \ref{sec:1}, it is revealed that neutrino physics is becoming a rising field in many branches of science. Recently, many studies suggested that neutrinos may have tiny electromagnetic interactions \cite{Giunti2008,Giunti2015,Giunti2016}, and they may have millicharge as well as magnetic moment. Theoretical and experimental explorations on neutrino millicharge and magnetic moment is becoming more and more attractive, and a number of researches on this area emerge in recent years \cite{Chen2013,Chen2014,Chen2014a}.

As discussed in section \ref{sec:4}, the RIA approach we developed in this work is irrelevant to the underling nature and mechanism of millicharged particles. In principle, our approach can also be applied to the study of millicharged neutrinos. In this section, we use our RIA approach to study atomic ionization process induced by millicharged neutrinos. The numerical results on differential cross section $d\sigma/dT$, differential reaction event rate $dR/dT$, and detecting sensitivity on neutrino millicharge $\delta_{\nu}$ in next-generation direct detection experiments are presented similar to the cases of millicharged dark matter particles discussed in section \ref{sec:5}.

\begin{figure}
\centering
\includegraphics[scale=0.55]{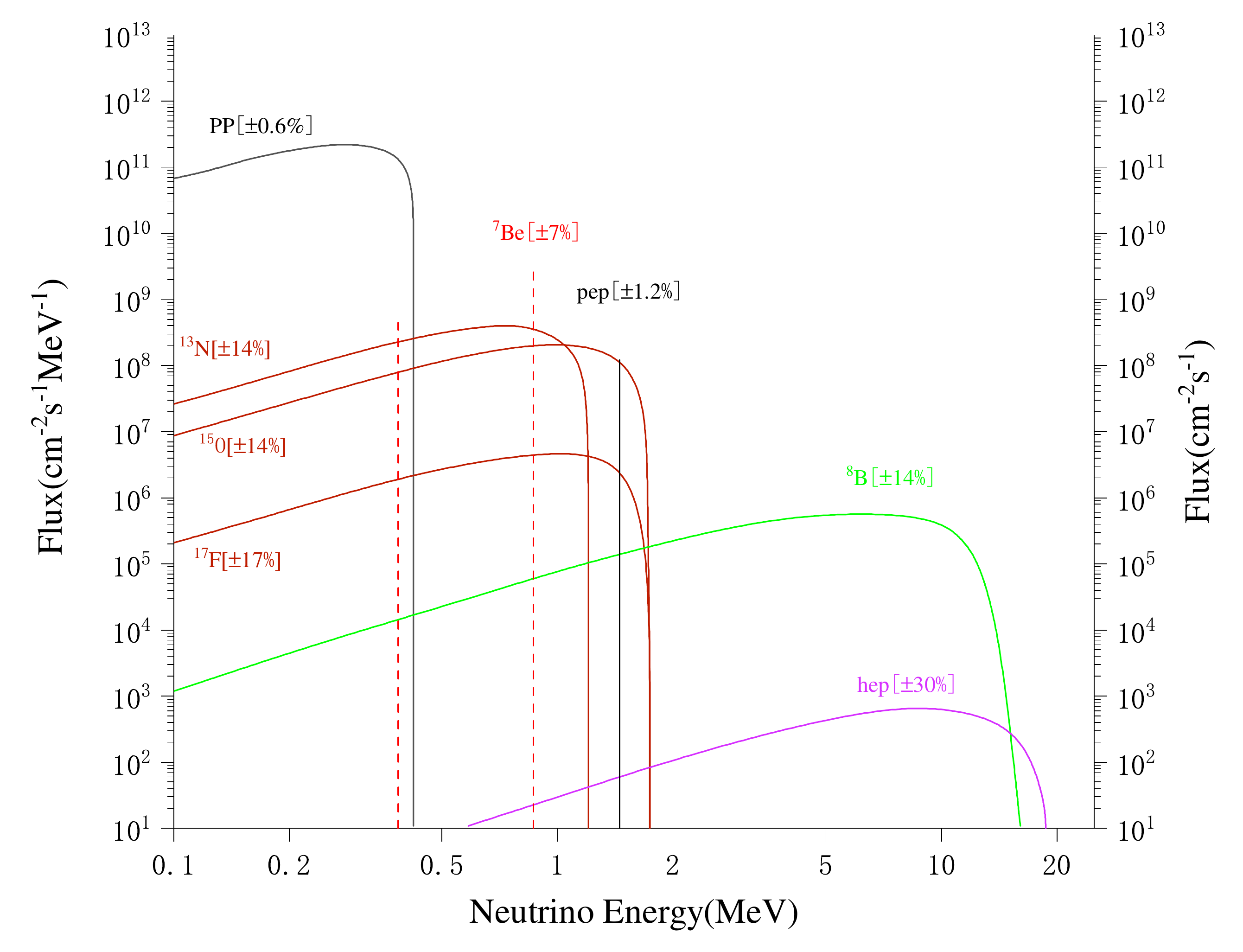}
\caption{Flux spectrum of solar neutrinos. This figure gives the flux spectra in earth surface for electron neutrino $\nu_{e}$ associated with different channels. The horizontal axis gives the energy of neutrino in unit of MeV. The $^{7}$Be neutrinos and $pep$ neutrinos have discrete spectra, the unit in vertical axis is $cm^{-2}s^{-1}$. The $pp$, $hep$, $^{8}$B, $^{13}$N, $^{15}$O, $^{17}$F channels give rise to continuous spectra, and the unit in vertical axis is $cm^{-2}s^{-1}MeV^{-1}$. In this figure, the solar neutrino flux spectrum is plotted based on the results in references \cite{Haxton,Serenelli}.}
\label{Solar Neutrino}
\end{figure}

There are several kinds of sources which may contribute to millicharged neutrinos: reactor neutrinos, cosmological neutrinos, solar neutrinos, atmospheric neutrinos, supernova neutrinos, cosmogenic neutrinos, and active galactic nucleus (AGN) produced neutrinos \cite{Perkins2003,Katz,Haxton}. The reactor neutrinos become dominant only when laboratory is near the nuclear reactors, and supernova neutrinos become notable when supernova is activated, i.e. supernova 1987A burst. For other sources, the cosmological neutrinos mainly appear in ultra-low energy range (below 1 eV), while atmospheric neutrinos, cosmogenic neutrinos and AGN produced neutrinos all centered in ultra-high energy range (above GeV). More details of neutrino sources and their flux can be found in reference \cite{Katz,Vitagliano}. Therefore, in energy range 100 eV $\leq E_{\nu} \leq$ GeV, which is sensitive to HPGe and LXe detectors and is of great interests in direct detection experiments, solar neutrino have the maximal flux and can be viewed as the main source of millicharged neutrinos. In this work, for simplicity, we only consider solar neutrinos as the source of millicharged neutrinos. Contributions from other sources are leaving for future studies.

There are several channels which can produce solar neutrinos \cite{Haxton,Bahcall2001}:
\begin{eqnarray}
pp\ \text{channel:} && p + p \rightarrow d + e^{+} + \nu_{e} \nonumber
\\
pep\ \text{channel:} && p + e^{-} + p \rightarrow d + \nu_{e} \nonumber
\\
hep\ \text{channel:} && \ ^{3}He + p \rightarrow ^{4}He + e^{+} + \nu_{e} \nonumber
\\
^{7}Be\ \text{channel:} && \ ^{7}Be + e^{-} \rightarrow ^{7}Li + \nu_{e} \nonumber
\\
^{8}B\ \text{channel:} && \ ^{8}B + e^{-} \rightarrow ^{8}Be + \nu_{e} \nonumber
\\
^{13}N\ \text{channel:} && \ ^{13}N \rightarrow ^{13}C + e^{+} + \nu_{e} \nonumber
\\
^{15}O\ \text{channel:} && \ ^{15}O \rightarrow ^{15}N + e^{+} + \nu_{e} \nonumber
\\
^{17}F\ \text{channel:} && \ ^{17}F \rightarrow ^{17}O + e^{+} + \nu_{e} \nonumber
\end{eqnarray}
Among these channels, the $^{7}$Be neutrinos and $pep$ neutrinos have discrete spectra, and other channels give rise to continuous spectra. The flux spectra for various channels are displayed in figure \ref{Solar Neutrino}. Furthermore, it should be mentioned that the $pp$ neutrinos and $^{7}$Be neutrinos give predominant contributions to solar neutrino flux, and this two channels contribute to 98\% of solar neutrinos \cite{Perkins2003,Haxton,Hsieh2019}.

\subsection{Differential Cross Section \label{sec:6a}}

\begin{figure*}
\centering
\includegraphics[scale=0.4]{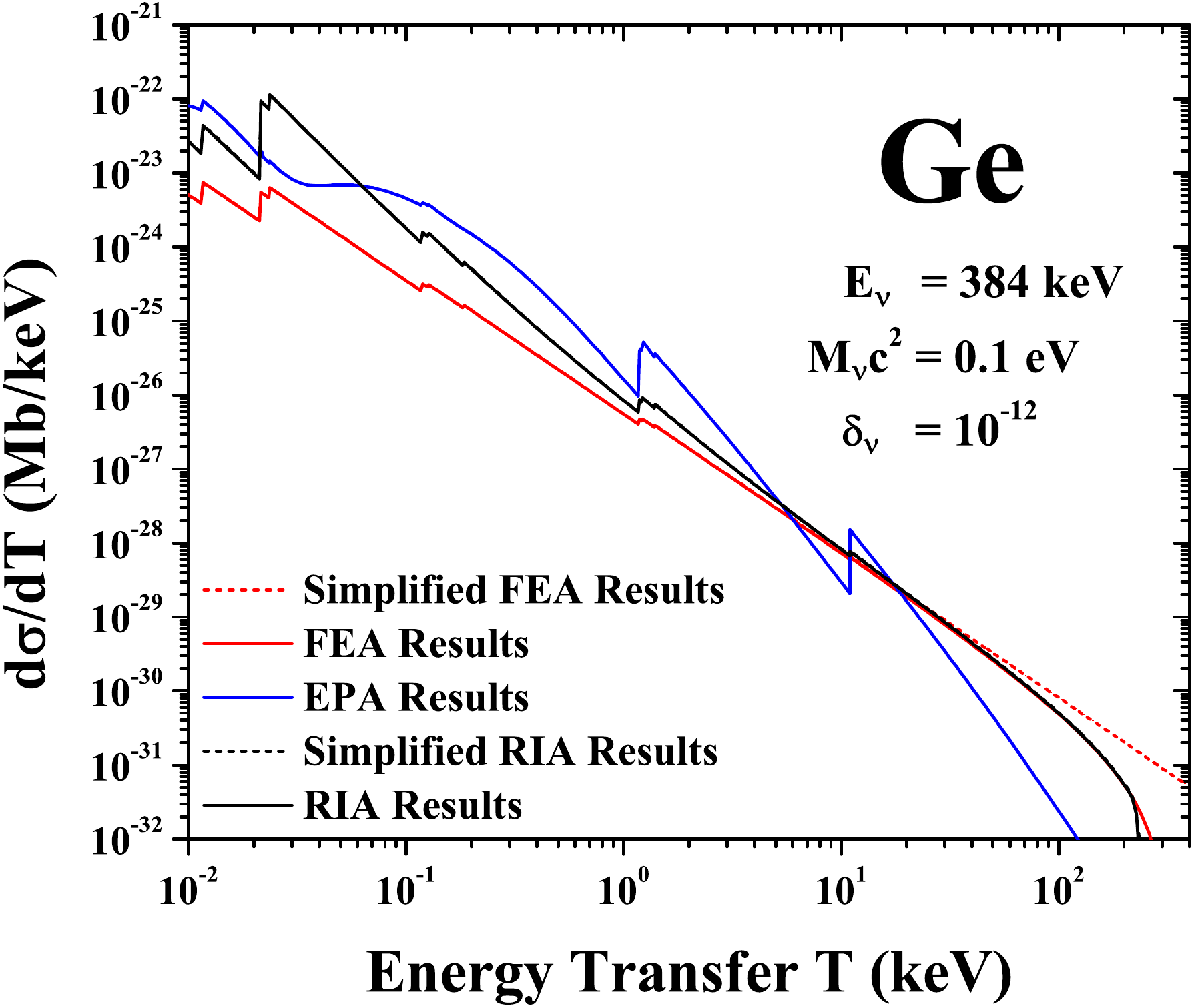}
\includegraphics[scale=0.4]{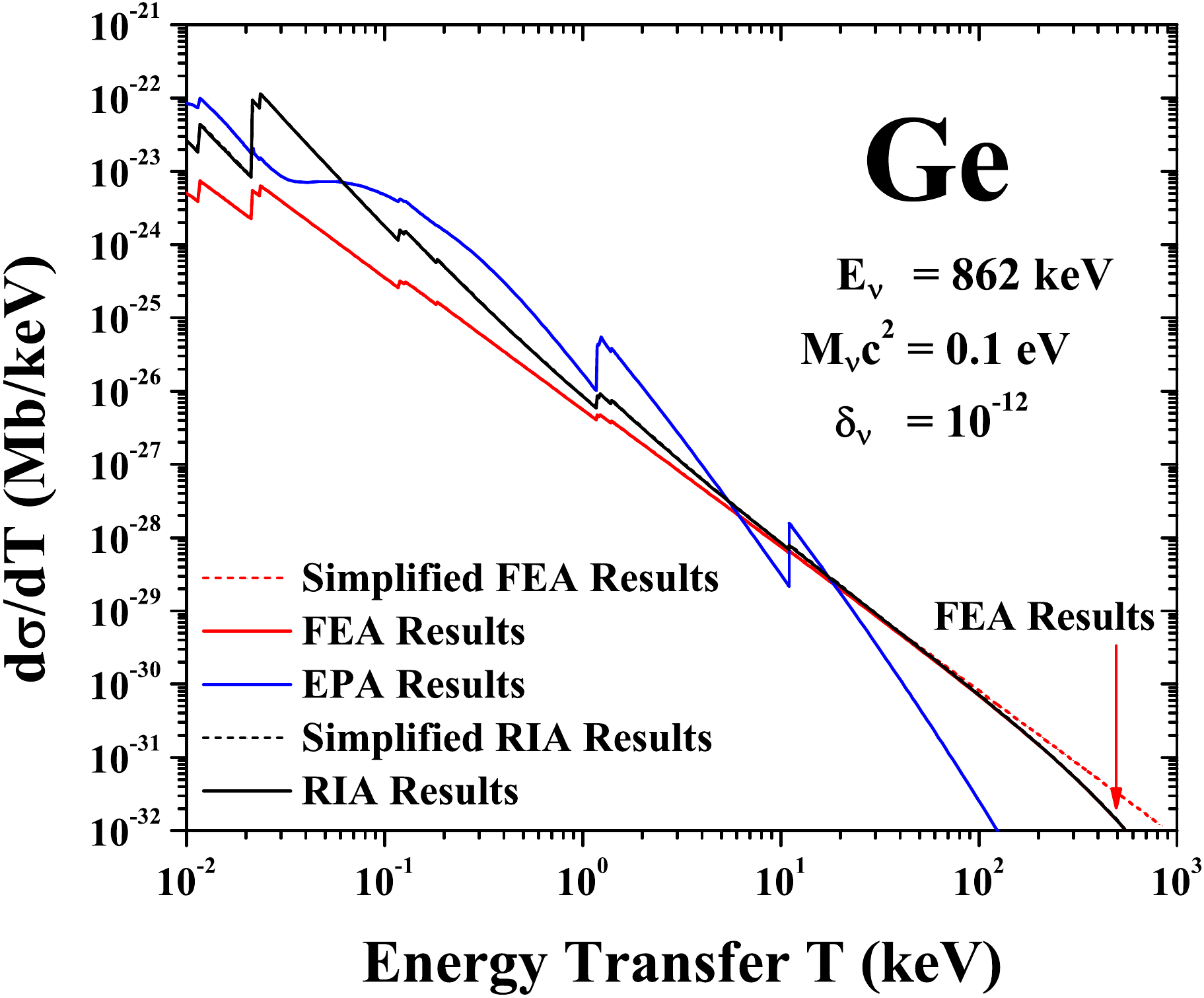}
\includegraphics[scale=0.4]{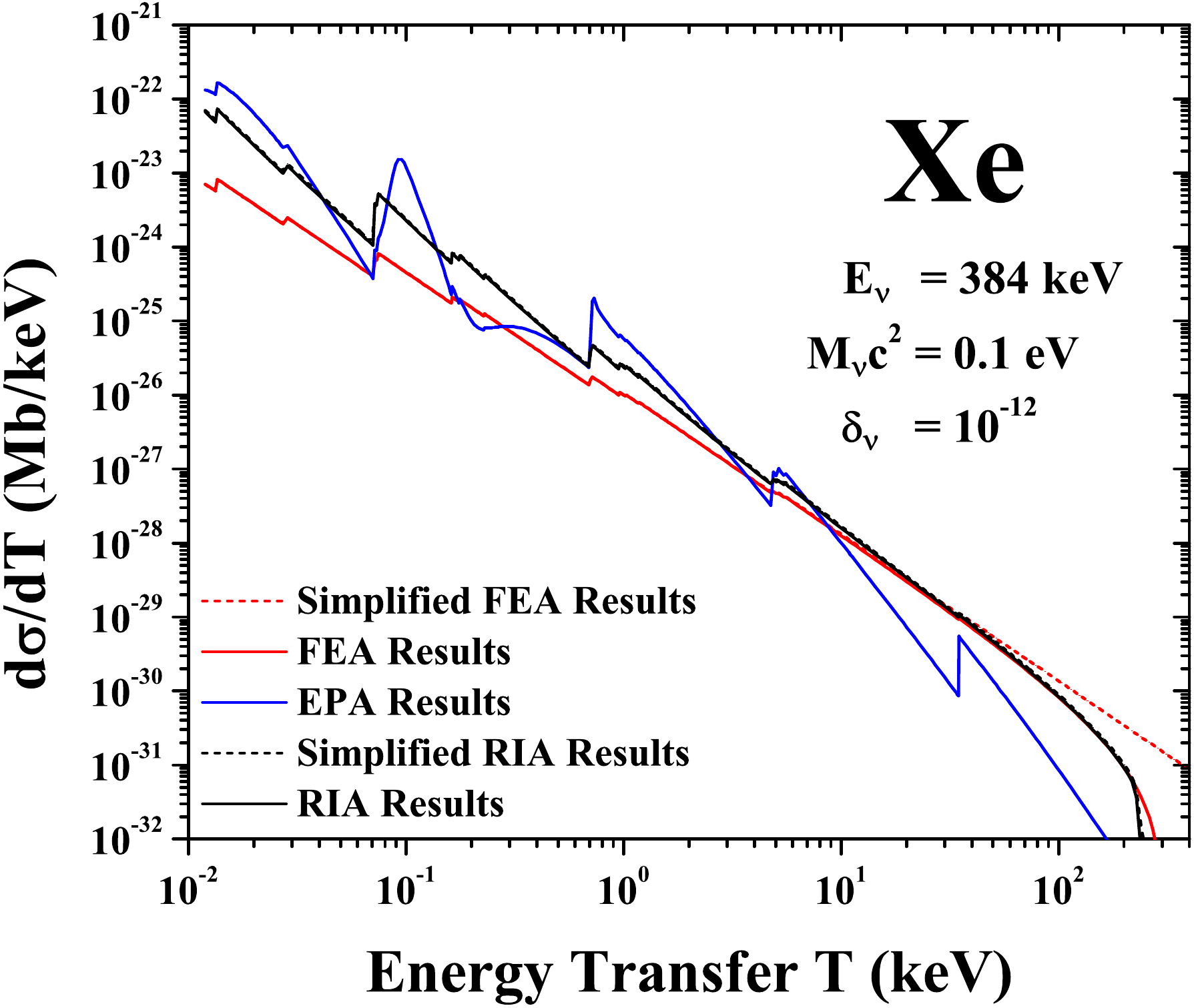}
\includegraphics[scale=0.4]{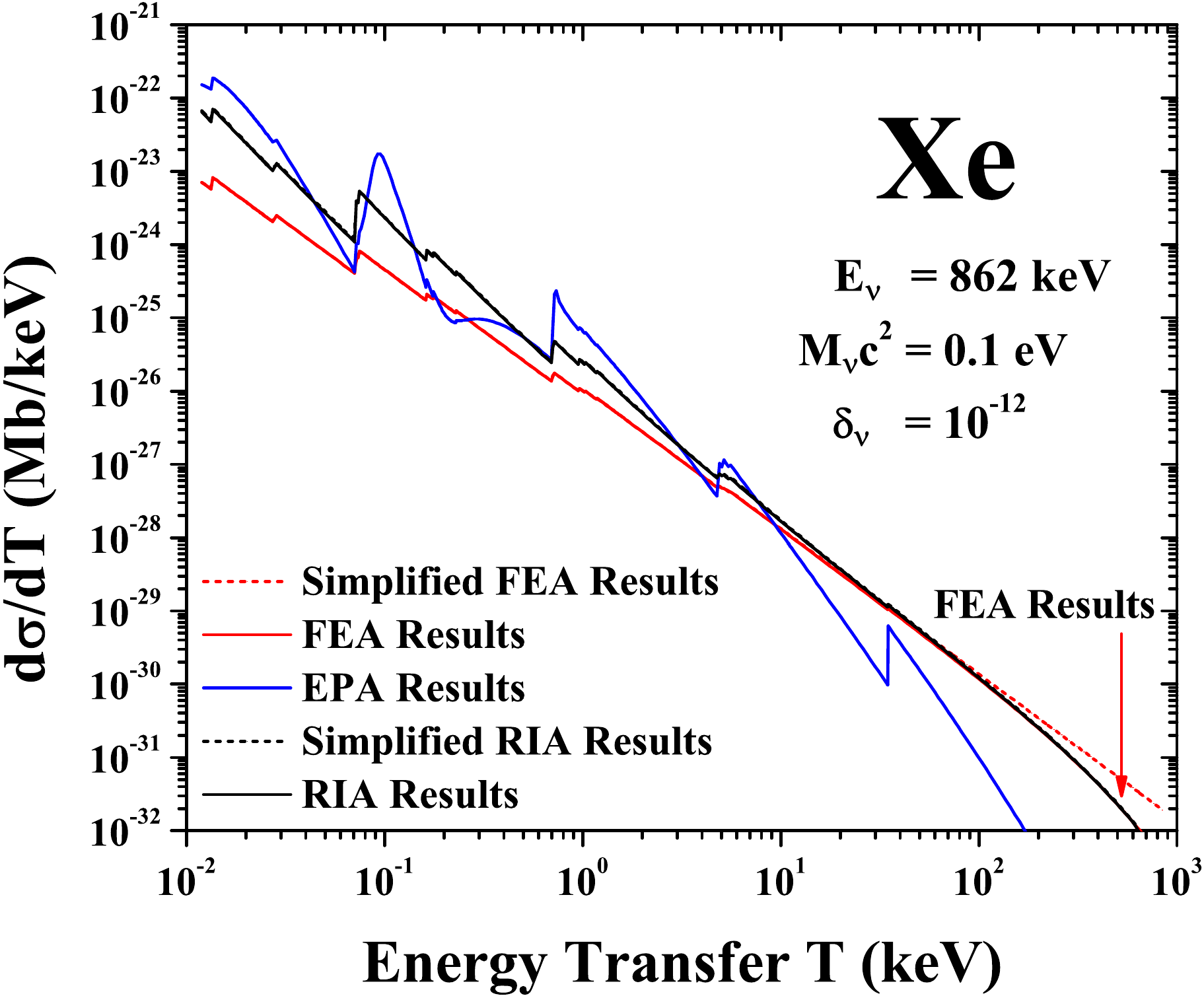}
\caption{Differential cross sections of atomic ionization process for Ge and Xe atoms induced by millicharged neutrinos. In this figure, the mass and millicharge of neutrino are chosen as $m_{\nu}c^{2}=0.1$ eV and $\delta_{\nu}=10^{-12}$, respectively. The energies of $^{7}$Be solar neutrinos are $E_{\nu}=384$ KeV and $E_{\nu}=862$ KeV. The same as in figure \ref{Millicharge cross section figure} and figure \ref{Millicharge cross section figure0}, the numerical results from FEA, EPA and RIA approaches are given for comparisons. The red solid lines correspond to the FEA results calculated through Eq. (\ref{millicharge FEA}); red dashed lines represent the simplified FEA results calculated through Eq. (\ref{millicharge FEA2}); blue lines stand for the EPA results calculated from Eq. (\ref{millicharge EPA}); black solid lines show the RIA results calculated using Eq. (\ref{RIA millicharge full DCS}); and black dashed lines present the simplified RIA results calculated using Eq. (\ref{RIA simplified millicharge full DCS}).}
\label{Millicharge cross section figure2}
\end{figure*}

In this subsection, we take $^{7}$Be solar neutrinos as an example to study the differential cross section with respect to energy transfer in the atomic ionization process induced by millicharged neutrinos. The differential cross section $d\sigma/dT$ of atomic ionization process for Ge and Xe atoms is presented in figure \ref{Millicharge cross section figure2}. The $^{7}$Be solar neutrinos have discrete spectrum, and the incoming neutrino energy is located in $E_{\nu}=384$ keV and $E_{\nu}=862$ keV with branch ratios 89.5\% and 10.5\% \cite{Perkins2003}. Meanwhile, in this figure, the neutrino mass and millicharge are chosen to be $m_{\nu}c^{2}=0.1$ eV and $\delta_{\nu}=10^{-12}$, respectively. The numerical results obtained from FEA, EPA and RIA approaches are shown in this figure for comparisons.

Form figure \ref{Millicharge cross section figure2}, it can be clearly shown that the results for millicharged neutrinos are very similar to the results for millicharged dark matter particles given in subsection \ref{sec:5a}. The differential cross section of atomic ionization process induced by millicharged neutrinos diminish as energy transfer $T$ increases, and FEA, EPA, RIA results show the same tendency. When the energy transfer $T$ is smaller than the atomic binding energy for 1s electron (which is 11.1 keV for Ge atom and 34.5 keV for Xe atom), the differential cross sections calculated using RIA and EPA methods are larger than those in FEA results, indicating that the atomic many-body effects tend to intensify the electromagnetic interaction for millicharged neutrinos in low-energy transfer region. In the ultra-low-energy transfer region, namely the $T \rightarrow 0$ limit, our RIA results are near the EPA results, which shows the validity of our methods in this region. When the energy transfer $T$ is sufficiently large, the EPA approach breaks down and it underestimate the differential cross sections, while the RIA results converge to the FEA results because the atomic many-body effects are negligible in high-energy transfer region. Furthermore, from figure \ref{Millicharge cross section figure2}, it is also indicated that the simplified RIA results converge to the RIA results in the entire region of energy transfer $T$. Therefore, for $^{7}$Be solar neutrinos with energies $E_{\nu}=384$ keV and $E_{\nu}=862$ keV, among the approximations of function $X$ in the integrand of Eq. (\ref{millicharge doubly differential cross section}), the more simplified approximation $X \approx X_{\text{sim}}$ is adequate and does not lead to large deviations compared with the more accurate approximation $X \approx \overline{X}(\overline{s}(p_{z}),\overline{t}(p_{z}),\overline{u}(p_{z}))$. This is similar to the cases of high-energy millicharged dark matter particles presented in figure \ref{Millicharge cross section figure}.

Similarly, the results in figure \ref{Millicharge cross section figure2} can also reflect some sort of generality. In various approaches, i.e. FEA, EPA and RIA, the differential cross section $d\sigma/dT$ of the atomic ionization process induced by millicharged neutrinos is proportional to $\delta_{\nu}^{2}$. For the same incoming energy $E_{\nu}$, results correspond to other neutrino millicharge $\delta_{\nu}$ can be obtained by proportional magnifying or shrinking the results in figure \ref{Millicharge cross section figure2} by $\delta_{\nu}^{2}$ times.

From the numerical calculations in subsection \ref{sec:5a} and subsection \ref{sec:6a}, we can draw a conclusion that the atomic ionization processes, whether induced by millicharged dark matter particles or millicharged neutrinos, exhibit similar tendency. The differential cross section $d\sigma/dT$ of atomic ionization process induced by millicarged particles drops rapidly as energy transfer becomes higher. In the low-energy transfer region, atomic binding, electron shielding and electron correlation effects could greatly enhance the atomic ionization process induced by millicharged particles. Our RIA approach developed in this work is appropriate in the entire region of energy transfer. For high energy millicharged particles, our RIA results on differential cross section show small discrepancies with EPA results in the ultra-low energy region, and our RIA results successfully converge to the FEA results in the high-energy transfer region, where the atomic effects are weak and atomic electron can be treated as free electron approximately \footnote{However, for low-energy millicharged dark matter particles, the FEA results and RIA results do not converge to each other as presented in figure \ref{Millicharge cross section figure0}. For solar neutrinos, the high-energy neutrinos give a major contribution in the flux spectrum, as shown in figure \ref{Solar Neutrino}, thus we do not give a discussion on the cases of low-energy millicharged neutrinos in this subsection.}.

\subsection{Reaction Event Rate in HPGe and LXe Detectors and Detecting Sensitivity on Neutrino Millicharge in Next-Generation HPGe and LXe Based Experiments \label{sec:6b}}

Similar to the calculations in subsection \ref{sec:5b} for millicharged dark matter particles, in a typical experiment environment, the differential reaction event rate in HPGe and LXe detectors for atomic ionization process induced by millicharged neutrinos can be expressed similar to Eq. (\ref{millicharge count rate}):
\begin{equation}
\frac{dR}{dT} = \rho_{A} \int_{E_{\nu}^{\text{min}}}^{E_{\nu}^{\text{max}}} dE_{\nu} \frac{d\sigma}{dT} \frac{d\phi_{\nu}}{dE_{\nu}}
\label{millicharge count rate2}
\end{equation}
where $d\phi_{\nu}/dE_{\nu}$ is the neutrino flux spectrum. As we have discussed in the beginning of this section, in the energy range relevant to direct detection experiments, which is from keV to GeV, solar neutrino is the main source for millicharged neutrinos. Since the pp channel and $^{7}$Be channel contribute to 98\% of solar neutrinos, we can omit contributions from other channels. Therefore, the solar neutrino flux spectrum can be simplified as:
\begin{equation}
\frac{d\phi_{\nu}}{dE_{\nu}} \approx \frac{d\phi_{\nu}^{pp}}{dE_{\nu}} + \frac{d\phi_{\nu}^{\text{Be}}}{dE_{\nu}}
\end{equation}
For $^{7}$Be reaction neutrino, the flux spectrum $d\phi_{\nu}^{\text{Be}}/dE_{\nu}$ is discrete with energy located at 384 keV and 862 keV. For pp reaction neutrino, the flex spectrum $d\phi_{\nu}^{pp}/dE_{\nu}$ is continuous. The flux spectra $d\phi_{\nu}^{pp}/dE_{\nu}$ and $d\phi_{\nu}^{\text{Be}}/dE_{\nu}$ and can be obtained either by fitting the corresponding curves in figure \ref{Solar Neutrino}, or from the solar neutrino databases \cite{Bahcall1997,Bahcall2004,Bahcall2005,solar-neutrino-datebase1,solar-neutrino-datebase2,solar-neutrino-datebase3}.

\begin{figure*}
\centering
\includegraphics[scale=0.6]{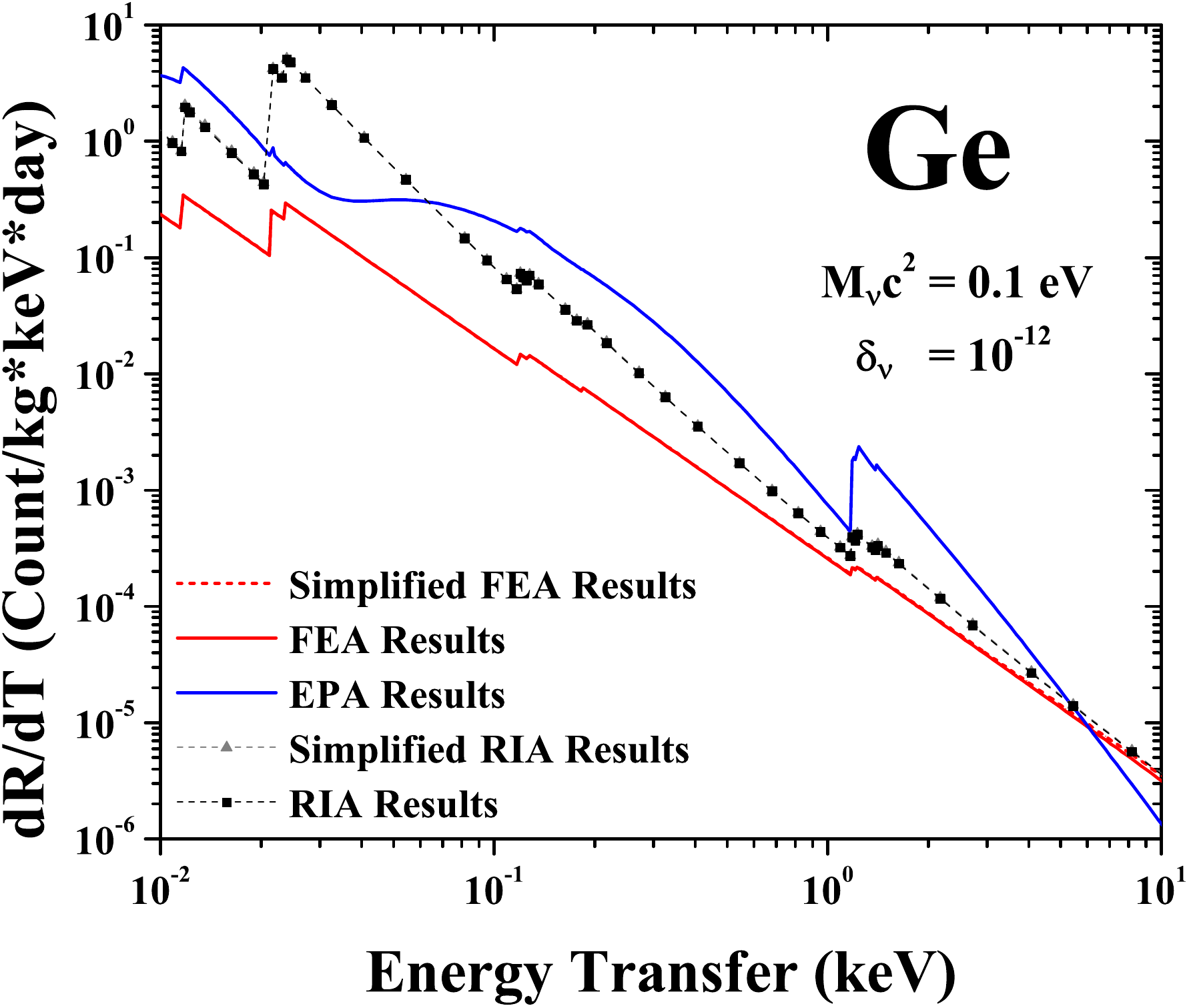}
\includegraphics[scale=0.6]{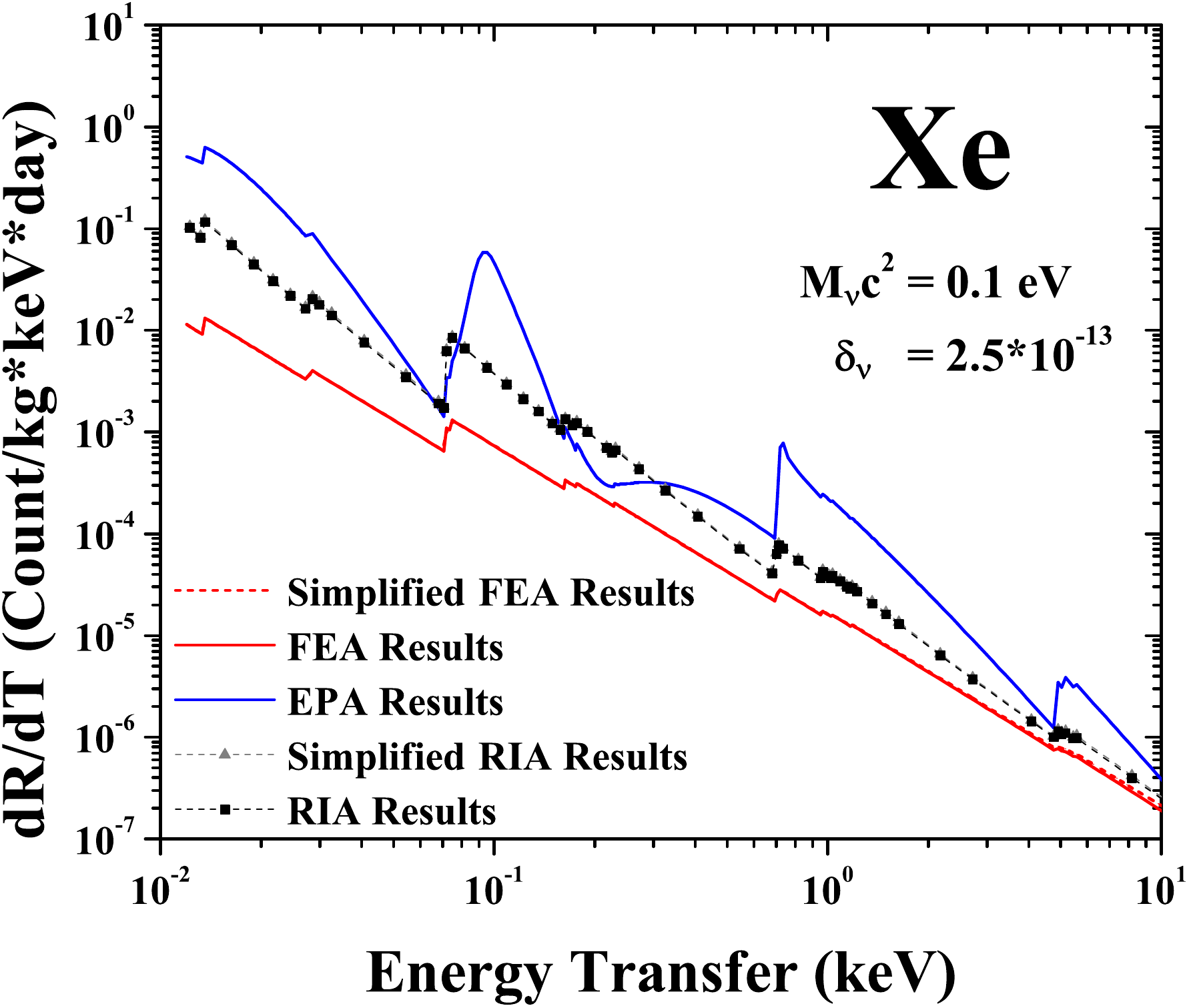}
\caption{The differential reaction event rates for atomic ionization process induced by millicharged neutrinos for HPGe and LXe detectors. The neutrino mass is chosen to be $m_{\nu}c^{2}=0.1$ eV, and the neutrino millicharge is set as $\delta_{\nu}=10^{-12}$ for HPGe detectors and $\delta_{\nu}=2.5\times10^{-13}$ for LXe detectors. In this figure, the horizontal axis represents the energy transfer $T$, and the vertical axis represents the differential event rate $dR/dT$ in unit of cpkkd. The same as in figure \ref{Millicharge count rate figure}, the red solid lines correspond to the FEA result; red dashed lines represent the simplified FEA results; blue lines stand for the EPA results; black squares display the RIA results; gray triangles show the simplified RIA results.}
\label{Millicharge count rate figure2}
\end{figure*}

Figure \ref{Millicharge count rate figure2} shows the differential reaction event rates $dR/dT$ for atomic ionization process induced by millicharged neutrinos for HPGe and LXe detectors in typical super-terranean or underground experimental environments. The neutrino mass is chosen to be $m_{\nu}c^{2}=0.1$ eV, and the neutrino millicharge is set as $\delta_{\nu}=10^{-12}$ for HPGe detectors and $\delta_{\nu}=2.5\times10^{-13}$ for LXe detectors \footnote{The neutrino millicharge $\delta_{\nu}$ is adjusted such that the reaction event rates in HPGe and LXe detectors are comparable to the experimental background levels in next-generation HPGe and LXe based experiments.}. The numerical results from FEA, EPA and RIA approaches are given in this figure for comparison. Similar to the cases for millicharged dark matter particles, figure \ref{Millicharge count rate figure2} also indicates that the differential event rates for atomic ionization process induced by millicharged neutrinos reduce significantly as energy transfer $T$ increases, both in HPGe and LXe detectors. Therefore, to search the millicharged neutrino in direct detection experiments, we should focus on the low-energy transfer region. Furthermore, figure \ref{Millicharge count rate figure2} shows that, in the low-energy transfer region, the differential event rates $dR/dT$ calculated using RIA and EPA approaches are larger than those from FEA results, indicating the atomic many-body effects could greatly enhance the atomic ionization process induced by millicharged neutrino in the low-energy transfer region. This totally agree with the conclusions for millicharged dark matter particles discussed in subsection \ref{sec:5b}. For our RIA results, they converge to the FEA results as energy transfer $T$ increases. Meanwhile, our RIA results slowly approach to the EPA results when energy transfer $T$ becomes very small, but the convergence between RIA results and EPA results in the ultra-low-energy region (in $T\rightarrow0$ limit) is not as good as those of millicharged dark matter particles presented in the figure \ref{Millicharge count rate figure} in subsection \ref{sec:5b}, as well as in figures \ref{Millicharge count rate figure more1}-\ref{Millicharge count rate figure more5} in appendix \ref{appendix4}.

Similarly, according to the calculated reaction event rates in HPGe and LXe detectors, we can give an estimation of the detecting sensitivity on neutrino millicharge $\delta_{\nu}$ in next-generation HPGe and LXe based experiments. In the next-generation direct detection experiments, we assume that the energy threshold and background level for HPGe based experiments would be 100 eV and 0.1 cpkkd. Meanwhile, the energy threshold and background level for LXe based experiments would reach 500 eV and 10$^{-4}$ cpkkd. From figure \ref{Millicharge count rate figure2}, it is clearly manifested that, for neutrino mass $m_{\nu}c^{2}=0.1$ eV, the reaction event rates in HPGe and LXe detectors in energy range above the experimental thresholds successfully suppress the experimental background levels. Therefore, the next-generation direct detection experiments have the ability to push the detecting sensitivity of neutrino millicharge to $\delta_{\nu} \sim 10^{-12}$ for HPGe based experiments and $\delta_{\nu} \sim 2.5\times10^{-13}$ for LXe based experiments.

There is one important point should be noted. Since the flux spectrum of solar neutrino is irrelevant to the neutrino mass $m_{\nu}$, both the reaction event rates in HPGe and LXe detectors and the estimated detecting sensitivities on neutrino millicharge $\delta_{\nu}$ in next-generation experiments do not have an obvious dependency on neutrino mass $m_{\nu}$ \footnote{Furthermore, the differential cross section of atomic ionization process induced by millicharged neutrinos does not show obvious differences when neutrino mass $m_{\nu}$ varies. The mass of neutrino is much too small compared with its energy $E_{\nu}$ in solar neutrino spectrum, and it is approximately massless in these cases.}. This is different with the cases of millicharged dark matter particles. Firstly, the dark matter particle flux spectrum in Eq. (\ref{dark cosmic ray}) is manifestly mass and energy dependent. Secondly, if the mass of millicharge dark matter particle is smaller, then the minimal energy $E_{\chi}^{\text{min}}$ of imcoming dark matter particles, which has been set as $E_{\chi}^{\text{min}}=10m_{\chi}c^{2}$ to ensure the ultrarelativistic property of millicharge dark matter particles, becomes lower and give rise to large numbers of low-energy millicharged dark matter particles according to the power law in the flux spectrum in Eq. (\ref{dark cosmic ray}). These points make the reaction event rates in HPGe and LXe detectors and the estimated detecting sensitivity on dark matter particle millicharge $\delta_{\chi}$ in next-generation HPGe and LXe based experiments highly depend on dark matter particle mass $m_{\chi}$.

\begin{table}
\centering
\caption{Detecting sensitivity on neutrino millicharge $\delta_{\nu}$ in the next-generation direct detection experiments. The results for HPGe and LXe based experiments in the FEA, EPA and RIA calculations are given in this table. In the HPGe based next-generation experiments, the energy threshold and background level would be 100 eV and 0.1 cpkkd, respectively. The energy threshold and background level for LXe based next-generation experiments could reach 500 eV and $10^{-4}$ cpkkd. The current experimental bounds on neutrino millicharge $\delta_{\nu}$ obtained from direct detection experiments are also given for comparisons. In this table, for the XENON1T, PandaX, projected DARWIN and projected LX experimental bounds, we present the results in reference \cite{Hsieh2019} analyzed using the experimental data.}
\label{table2}
\vspace{5mm}
%\begin{ruledtabular}
%\begin{tabular}{lccccccccc}
\begin{tabular}{|cccc|}
\hline
& \multicolumn{2}{c}{HPGe Detector} &
\\
\hline
& Group & detecting sensitivity on neutrino millicharge\ $\delta_{\nu}$ &
\\
\hline
& Our Results & $2.5\times10^{-12}$ (FEA Results) &
\\
&             & $7\times10^{-13}$ (EPA Results) &
\\
&             & $1\times10^{-12}$ (RIA Results) &
\\
& TEXONO \cite{Chen2014} & $2.1\times10^{-12}$  &
\\
& GEMMA \cite{GEMMA}     & $1.5\times10^{-12}$ / $2.7\times10^{-12}$ (based on different methods) &
\\
\hline
& \multicolumn{2}{c}{LXe Detector} &
\\
\hline
& Group & detecting sensitivity on neutrino millicharge\ $\delta_{\nu}$ &
\\
\hline
& Our Results & $4\times10^{-13}$ (FEA Results) &
\\
&             & $9\times10^{-14}$ (EPA Results) &
\\
&             & $2.5\times10^{-13}$ (RIA Results) &
\\
& XENON1T \cite{XENON}           & $6.4\times10^{-13}$ &
\\
& PandaX \cite{PandaX2017}       & $2.06\times10^{-12}$ &
\\
& Projected DARWIN \cite{DARWIN} & $2.4\times10^{-13}$ &
\\
& Projected LZ \cite{LZ}         & $2.8\times10^{-13}$ &
\\
\hline
\end{tabular}
%\end{ruledtabular}
\end{table}

Based on the numerical calculations presented in this subsection, in table \ref{table2}, we give the comparison between our estimated detecting sensitivities in next-generation experiments and current experimental bounds on neutrino millicharge $\delta_{\nu}$ in the direct detection experiments. From this table, the estimated detecting sensitivity on neutrino millicharge $\delta_{\nu}$ in next-generation experiments is roughly 2-3 times smaller than the current best experiment bound \footnote{Very recently, an excess of electron recoil events was reported in the XENON1T experiment \cite{XENON2}. Amir N. Khan interpreted these signals to be nonstandard neutrino interactions, and a constrain was given on neutrino millicharge $\delta_{\nu}$ based on the electron recoil excess in XENON1T experiment \cite{Khan2020}. Khan's results indicated that neutrino millicharge would be $\delta_{\nu}=(1.7-2.3)\times10^{-12}$. However, other studies also suggested that the electron recoil excess maybe caused by some experimental backgrounds, which were ignored in the experimental analysis \cite{Bhattacherjee,Shitov}. Therefore, more experimental data are needed to confirm this excess.}. Therefore, the next-generation HPGe and LXe based experiments have the potential to make a great progress on the detecting ability of millicharged neutrinos.

Similar to the cases of millicharged dark mater particles presented in subsection \ref{sec:5c}, the numerical calculations of reaction event rates in HPGe and LXe detectors as well as the calculations of detection sensitivities on neutrino millicharge $\delta_{\nu}$ in next-generation HPGe and LXe based experiments are just a leading order estimation. In our calculations, the interactions between millicharged neutrinos and charged particles in earth atmosphere as well as the interactions between millicharged neutrinos and atoms and molecules in environmental rocks are not taken into considerations. These interactions may lead to an upper bound in the parameter space of millicharged neutrinos, as discussed in subsection \ref{sec:5c}.

\section{Summary and Conclusion \label{sec:7}}

In this work, we develop the RIA approach in the atomic ionization process induced by millicharged particles. Our approach is inspired and benefited from many-body physics, because the RIA approach was originally invented in atomic and molecular physics to study the various electromagnetic interactions in atoms and molecules. In the experimental detections for millicharged particles, atomic many-body effects play a crucial role and they cannot be arbitrarily neglected in theoretical calculations, especially when energy transfer in the scattering process is not very high (comparable to atomic binding energies). Our new developed RIA approach could effectively handle with atomic many-body effects in the entire energy region, therefore it can give a more precise results than traditional FEA approach in the study of millicharged particles. With this superiority, our new developed RIA approach would be helpful and have impacts on the theoretical predictions and experimental investigations for millicharged particles. In the present work, the formulation of RIA is derived for atomic ionization process induced by millicharged particles, and a numerical program is developed based on our RIA approach. The numerical results obtained using our RIA approach are compared with those from FEA and EPA approaches. Concretely, we study the atomic ionization process induced by millicharged dark matter particles as well as millicharged neutrinos in HPGe and LXe detectors. The differential cross section with respect to energy transfer, the differential reaction event rate in HPGe and LXe detectors, and the estimated detecting sensitivity for next-generation HPGe and LXe based experiments are presented in this work. In addition, to incorporate relativistic effects of atomic electrons from the beginning, the fully relativistic Dirac-Fock theory is used to obtain the ground state wavefunctions, electron momentum distributions and atomic Compton profiles for atomic systems.

For the differential cross section $d\sigma/dT$, when energy transfer is smaller than the binding energies $E_{1s}^{B}$ for $1s$ electron (which is 11.1 keV for Ge atom and 34.5 keV for Xe atom), our RIA results present large discrepancies with respect to the FEA results. In this region, the atomic many-body effects play a significant role in the atomic ionization process induced by millicharged particles, which lead to the breaking down of FEA approach. On the other hand, in high-energy transfer region, where electrons can be treated as free electrons approximately, there are no notable differences between our RIA results and FEA results. In the ultra-low-energy transfer region where EPA approach is derived (namely in $T \rightarrow 0$ limit), our RIA results does not exhibit large differences with respect to EPA results. The above conclusions show that our RIA approach developed in this work is valid and could treat the atomic many-body effects efficiently in the entire region of energy transfer $T$. Another important point is that both RIA and EPA results give larger cross sections than FEA results in the low-energy transfer region, indicating that atomic many-body effects could greatly enhance the atomic ionization process induced by millicharged particles. Furthermore, for the reaction event rates $dR/dT$ in HPGe and LXe detectors, numerical calculations present the similar phenomenon and tendency with those obtained from differential cross sections.

According to the calculated reaction event rates in HPGe and LXe detectors, we give an estimation of the detecting sensitivity on dark matter particle and neutrino millicharge $\delta_{\chi}$ and $\delta_{\nu}$ in next-generation HPGe and LXe based experiments. The energy threshold and background level for next-generation HPGe based experiments are postulated as 100 eV and 0.1 cpkkd, and the energy threshold and background level for next-generation LXe based experiments are postulated as 500 eV and $10^{-4}$ cpkkd, respectively. Our numerical results show that, with relatively lower backgrounds, the next-generation LXe based experiments probably could give a better constrain on neutrino and dark matter particle millicharge. For millicharged dark matter particles, the next-generation LXe based experiments would give a tremendous improvement on dark matter particle millicharge $\delta_{\chi}$ in mass range $0.01\leq m_{\chi}c^{2} \leq 100$ keV for direct detection experiments, with estimated detecting sensitivity 2-3 orders of magnitude smaller than the current best experimental bound in direct detection experiments. For millicharged neutrino, the estimated detection sensitivity of neutrino millicharge $\delta_{\nu}$ for next-generation LXe based experiments would reach $\delta_{\nu} \sim 2.5 \times 10^{-13}$, which is improved by roughly 2-3 times than the current best experimental bound.

In particular, our RIA approach developed in the present work is quite general, neither depends on the underling nature and mechanisms for millicharged particles, nor on the atomic and molecular composition of detector materials. In this work, we choose HPGe and LXe detectors to given a comprehensive study on the atomic ionization process induced by millicharged particles. Other detector materials, such as liquid argon (LAr), sodium iodide (NaI) and cesium iodide (CsI), still deserve to study in the following works. Furthermore, the physical ideas and formulation for RIA approach can also be applied to other electromagnetic interactions for millicharged particles, e.g. the atomic Compton scattering between millicharged particles and dark photons, the magnetic moment interactions for millicharged neutrinos, as well as other processes relevant to the millicharged particles and detector atoms or molecules. We wish our work could enlarge the understanding of millicharged particles and push forward studies in the related fields.

\appendix
%\section{Some title}
%Please always give a title also for appendices.

\section{Free Electron Approximation \label{appendix1}}

In this Appendix, we give a brief description of the free electron approximation (FEA) in the atomic ionization process induced by millicharged particles. In the FEA formulation, atomic electrons are treated as free electrons, and atomic bindings, electron shielding, electron correlation as well as other many-body effects are neglected.

In the atomic ionization process induced by millicharged particles $\chi + A \rightarrow \chi + A^{+} + e^{-}$, assuming the electric charge of millicharged particle is $q_{\chi}=\delta_{\chi}e$, the differential cross section for this process in the FEA formulation can be expressed as \cite{Singh2019}:
\begin{equation}
\bigg( \frac{d\sigma}{dT} \bigg)_{\text{FEA}}
= \frac{\pi r_{0}^{2} \delta_{\chi}^{2}}
       {T^{2}(E_{\chi}^{2}-m_{\chi}^{2}c^{4})}
  \bigg[
    m_{e}c^{2} \big( E_{\chi}^{2}+(E_{\chi}-T)^{2} \big)
    -T(m_{e}^{2}c^{4}+m_{\chi}^{2}c^{4})
  \bigg]
\label{millicharge FEA0a}
\end{equation}
where $m_{e}$ and $m_{\chi}$ are the mass of electron and millicharged particle, $E_{\chi}$ is the incoming energy of millicharged particle, and $T=E_{\chi}-E_{\chi}'$ is the energy transfer in the atomic ionization process. When the millicharged particle mass $m_{\chi}$ and energy transfer $T$ are both sufficiently small, namely the conditions $m_{\chi} \ll m_{e}$, $T \ll m_{e}c^{2}$, $T \ll E_{\chi}$ are satisfied, the above differential cross section in Eq. (\ref{millicharge FEA0a}) for ultra-relativistic millicharged particles (with $m_{\chi}c^{2} \ll E_{\chi}$) can be simplified as \cite{Chen2014}:
\begin{equation}
\bigg( \frac{d\sigma}{dT} \bigg)_{\text{FEA}}
= 2\pi r_{0}^{2}\delta_{\chi}^{2}\frac{m_{e}c^{2}}{T^{2}}
\label{millicharge FEA0b}
\end{equation}

The above results in Eqs. (\ref{millicharge FEA0a}) and (\ref{millicharge FEA0b}) only correspond to the single-electron system. To calculate the differential cross section for multi-electron atomic systems, contributions from all subshell electrons should be summed over to give the following results
\begin{eqnarray}
\bigg( \frac{d\sigma}{dT} \bigg)_{\text{FEA}}
& = &
\sum_{njl} Z_{njl} \bigg( \frac{d\sigma_{njl}}{dT} \bigg)_{\text{FEA}} \nonumber
\\
& = &
\sum_{njl} Z_{njl} \Theta(T-E^{B}_{njl})
\frac{\pi r_{0}^{2}\delta_{\chi}^{2}}
     {T^{2}(E_{\chi}^{2}-m_{\chi}^{2}c^{4})} \nonumber
\\
&   &
\times
\bigg[
  m_{e}c^{2} \big( E_{\chi}^{2}+(E_{\chi}-T)^{2} \big)
  -T(m_{e}^{2}c^{4}+m_{\chi}^{2}c^{4})
\bigg]
\label{millicharge FEA}
\end{eqnarray}
with $E^{B}_{njl}$ and $Z_{njl}$ to be the atomic binding energy and number of electron in $(njl)$ subshell. Similarly, when $m_{\chi} \ll m_{e}$, $T \ll m_{e}c^{2}$, $T \ll E_{\chi}$ are satisfied, the differential cross section can be simplified as:
\begin{equation}
\bigg( \frac{d\sigma}{dT} \bigg)_{\text{FEA}}
= \sum_{njl} Z_{njl} \Theta(T-E^{B}_{njl})
  \times 2\pi r_{0}^{2}\delta_{\chi}^{2}\frac{m_{e}c^{2}}{T^{2}}
\label{millicharge FEA2}
\end{equation}

The FEA formulation works well when the energy transfer $T$ is much larger than the atomic binding energy, in which cases the atomic many-body effects can be neglected and electrons are approximately free. Previous studied have confirmed that when energy transfer $T$ is comparable to the atomic binding energy, FEA results underestimate the differential cross section in the atomic ionization process induced by millicharged particles \cite{Chen2013,Chen2015,Singh2019}, which implies atomic effects could intensify the atomic ionization process $\chi + A \rightarrow \chi + A^{+} + e^{-}$. This conclusion is consistent with our numerical results presented in section \ref{sec:5} and section \ref{sec:6}.

\section{Equivalent Photon Approximation \label{appendix2}}

Equivalent Photon Approximation (EPA) is an approaches widely used in nuclear and elementary particle physics, especially in Quantum Electrodynamics (QED) and Quantum Chromodynamics (QCD) calculations \cite{Peskin,Greiner}.

In the EPA formulation, considering the process $e^{-} + X \rightarrow e^{-} + Y$. The electron scatters with particle $X$ by exchanging virtual photons. In the low-momentum transfer limit $q \rightarrow 0$, the contribution from the longitudinal polarized virtual photons vanishes, and the contribution coming from virtual photons can be equivalent to those from real photons. Therefore, the cross section of the whole process $e^{-} + X \rightarrow e^{-} + Y$ can be calculated from the cross section of its subprocess $\gamma + X \rightarrow Y$ \cite{Schwartz,Peskin}:
\begin{eqnarray}
\sigma(e^{-} + X \rightarrow e^{-} + Y)
& = &
\sigma(\gamma + X \rightarrow Y)
\times
\int_{0}^{1}dz
\frac{\alpha_{\text{em}}}{2\pi}
\log \bigg( \frac{s}{m_{e}^{2}c^{4}} \bigg)
\bigg[ \frac{1+(1-z)^{2}}{z} \bigg] \nonumber
\\
& = &
\sigma(\gamma + X \rightarrow Y)
\times
\int_{0}^{1}f_{\gamma}(z)dz
\label{Equivalent Photon Approximation0}
\end{eqnarray}
where $\alpha_{\text{em}}$ is the conventional fine-structure constant, $s=(p_{e}^{\mu}+p_{X}^{\mu})^{2}$ is the total energy square in the center-of-mass frame, $z=q/p_{e}$ is the ratio between and virtual photon momentum $q$ (the momentum transfer in the whole process $e^{-} + X \rightarrow e^{-} + Y$) and electron momentum $p_{e}$, and $\sigma(\gamma + X \rightarrow Y)$ is the cross section for subprocess $\gamma + X \rightarrow Y$. In the Eq. (\ref{Equivalent Photon Approximation0}), $f_{\gamma}(z)$ is the Weizsacker-Williams distribution function defined as:
\begin{equation}
f_{\gamma}(z)
= \frac{\alpha_{\text{em}}}{2\pi}
 \log \bigg( \frac{s}{m_{e}^{2}c^{4}} \bigg)
 \bigg[ \frac{1+(1-z)^{2}}{z} \bigg]
\end{equation}
It can be viewed as the probability of finding a photon with momentum $q=p_{z}$ from the incident electron beam \cite{Peskin}.

Similarly, in the atomic ionization process induced by millicharged particles $\chi + A \rightarrow \chi + A^{+} + e^{-}$, the EPA approach connects its cross section with the cross section of photoionization process $\gamma + A \rightarrow A^{+} + e^{-}$. In analogy with Eq. (\ref{Equivalent Photon Approximation0}), the cross section of atomic ionization process $\chi + A \rightarrow \chi + A^{+} + e^{-}$ can be expressed as:
\begin{eqnarray}
\sigma(\chi + A \rightarrow \chi + A^{+} + e^{-})
& = &
 \sigma(\gamma + A^{+} \rightarrow A^{+} + e^{-})
 \times
 \int_{0}^{1}dz
 \frac{\alpha_{\chi}}{2\pi}
 \log \bigg( \frac{s}{m_{\chi}^{2}c^{4}} \bigg)
 \bigg[ \frac{1+(1-z)^{2}}{z} \bigg] \nonumber
\\
\label{Equivalent Photon Approximation}
\end{eqnarray}
Here, $\alpha_{\chi}$ is the ``fine structure constant'' in the electromagnetic interactions induced by millicharged particles. It is defined as:
\begin{equation}
\alpha_{\chi}=\frac{q_{\chi}^{2}}{4\pi\epsilon_{0}\hbar c}=\delta_{\chi}^{2}\alpha_{\text{em}}
\end{equation}
where $q_{\chi}=\delta_{\chi}e$ is the electric charge of the millicharged particle.

To extract differential cross section in the EPA formulation, we consider the case that both energy transfer $T$ and momentum transfer $q$ is sufficiently small, namely in the limit
\begin{equation}
q^{2}=2m_{e}T\rightarrow 0,\ \ \ z = \frac{q}{p_{\chi}} \rightarrow 0  \label{EPA low energy transfer limit}
\end{equation}
In this case, the total cross section for photoionization process can be simplified as \cite{Chen2013,Chen2015,Singh2019}:
\begin{eqnarray}
\sigma(\gamma + A^{+} \rightarrow A^{+} + e^{-})
\approx
\sigma_{\text{abs}}^{\gamma}(T)
\approx
\frac{2\pi^{2}\alpha_{\text{em}}}{T}R_{T}^{0}(q^{2}=0)
\end{eqnarray}
with $\sigma_{\text{abs}}^{\gamma}(T)$ to be the total cross section for photoabsorption process at incident photon energy $T$, and $R_{T}^{0}(q^{2}=0)$ to be the atomic transverse response function for on-shell real photons in zero-momentum transfer cases.
In this work, we only duel with the case that the mass of millicharged particle $m_{\chi}$ is tiny and much smaller than the mass of detector atom $m_{A}$. In this case, laboratory frame can be viewed as the center-of-mass frame approximately, and the total energy square in the center-of-mass frame can be simplified as $s=(p_{e}^{\mu}+p_{X}^{\mu})^{2} \approx E_{\chi}^{2}$ \footnote{It has been assumed that, before the scattering, the atom A in detector materials is at rest in the laboratory frame.}.

In Eq. (\ref{Equivalent Photon Approximation}), a divergent part arise in the $z \rightarrow 0$ limit. This divergent part can be cancelled by the higher order corrections with the help of renormalization. Finally, after tidies calculations, the differential cross section of the atomic ionization process induced by millicharged particles in the EPA approach can be expressed as \cite{Chen2015}:
\begin{equation}
\bigg( \frac{d\sigma}{dT} \bigg)_{\text{EPA}}
= \delta_{\chi}^{2} \frac{2\alpha_{\text{em}}}{\pi} \frac{\sigma_{\text{abs}}^{\gamma}(T)}{T}
  \log \bigg( \frac{E_{\chi}}{m_{\chi}c^{2}} \bigg)
\label{millicharge EPA}
\end{equation}

From the discussions above, it can be clearly shown that the derivation of Eq. (\ref{millicharge EPA}) requires that the energy transfer $T$ and the momentum transfer $q$ to be sufficiently small (in the $q \rightarrow 0$ and $T \rightarrow 0$ limits). Therefore, the EPA formulation only works well in the ultra-low-energy transfer region ($T \rightarrow 0$). When energy transfer $T$ is large, EPA results would bring about large discrepancies, and this point has been confirmed by recent researches \cite{Chen2015}. This conclusion is also consistent with our numerical results presented in section \ref{sec:5} and section \ref{sec:6}.

\section{Dirac-Fock Theory \label{appendix3}}

In this appendix, we give an introduction of the relativistic Dirac-Fock theory. We will focus on the construction of Dirac-Fock Hamiltonian for atomic systems, and how to calculate ground state wavefunctions, electron momentum distributions and atomic Compton profiles using the Dirac-Fock theory.

The Dirac-Fock theory \cite{Grant1961,Desclaux1971,Desclaux,Visscher}, which is a relativistic extension of the nonrelativistic Hartree-Fock self-consistent method, is commonly used in \emph{ab initio} calculations in atomic and molecular physics. In the last few decades, it has been confirmed by a number of experiments and has become a milestone in atomic and molecular physics \cite{Grant2007,Zanna,Grant,grasp2K}. In this work, the Dirac-Fock theory is used to obtain the ground state wavefunctions, electron momentum distributions and Compton profiles for atomic systems.

In the Dirac-Fock theory, the total Hamiltonian for atomic systems is given by \cite{Desclaux,Grant,grasp2K}:
\begin{equation}
H_{\text{atom}}^{\text{Dirac-Fock}} = \sum_{a=1}^{Z}h_{\text{(a)}} + \sum_{a=1}^{Z}\sum_{a<b}h_{\text{(ab)}}
\end{equation}
Here, $h_{\text{(a)}}$ is the single-particle Hamiltonian for the $a$-th election, which includes the kinetic energy for $a$-th electron and the Coulomb potential between the atomic nuclei and this electron. The $h_{\text{(ab)}}$ is the two-particle Hamiltonian, which is the interaction between the $a$-th and $b$-th electrons. In the relativistic case, the single-particle Hamiltonian is given by:
\begin{eqnarray}
h_{\text{(a)}}^{\text{relativistic}} & = & \boldsymbol{\alpha}_{\text{(a)}} \cdot p_{\text{(a)}}c+\beta m_{e}c^{2}+V_{\text{nucl}}(r) \nonumber
\\
                                     & = & -ic\hbar \boldsymbol{\alpha}_{\text{(a)}} \cdot \nabla_{\text{(a)}}
                                           +\beta m_{e}c^{2}
                                           -\frac{1}{4\pi\epsilon_{0}}\frac{Ze^{2}}{r_{\text{(a)}}}
\label{single-electron interaction}
\end{eqnarray}
where $\boldsymbol{\alpha}_{\text{(a)}}$ is the conventional Dirac--$\alpha$ matrices for $a$-th electron, $\beta$ is the Dirac--$\beta$ matrix, the symbol $\nabla_{\text{(a)}}$ represent the gradient operator for $a$-th electron, and $r_{\text{(a)}}$ is the radius between atomic nuclei and the $a$-th electron. In this work, we only consider the leading order Coulomb interactions between electrons. Therefore, the two-particle Hamiltonian can be simply expressed as \footnote{More generally, in relativistic cases, the interactions between two electrons contain the Coulomb interaction, which is expressed in Eq. (\ref{two-electron interaction}), and the Breit interaction \cite{Grant2007,Johnson,Chantler}. For simplicity, the Dirac-Fock Hamiltonian written here does not include the Breit interaction.}:
\begin{equation}
h_{\text{(ab)}} = \frac{1}{4\pi\epsilon_{0}}\frac{e^{2}}{r_{\text{(ab)}}}
                = \frac{1}{4\pi\epsilon_{0}}\frac{e^{2}}{|r_{\text{(a)}}-r_{\text{(b)}}|}
\label{two-electron interaction}
\end{equation}
where $r_{\text{(ab)}}$ is the distance between the $a$-th and $b$-th electron.

With the single-particle Hamiltonian $h_{\text{(a)}}$ and two-particle Hamiltonian $h_{\text{(ab)}}$ given in Eqs. (\ref{single-electron interaction}) and (\ref{two-electron interaction}), the total Hamiltonian for atomic systems in the Dirac-Fock theory can be expressed as \cite{Desclaux,Grant,grasp2K}:
\begin{eqnarray}
H_{\text{atom}}^{\text{Dirac-Fock}} & = & \sum_{a=1}^{Z}h_{\text{(a)}} + \sum_{a=1}^{Z}\sum_{a<b}h_{\text{(ab)}} \nonumber
\\
                                    & = & \sum_{a=1}^{Z}
                                          \bigg[
                                            -ic\hbar \boldsymbol{\alpha}_{\text{(a)}} \cdot \nabla_{\text{(a)}}
                                            +\beta m_{e}c^{2}
                                            -\frac{1}{4\pi\epsilon_{0}}\frac{Ze^{2}}{r_{\text{(a)}}}
                                          \bigg]
                                          +\sum_{a=1}^{Z}\sum_{a<b}
                                           \frac{1}{4\pi\epsilon_{0}}\frac{e^{2}}{|r_{\text{(a)}}-r_{\text{(b)}}|} \nonumber
\\
\label{Dirac-Fock Hamiltonian}
\end{eqnarray}

In the following part, we give a description on how to calculate the ground state energies and wavefunctions for atomic systems in the Dirac-Fock theory. To satisfy the Pauli exclusion principle, in Dirac-Fock theory, the total ground state wavefunctions for atomic systems can be constructed through the Slater determinant of single-electron wavefunctions
\begin{eqnarray}
\Psi(\boldsymbol{r}_{\text{(1)}},\boldsymbol{r}_{\text{(2)}},\cdots,\boldsymbol{r}_{\text{(Z)}})
= \frac{1}{\sqrt{Z!}}
  \left|
    \begin{array}{cccc}
      u_{\text{(1)}}(\boldsymbol{r}_{\text{(1)}}) & u_{\text{(1)}}(\boldsymbol{r}_{\text{(2)}}) & \cdots & u_{\text{(1)}}(\boldsymbol{r}_{\text{(N)}}) \\
      u_{\text{(2)}}(\boldsymbol{r}_{\text{(1)}}) & u_{\text{(2)}}(\boldsymbol{r}_{\text{(2)}}) & \cdots & u_{\text{(2)}}(\boldsymbol{r}_{\text{(N)}}) \\
      \vdots & \vdots & \ddots & \vdots \\
      u_{\text{(N)}}(\boldsymbol{r}_{\text{(1)}}) & u_{\text{(N)}}(\boldsymbol{r}_{\text{(2)}}) & \cdots & u_{\text{(N)}}(\boldsymbol{r}_{\text{(N)}}) \\
    \end{array}
  \right|
\label{Atomic wavefunction}
\end{eqnarray}
In the expression, $u_{\text{(a)}}(\boldsymbol{r}_{\text{(a)}})$ is the single-electron wavefunction for $a$-th electron, and $\boldsymbol{r}_{\text{(a)}}$ is the position of $a$-th electron (with the center of atomic nucleus set as the coordinate origin).

In this work, we only consider the spherical symmetric atomic systems. Therefore, the single-electron wavefunction for atomic ground state with definite quantum number $(n\kappa m)=(njlm)$, which is also called as the Dirac orbital, has the following form \cite{Desclaux,Grant,HuangSpin}:
\begin{equation}
u_{n\kappa m}(\boldsymbol{r}) = u_{n\kappa m}(r,\theta,\phi)
                              = \frac{1}{r}
                                \left[
                                \begin{array}{cc}
                                  G_{n\kappa}(r)\Omega_{\kappa m}(\theta,\phi) \\
                                  iF_{n\kappa}(r)\Omega_{-\kappa m}(\theta,\phi)
                                \end{array}
                                \right]
\label{Dirac orbital}
\end{equation}
where $G_{n\kappa}(r)$ and $F_{n\kappa}(r)$ are the large and small components respectively, $\Omega_{\kappa m}(\theta,\phi)$ is normalized spherical spinor defined as:
\begin{equation}
\Omega_{\kappa m}(\theta,\phi) = \sum_{s_{z}=\mu}\langle l m-\mu; \frac{1}{2} \mu| jm \rangle Y_{lm}(\theta,\phi) \chi_{\mu}
\end{equation}
where $Y_{lm}(\theta,\phi)$ is the spherical harmonics, $\langle l m-\mu; \frac{1}{2} \mu| jm \rangle$ is the Clebsch-Gordan coefficient, and $\chi_{\mu}$ is a spinor with $s=1/2$ and $s_{z}=\mu$.

In many cases, only the radial part of Dirac orbital need to be focused, and the angular part can be separated and neglected for simplicity. Therefore, we can introduce the following two-component radial Dirac orbital:
\begin{equation}
u_{n\kappa}(r) \equiv u_{njl}(r) = \left[
                                   \begin{array}{c}
                                     G_{njl}(r) \\
                                     F_{njl}(r)
                                   \end{array}
                                   \right]
\label{bound state radial orbital}
\end{equation}
After the introduction of Dirac orbital as well as its large and small components $F_{n\kappa}=F_{njl}$, $G_{n\kappa}=G_{njl}$, the Dirac-Fock equations for atomic systems can be expressed and solved routinely. The total energies and ground state wavefunctions for  atomic systems as well as the energy eigenvalues and single-electron wavefunctions for individual electrons can be obtained.

In the Dirac-Fock theory, the total energy for atomic system is calculated by solving the eigen-equation
\begin{equation}
H_{\text{atom}}^{\text{Dirac-Fock}}\Psi(\boldsymbol{r}_{\text{(1)}},\boldsymbol{r}_{\text{(2)}},\cdots,\boldsymbol{r}_{\text{(Z)}})
=
E_{\text{atom}}^{\text{Dirac-Fock}}\Psi(\boldsymbol{r}_{\text{(1)}},\boldsymbol{r}_{\text{(2)}},\cdots,\boldsymbol{r}_{\text{(Z)}})
\label{energy eigen equation}
\end{equation}
Put the Dirac-Fock Hamiltonian in Eq. (\ref{Dirac-Fock Hamiltonian}), atomic total wavefunction in Eq. (\ref{Atomic wavefunction}) and single-electron wavefunction in Eq. (\ref{Dirac orbital}) into Eq. (\ref{energy eigen equation}). After separation of variables for the angular part $\theta$ and $\phi$, the total energy for atomic system can be expressed as \cite{Grant1961,Desclaux1971}:
\begin{equation}
E_{\text{atom}}^{\text{Dirac-Fock}} = \sum_{p}
                                      \bigg[
                                        Z_{p}I(pp)
                                        +\sum_{q\geq p}\sum_{k=0,2,\cdots}^{k_{0}}f^{k}(pq)R^{k}(ppqq)
                                        +\sum_{q>p}\sum_{k=k_{1},k_{1}+2,\cdots}^{k_{2}}g^{k}(pq)R^{k}(pqpq)
                                      \bigg]
\label{atomic total energy}
\end{equation}
where $p$ is the abbreviation for subshell $(n_{p}\kappa_{p})=(n_{p}j_{p}l_{p})$, $Z_{p}=2j_{p}+1$ is the number of electrons in subshell $p=(n_{p}\kappa_{p})=(n_{p}j_{p}l_{p})$. The $f^{k}(pq)$ and $g^{k}(pq)$ are angular coefficients, which are calculated through Wigner--$3j$ coefficients, and $k_{0}$, $k_{1}$, $k_{2}$ are defined in reference \cite{Grant}. In the Eq. (\ref{atomic total energy}), integral $I(pq)$ gives rise to the one-body interaction, and Slater integral $R^{k}(pqrs)$ represents the two-body interaction. The explicit expressions for $I(pq)$ and $R^{k}(pqrs)$ can be found in reference \cite{Grant}.

The total energy value for atomic system $E_{\text{atom}}^{\text{Dirac-Fock}}$ relies on the large and small components of single-electron wavefunctions through the integrals $I(p)$ and $R^{k}(pqrs)$. The ground state wavefunctions for atomic system in Eq. (\ref{Atomic wavefunction}) as well as the ground state wavefunction for individual electron in Eq. (\ref{Dirac orbital}) should minimize the total energy $E_{\text{atom}}^{\text{Dirac-Fock}}$ in Eq. (\ref{atomic total energy}). After the variational method, the differential equations for large and small components of single-electron wavefunctions become \cite{Grant1961,Desclaux1971,Desclaux}:
\begin{subequations}
\begin{eqnarray}
\frac{dG_{n\kappa}}{dr}+\frac{\kappa}{r}G_{n\kappa}(r)
+ \bigg[ \frac{2m_{e}c}{\hbar} - \frac{\varepsilon_{n\kappa}}{c\hbar} + \frac{Y_{n\kappa}(r)}{rc} \bigg]
  F_{n\kappa}(r)
& = &
\frac{X^{(G)}_{n\kappa}(r)}{r}
\label{Dirac-Fock equation1}
\\
\frac{dF_{n\kappa}}{dr}-\frac{\kappa}{r}F_{n\kappa}(r)
+ \bigg[ \frac{\varepsilon_{n\kappa}}{c\hbar} - \frac{Y_{n\kappa}(r)}{rc} \bigg]
  G_{n\kappa}(r)
& = &
\frac{X^{(F)}_{n\kappa}(r)}{r}
\label{Dirac-Fock equation2}
\end{eqnarray}
\end{subequations}
where $\varepsilon_{n\kappa}$ is the energy eigenvalue for subshell $(n\kappa)=(njl)$. The $Y_{n\kappa}(r)$ is the direct potential acts on large and small components of Dirac orbital for $(n\kappa)$ subshell, while $X^{(G)}_{n\kappa}(r)$ and $X^{(F)}_{n\kappa}(r)$ are the exchange potentials act on the large and small components. The explicit expressions for $Y_{n\kappa}(r)$, $X^{(G)}_{n\kappa}(r)$ and $X^{(F)}_{n\kappa}(r)$ can be found in reference \cite{Desclaux1971,Desclaux,Grant}.

The above Eqs. (\ref{Dirac-Fock equation1})-(\ref{Dirac-Fock equation2}) are called the \textbf{Dirac-Fock equations}. The direct potential $Y_{n\kappa}(r)$ and exchange potential $X^{(G)}_{n\kappa}(r)$, $X^{(F)}_{n\kappa}(r)$ contain integral for large and small components $G_{n\kappa}$ and $F_{n\kappa}$, which makes Dirac-Fock equations a little more difficult to solve. In the numerical calculations, the Dirac-Fock equations can be solved by the following iterative method:
\begin{itemize}
\item First, pick the large component $G_{n\kappa}^{(0)}$, small component $F_{n\kappa}^{(0)}$ and energy eigenvalue $\varepsilon_{n\kappa}^{(0)}$ as trial solutions of Dirac-Fock equations. The direct potential $Y_{n\kappa}(r)$ and exchange potential $X^{(G)}_{n\kappa}(r)$, $X^{(F)}_{n\kappa}(r)$ can be calculated through these trial solutions $G_{n\kappa}^{(0)}$, $F_{n\kappa}^{(0)}$ and $\varepsilon_{n\kappa}^{(0)}$. Plug the calculated $Y_{n\kappa}(r)$, $X^{(G)}_{n\kappa}(r)$, $X^{(F)}_{n\kappa}(r)$ into Dirac-Fock equations (\ref{Dirac-Fock equation1})-(\ref{Dirac-Fock equation2}) and obtain the new solutions $G_{n\kappa}^{(1)}$, $F_{n\kappa}^{(1)}$, $\varepsilon_{n\kappa}^{(1)}$.
\item Take the solutions $G_{n\kappa}^{(1)}$, $F_{n\kappa}^{(1)}$, $\varepsilon_{n\kappa}^{(1)}$ in the first step as new trial solutions, then calculate the direct and exchange potentials $Y_{n\kappa}(r)$, $X^{(G)}_{n\kappa}(r)$, $X^{(F)}_{n\kappa}(r)$ with the help of $G_{n\kappa}^{(1)}$, $F_{n\kappa}^{(1)}$, $\varepsilon_{n\kappa}^{(1)}$. Plug the calculated $Y_{n\kappa}(r)$, $X^{(G)}_{n\kappa}(r)$, $X^{(F)}_{n\kappa}(r)$ into Dirac-Fock equations (\ref{Dirac-Fock equation1})-(\ref{Dirac-Fock equation2})and obtained the new solutions $G_{n\kappa}^{(2)}$, $F_{n\kappa}^{(2)}$, $\varepsilon_{n\kappa}^{(2)}$, the same as in the first step.
\item ......
\item Repeat the above procedures routinely. When the energy eigenvalue in the $i$ step $\varepsilon_{n\kappa}^{(i)}$ converges to the energy eigenvalue in the $i+1$ step $\varepsilon_{n\kappa}^{(i+1)}$, the correct energy eigenvalue $\varepsilon_{n\kappa}$ and the ground state wavefunction $u_{n\kappa}=(G_{n\kappa},F_{n\kappa})$ for each subshell electron are solved from Dirac-Fock equation, and this iterative algorithm is self-consistent.
\end{itemize}

Once Dirac-Fock equations are solved, the large component $G_{n\kappa}=G_{njl}$, small components $F_{n\kappa}=F_{njl}$ as well as the energy eigenvalue $\varepsilon_{n\kappa}=\varepsilon_{njl}$ for different subshell electrons are obtained. Therefore, the corresponding electron momentum wavefunctions are given by the following Fourier transformation \cite{Kahane}:
\begin{subequations}
\begin{eqnarray}
\phi_{njl}^{G}(p) & = & \sqrt{\frac{2}{\pi}} \int_{0}^{\infty} G_{nlj}(r)j_{l}(pr)r^{2}dr \\
\phi_{njl}^{F}(p) & = & \bigg\{
                          \begin{array}{cc}
                           \sqrt{\frac{2}{\pi}} \int_{0}^{\infty}F_{njl}(r)j_{l+1}(pr)r^{2}dr & j=l+\frac{1}{2} \\
                           \sqrt{\frac{2}{\pi}} \int_{0}^{\infty}F_{njl}(r)j_{l-1}(pr)r^{2}dr & j=l-\frac{1}{2}
                          \end{array}
\end{eqnarray}
\end{subequations}
where $\phi_{njl}^{G}$, $\phi_{njl}^{F}$ are the large and small components of electron momentum wavefunctions of $(njl)$ subshell, and $j_{l}(pr)$ is the spherical Bessel function. Based on electron momentum wavefunctions, the momentum distribution of electrons in atomic system is calculated as follows \cite{Desclaux,Qiao}:
\begin{subequations}
\begin{eqnarray}
\rho_{njl}(p) & = & |\phi_{njl}(\boldsymbol{p})|^{2}
                =   (\phi_{njl}^{G}(p))^{2}+(\phi_{njl}^{F}(p))^{2}
\\
\rho(p) & = & \sum_{a=1}^{Z}|\phi_{a}(\boldsymbol{p})|^{2}
          =   \sum_{njl}N_{njl}
              \bigg(
                (\phi_{njl}^{G}(p))^{2}+(\phi_{njl}^{F}(p))^{2}
              \bigg) \nonumber
\\
        & = & \sum_{njl}N_{njl}\rho_{njl}(p) \label{rho}
\end{eqnarray}
\end{subequations}

Finally, the atomic Compton profile defined in Eq. (\ref{Compton profile}) can be calculated through the integration of electron momentum distributions $\rho(p)$ and $\rho_{njl}(p)$.
\begin{subequations}
\begin{eqnarray}
J_{njl}(p_{z}) & = & 2\pi\int\limits_{|p_{z}|}^{\infty}p\rho_{njl}(p)dp
\\
J(p_{z}) & = & 2\pi\int\limits_{|p_{z}|}^{\infty}p\rho(p)dp
           =   \sum_{njl}N_{njl}J_{njl}(p_{z})
\end{eqnarray}
\end{subequations}
The atomic Compton profile, when plugged into Eqs. (\ref{RIA millicharge full})-(\ref{RIA simplified millicharge full}) and Eqs. (\ref{RIA millicharge full DCS})-(\ref{RIA simplified millicharge full DCS}), can give the differential cross section for the atomic ionization process induced by millicharged particles in RIA approach.

\section{Supplementary: More Figures on Reaction Event Rates in HPGe and LXe Detectors \label{appendix4}}

\begin{figure*}
\centering
\includegraphics[scale=0.4]{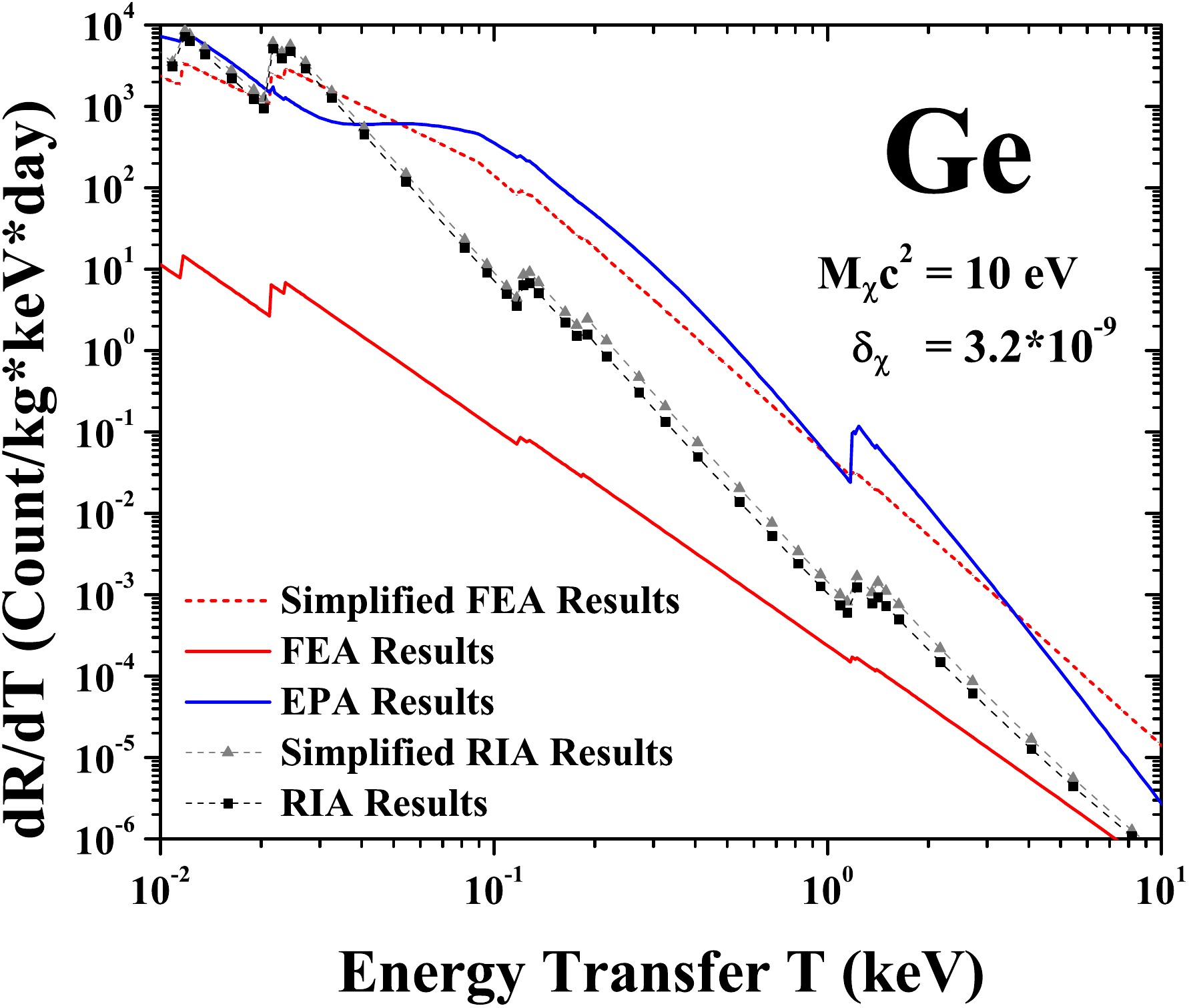}
\includegraphics[scale=0.4]{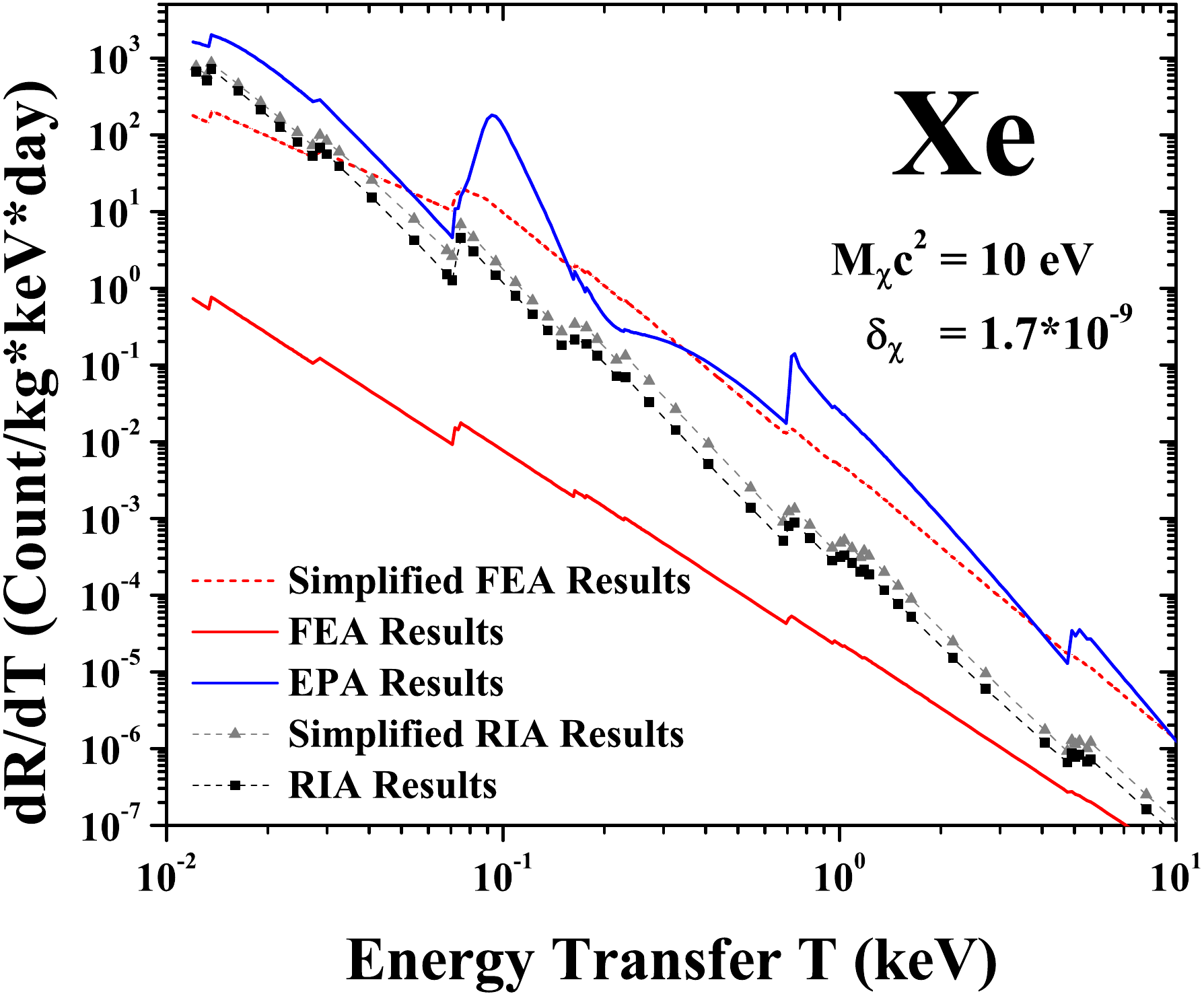}
\includegraphics[scale=0.4]{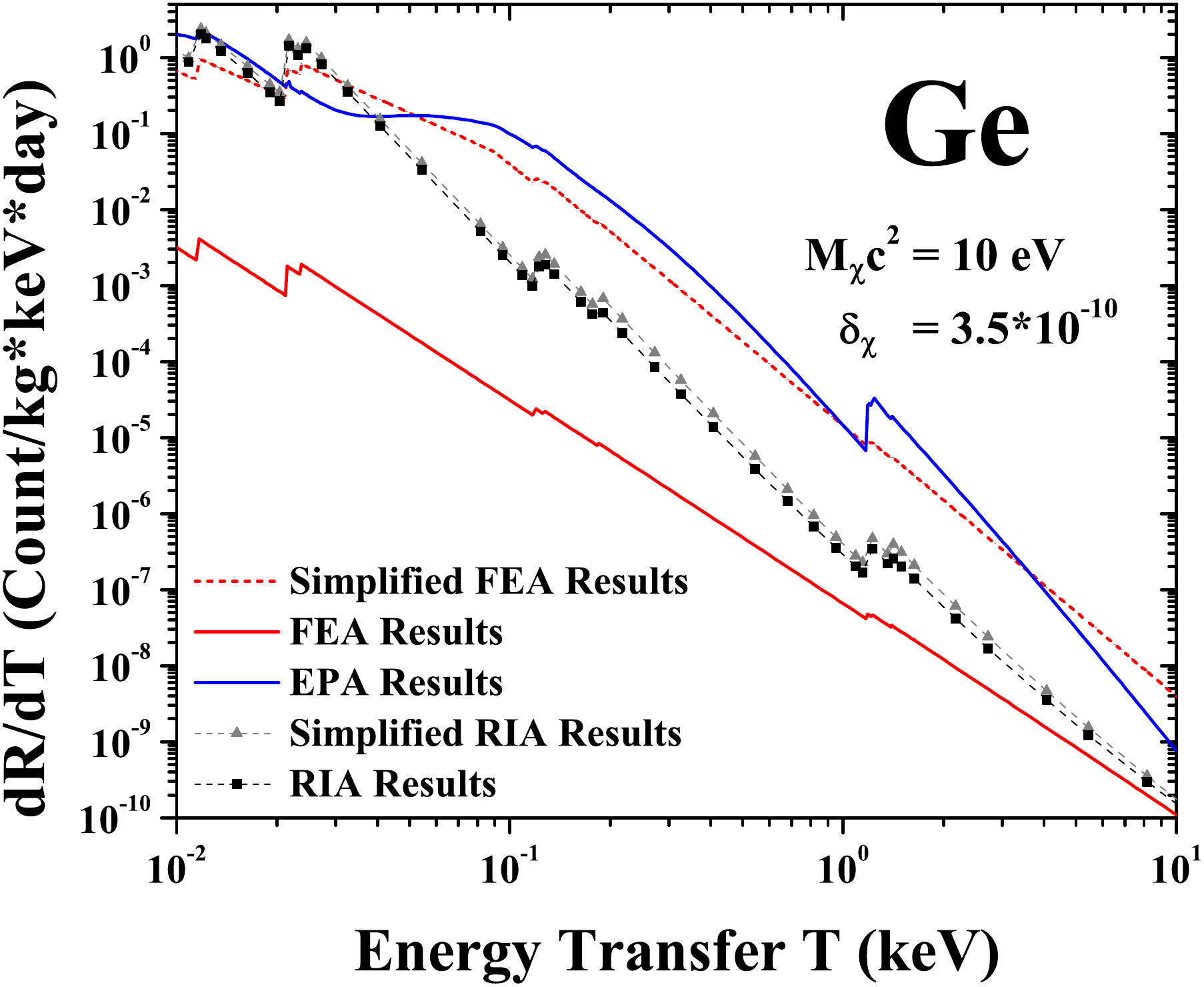}
\includegraphics[scale=0.4]{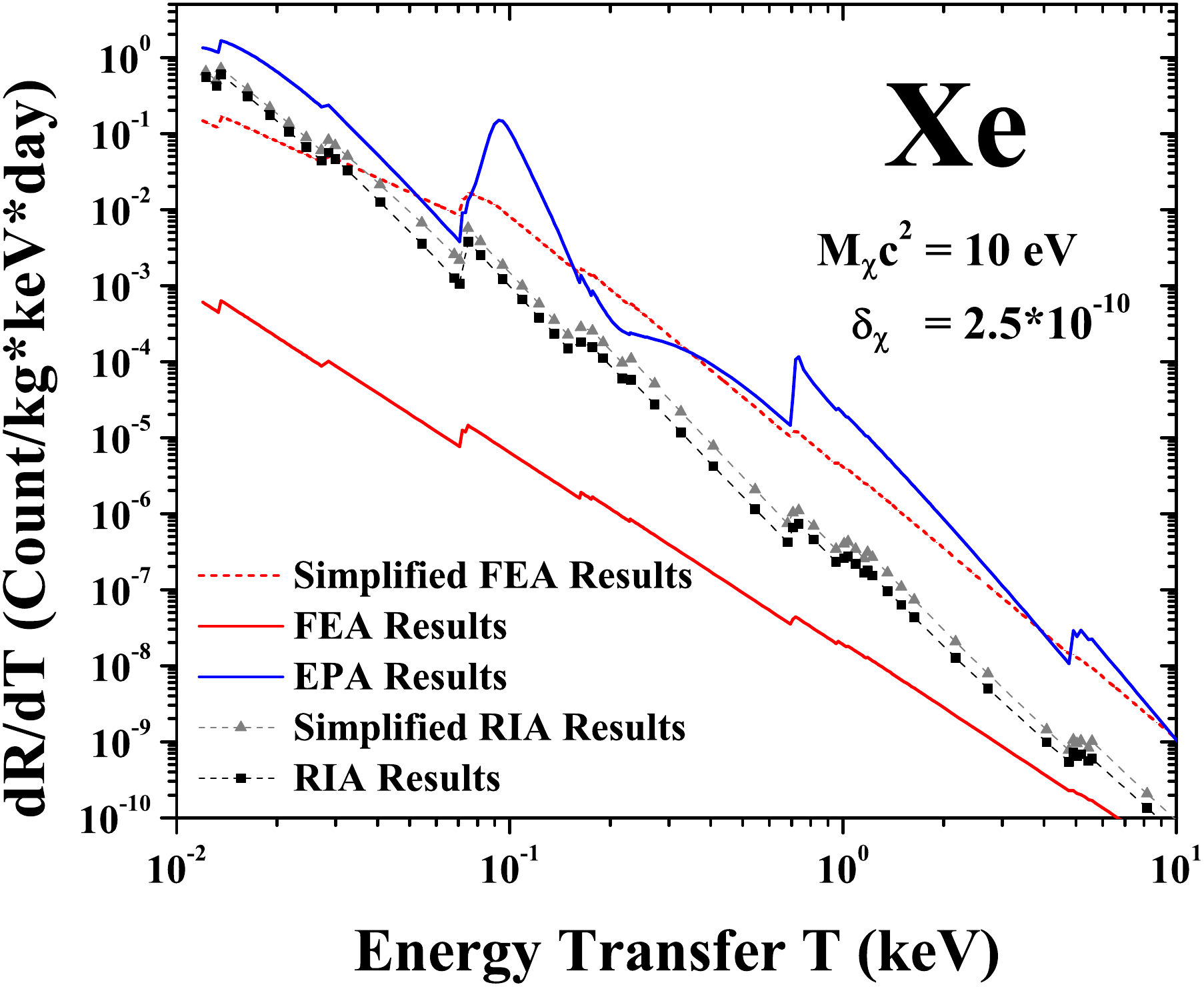}
\includegraphics[scale=0.4]{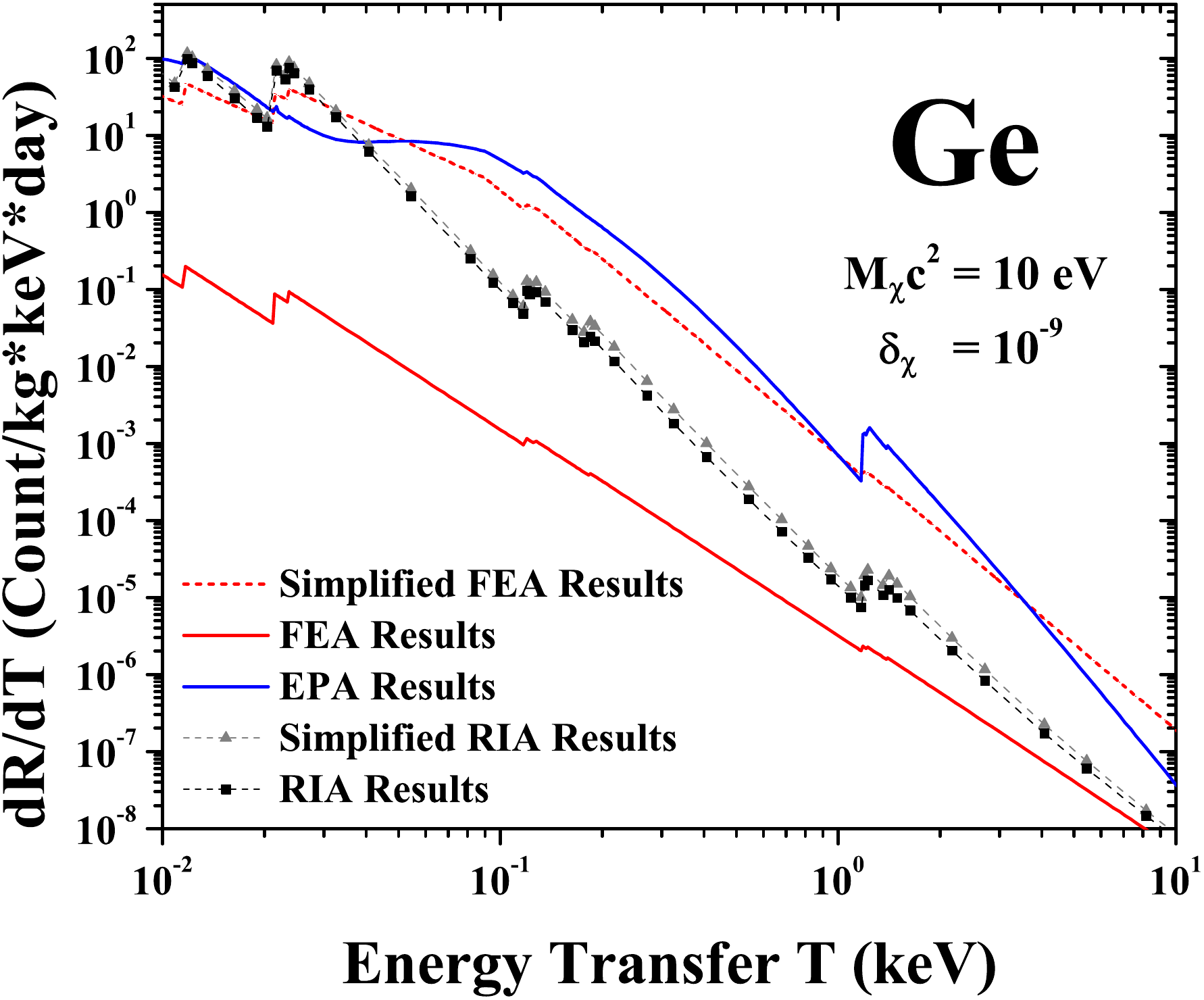}
\includegraphics[scale=0.4]{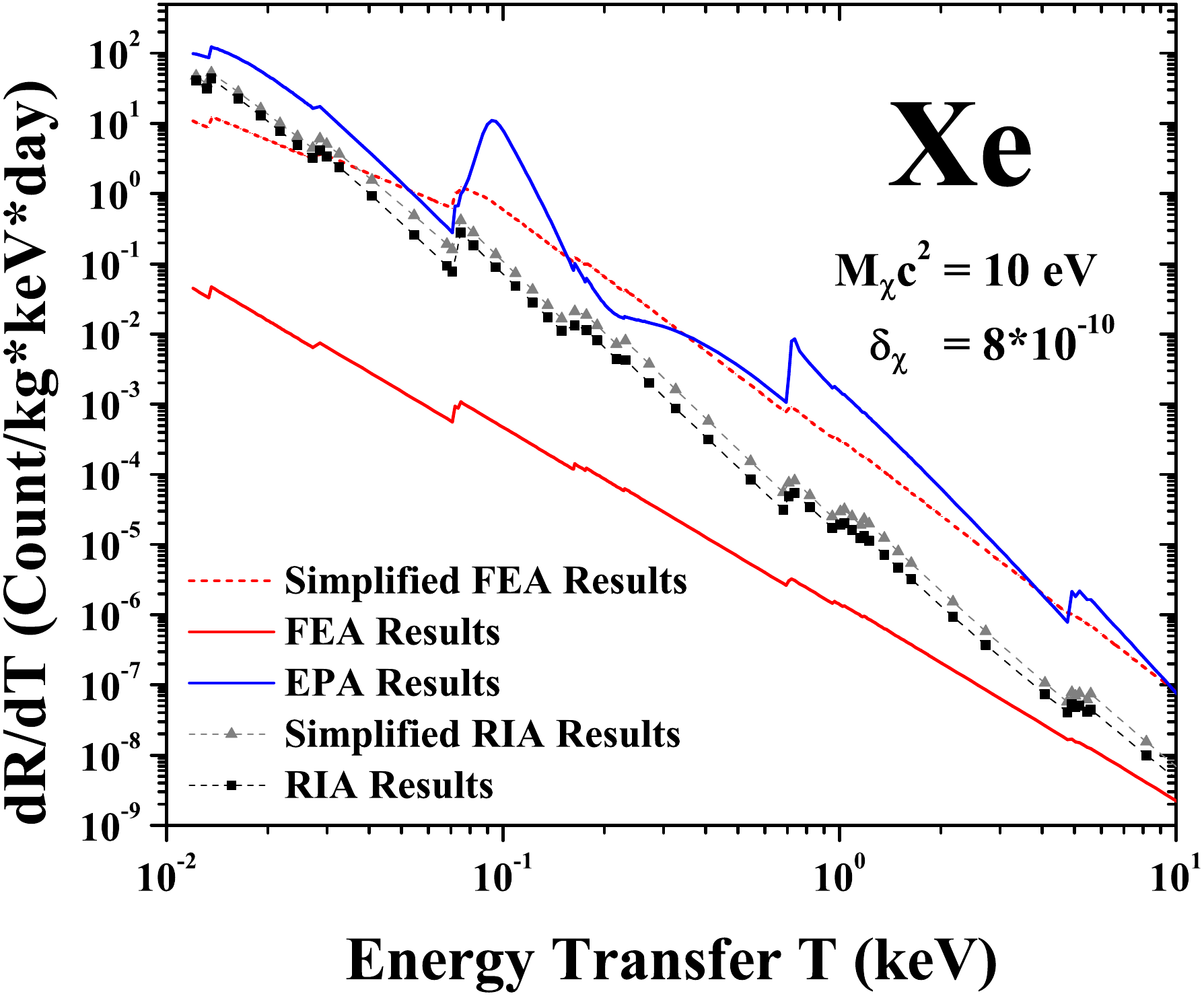}
\caption{The differential reaction event rates for the atomic ionization process induced by millicharged particles in HPGe and LXe detectors. The mass of millicharged dark matter particle is set as $m_{\chi}c^{2}=10$ eV, and its millicharge $\delta_{\chi}$ is chosen such that the reaction event rates in HPGe and LXe detectors match the experimental background levels for next-generation experiments, respectively. In this figure, the horizontal axis represents the energy transfer $T$, and the vertical axis represents the differential event rate $dR/dT$ in unit of cpkkd. The same as in figure \ref{Millicharge count rate figure} and figure \ref{Millicharge count rate figure2}, the red solid lines correspond to the FEA results; red dashed lines represent the simplified FEA results; blue lines stand for the EPA results; black squares display the RIA results; gray triangles show the simplified RIA results.}
\label{Millicharge count rate figure more1}
\end{figure*}

\begin{figure*}
\centering
\includegraphics[scale=0.4]{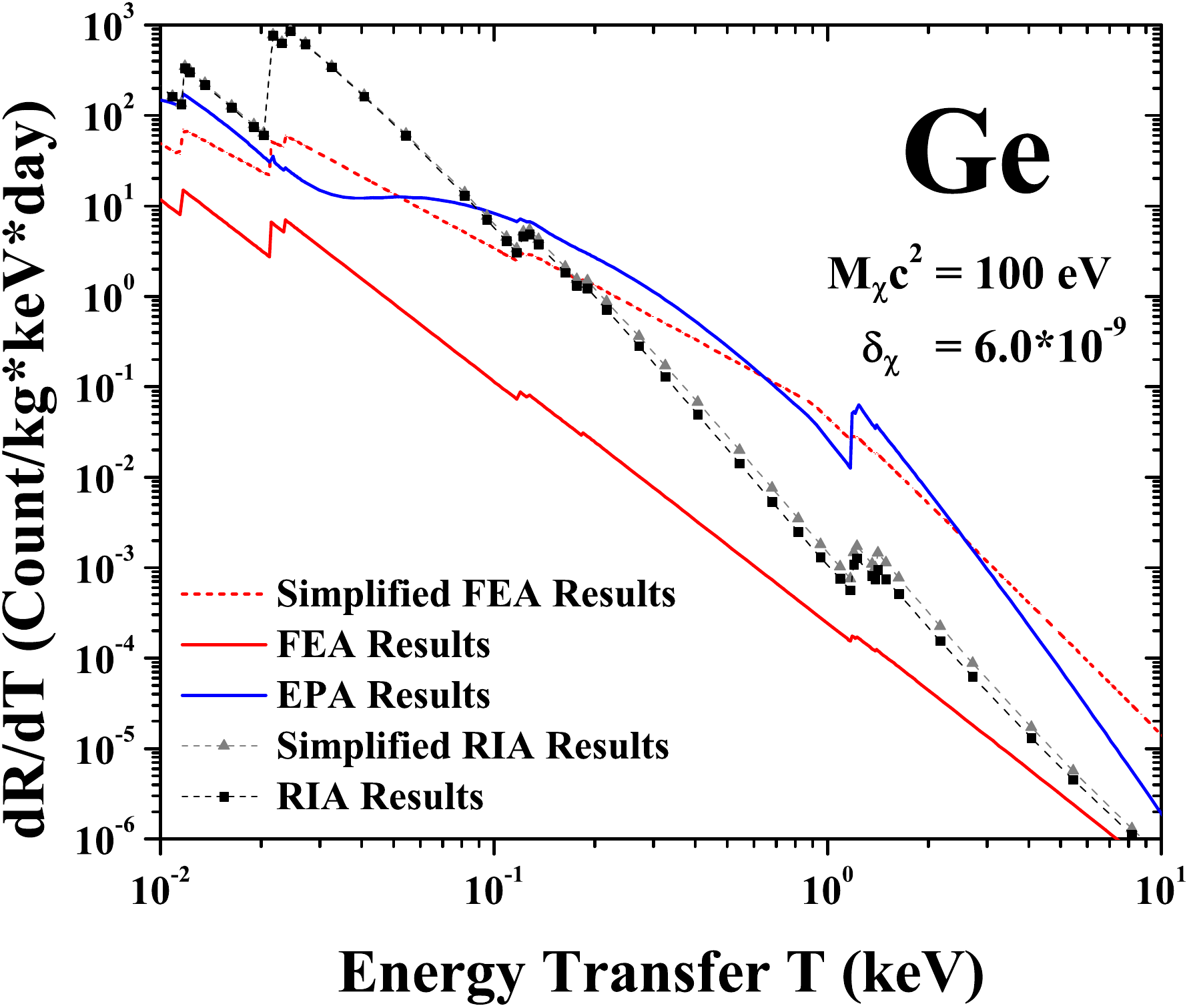}
\includegraphics[scale=0.4]{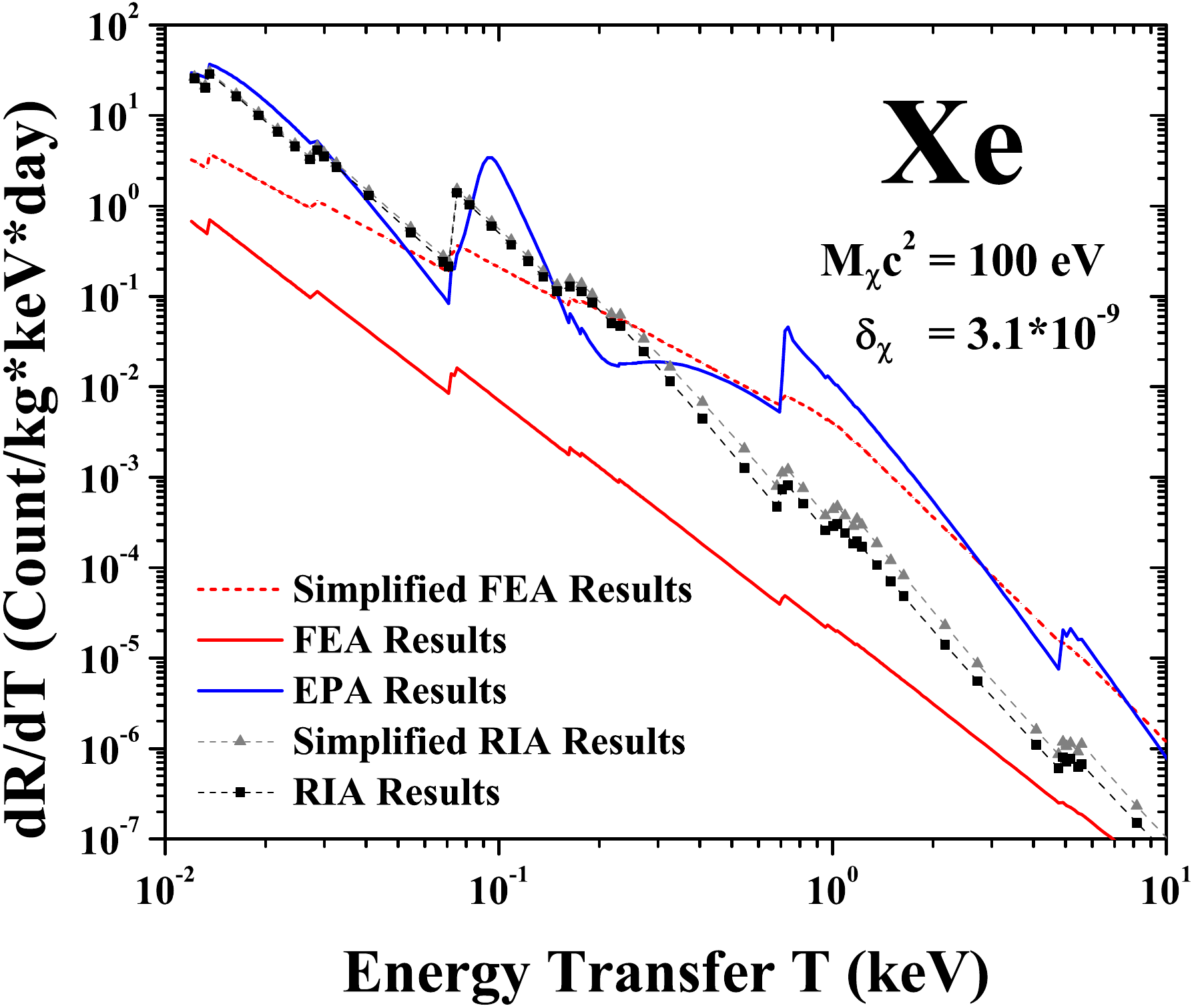}
\includegraphics[scale=0.4]{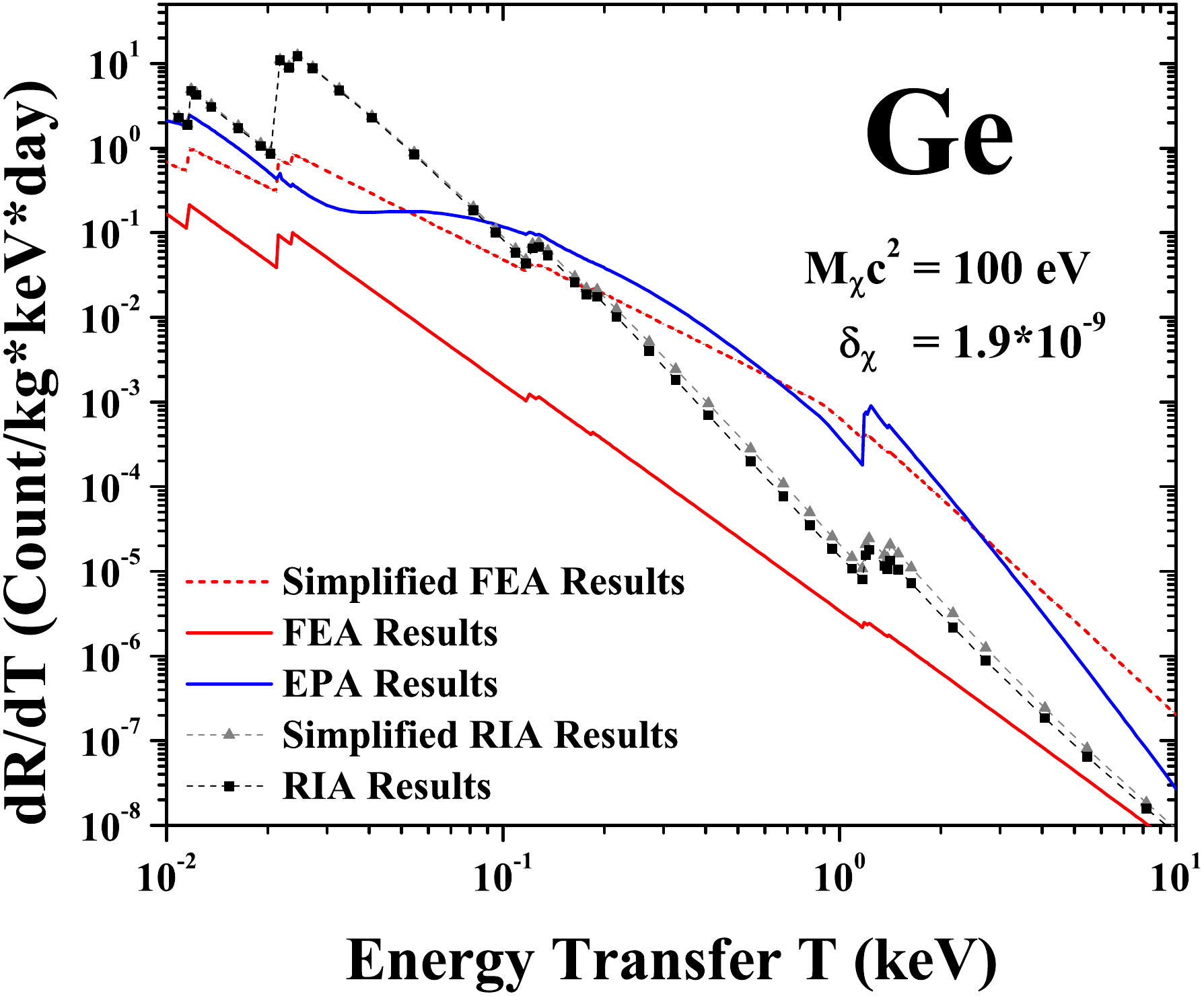}
\includegraphics[scale=0.4]{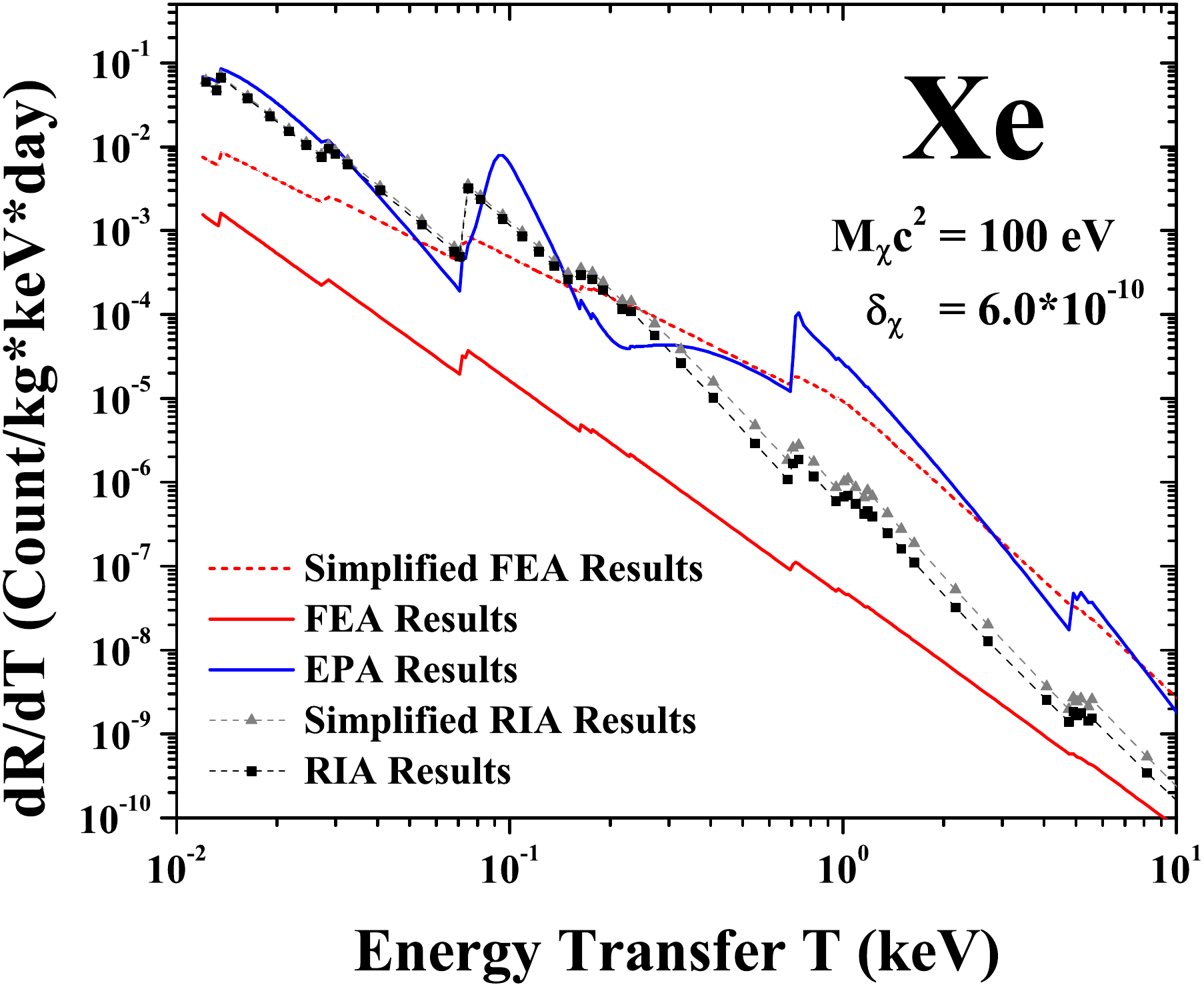}
\includegraphics[scale=0.4]{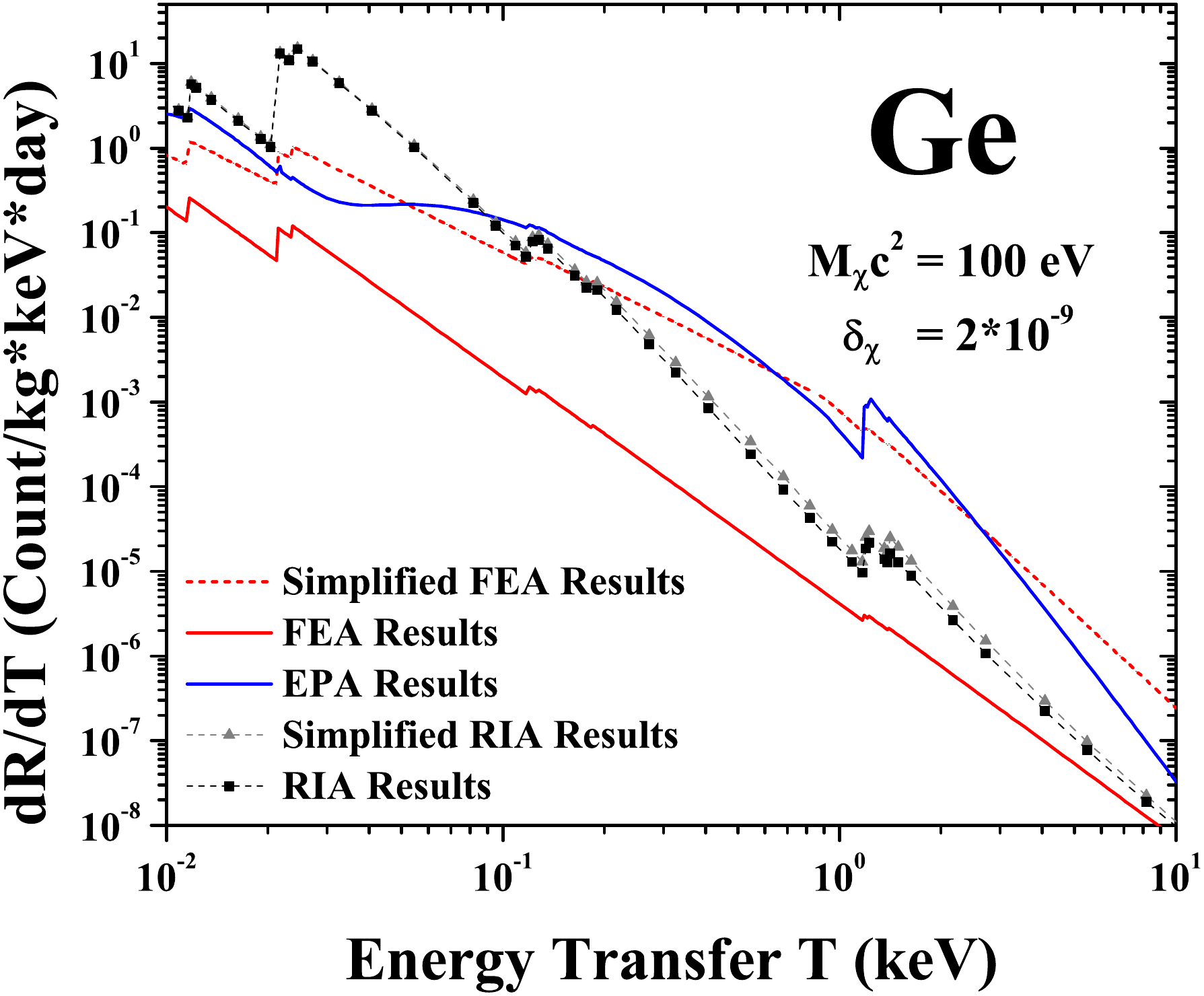}
\includegraphics[scale=0.4]{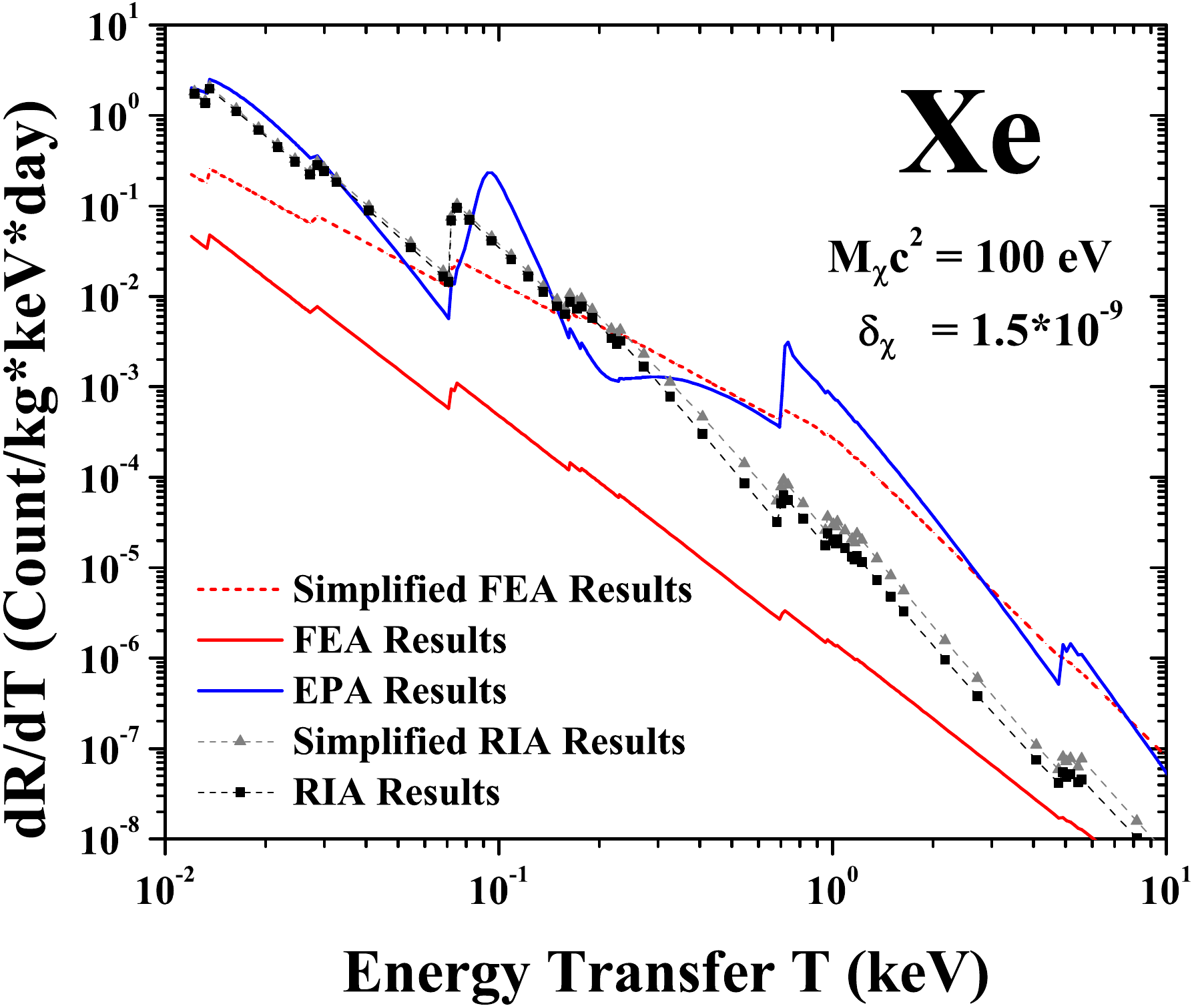}
\caption{The differential reaction event rates for the atomic ionization process induced by millicharged particles in HPGe and LXe detectors. The mass of millicharged dark matter particle is set as $m_{\chi}c^{2}=100$ eV, and its millicharge $\delta_{\chi}$ is chosen such that the reaction event rates in HPGe and LXe detectors match the experimental background levels for next-generation experiments, respectively. In this figure, the horizontal axis represents the energy transfer $T$, and the vertical axis represents the differential event rate $dR/dT$ in unit of cpkkd. The same as in figure \ref{Millicharge count rate figure} and figure \ref{Millicharge count rate figure2}, the red solid lines correspond to the FEA results; red dashed lines represent the simplified FEA results; blue lines stand for the EPA results; black squares display the RIA results; gray triangles show the simplified RIA results.}
\label{Millicharge count rate figure more2}
\end{figure*}

\begin{figure*}
\centering
\includegraphics[scale=0.4]{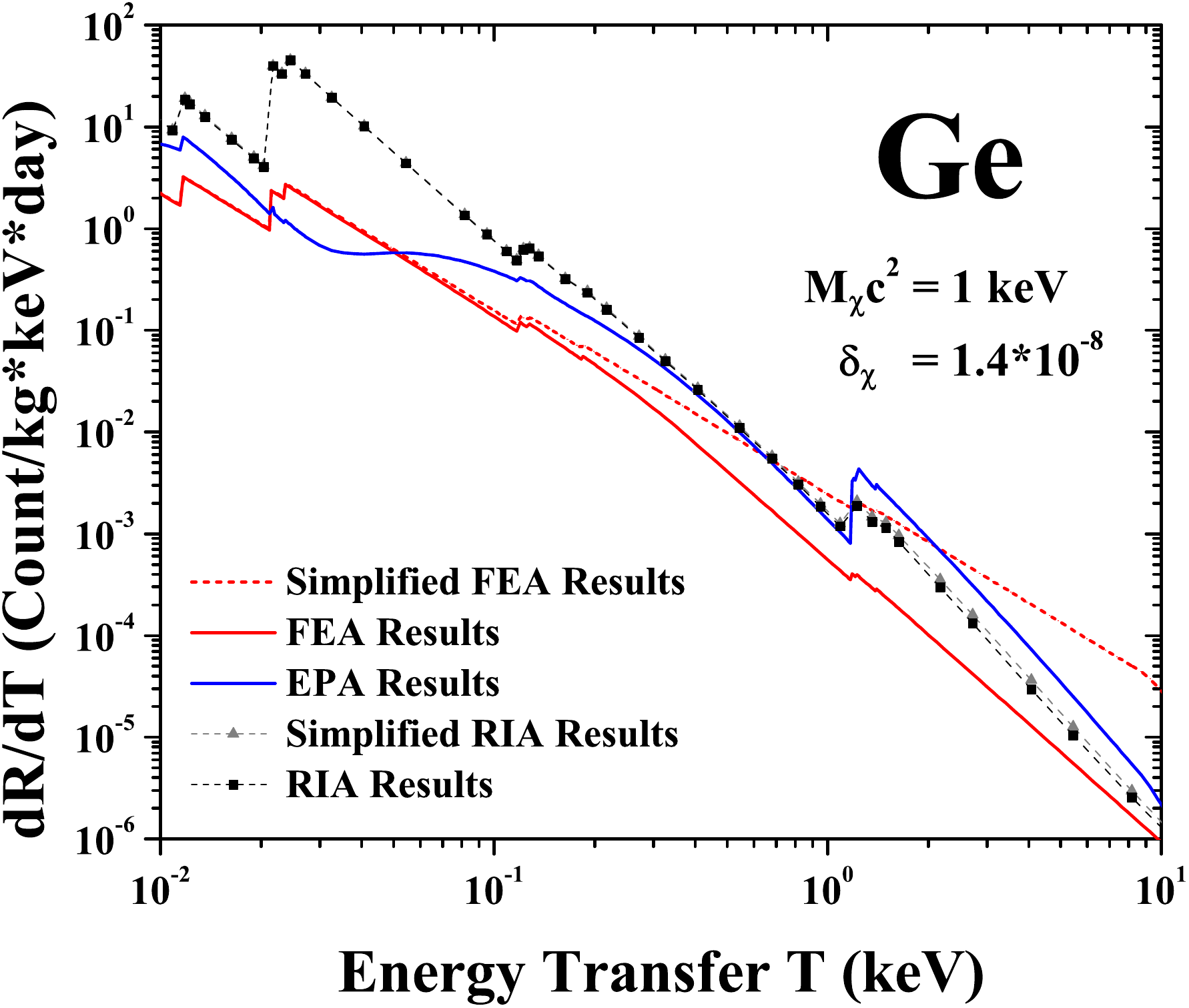}
\includegraphics[scale=0.4]{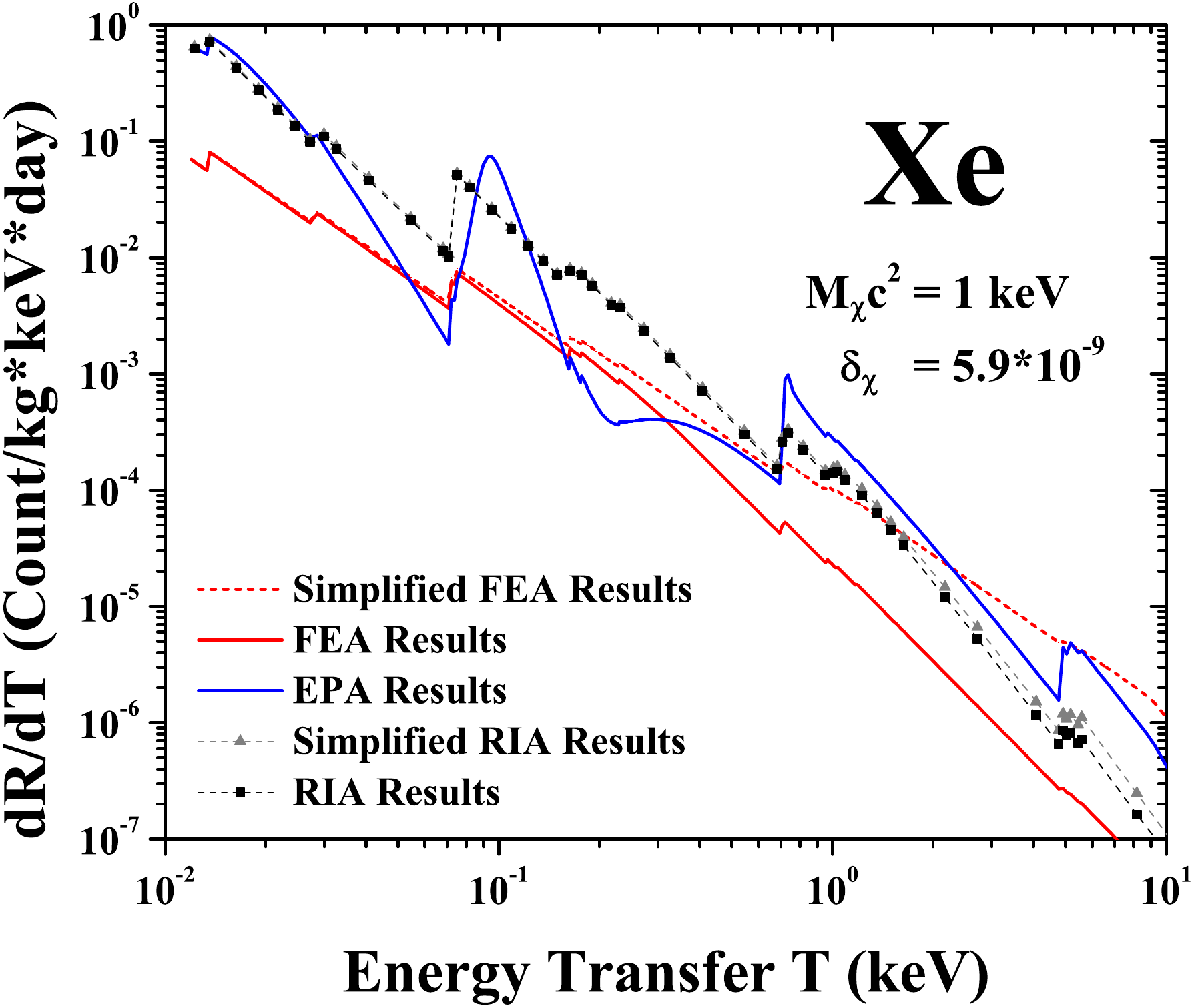}
\includegraphics[scale=0.4]{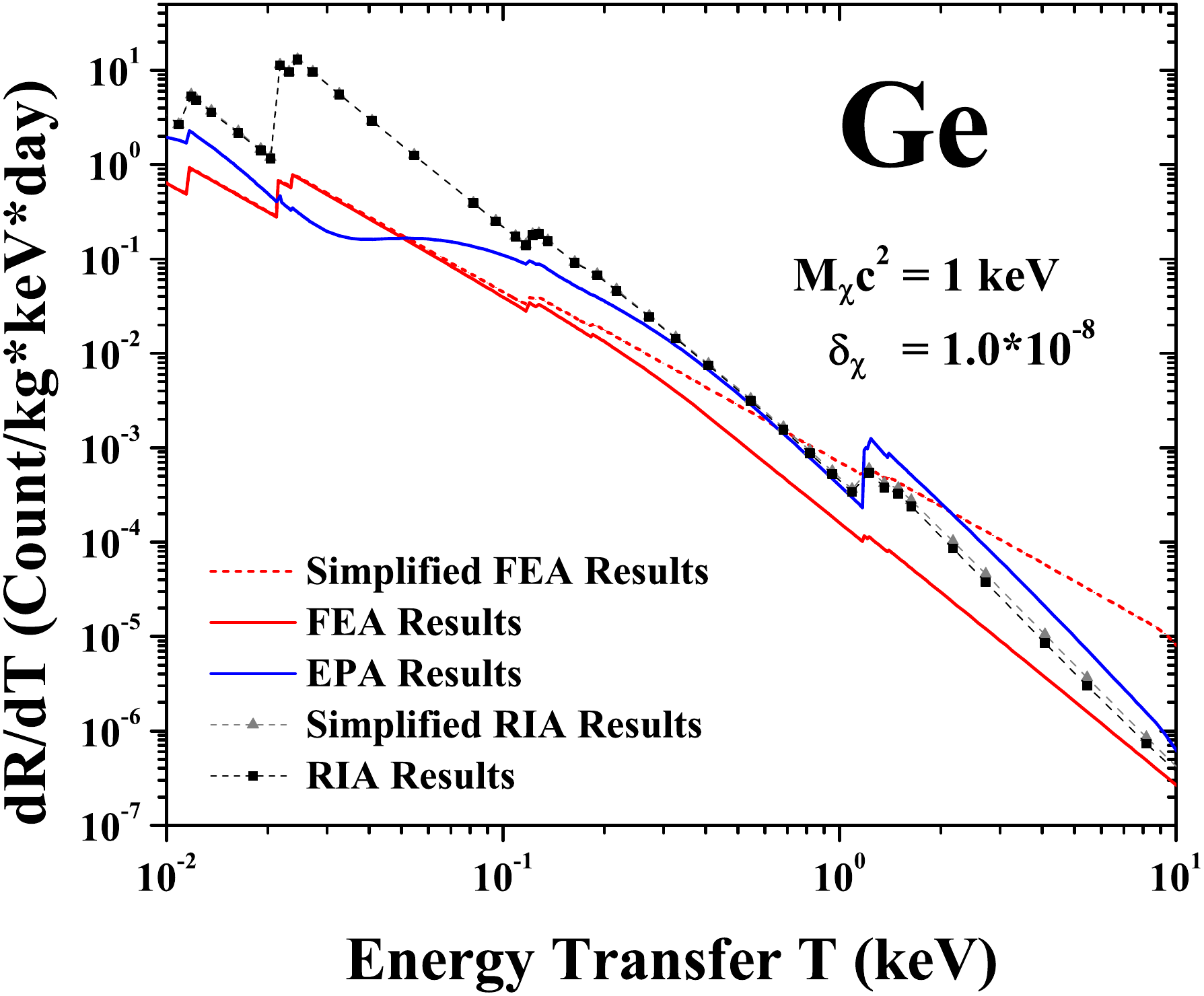}
\includegraphics[scale=0.4]{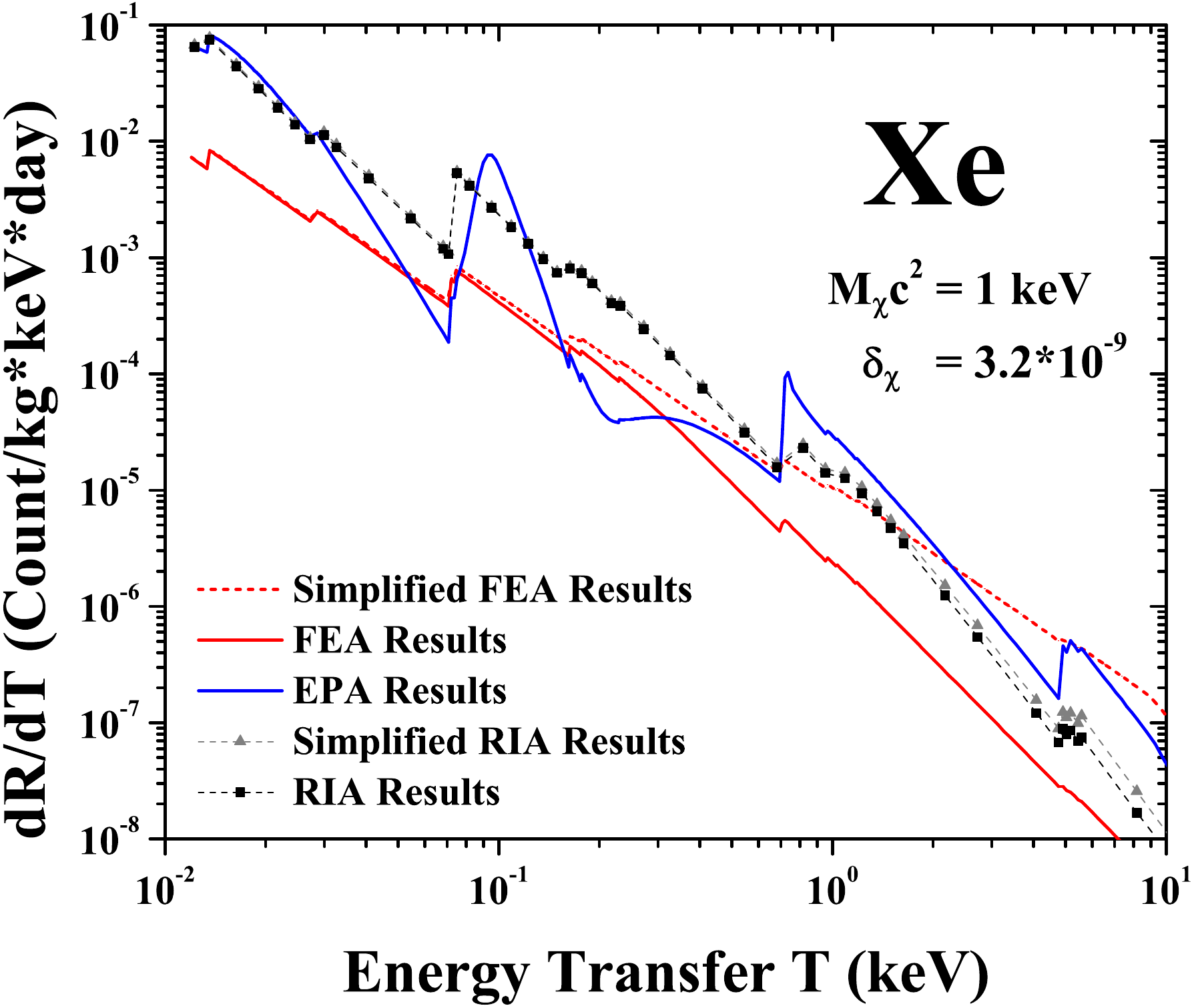}
\includegraphics[scale=0.4]{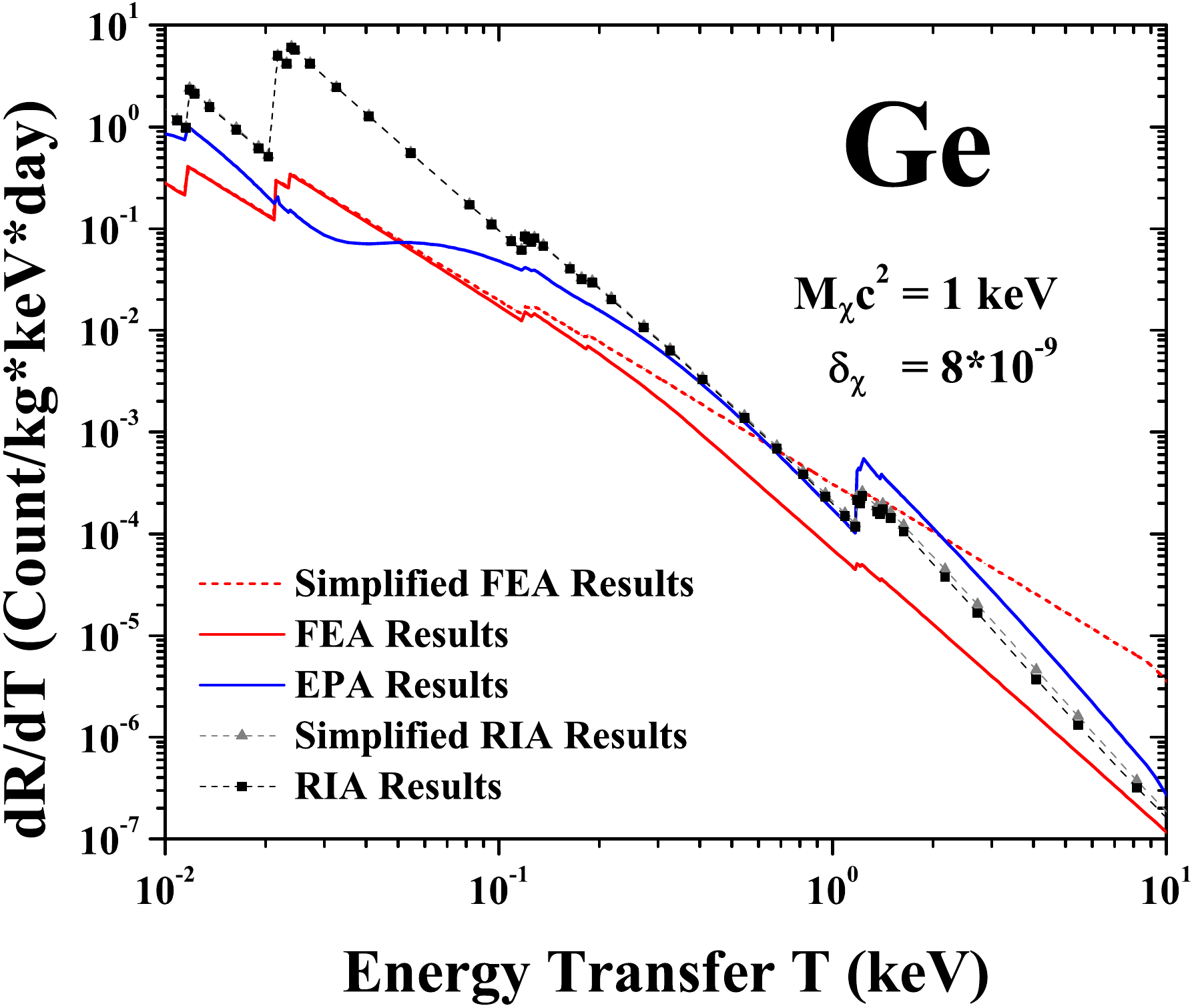}
\includegraphics[scale=0.4]{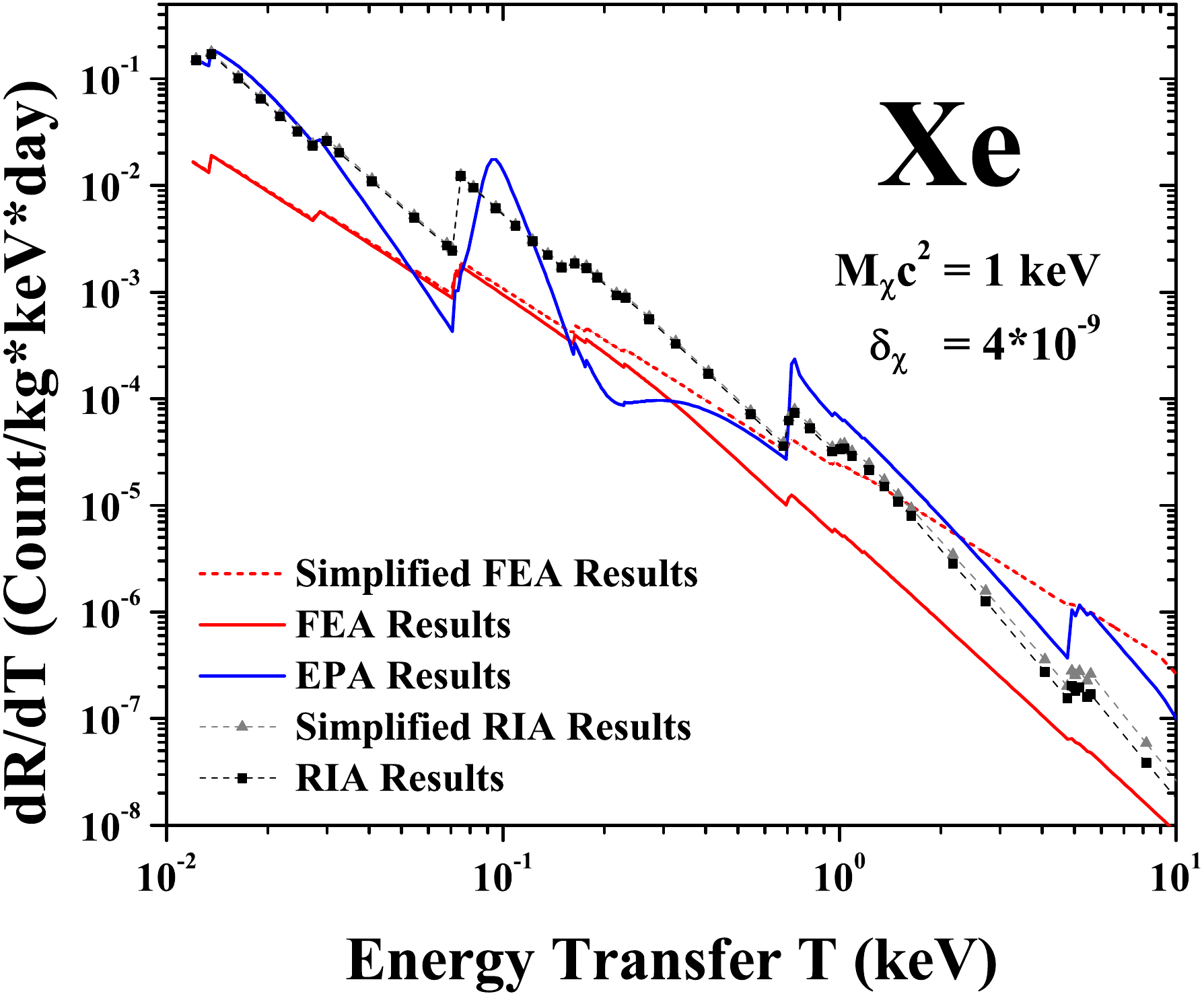}
\caption{The differential reaction event rates for the atomic ionization process induced by millicharged particles in HPGe and LXe detectors. The mass of millicharged dark matter particle is set as $m_{\chi}c^{2}=1$ keV, and its millicharge $\delta_{\chi}$ is chosen such that the reaction event rates in HPGe and LXe detectors match the experimental background levels for next-generation experiments, respectively. In this figure, the horizontal axis represents the energy transfer $T$, and the vertical axis represents the differential event rate $dR/dT$ in unit of cpkkd. The same as in figure \ref{Millicharge count rate figure} and figure \ref{Millicharge count rate figure2}, the red solid lines correspond to the FEA results; red dashed lines represent the simplified FEA results; blue lines stand for the EPA results; black squares display the RIA results; gray triangles show the simplified RIA results.}
\label{Millicharge count rate figure more3}
\end{figure*}

\begin{figure*}
\centering
\includegraphics[scale=0.4]{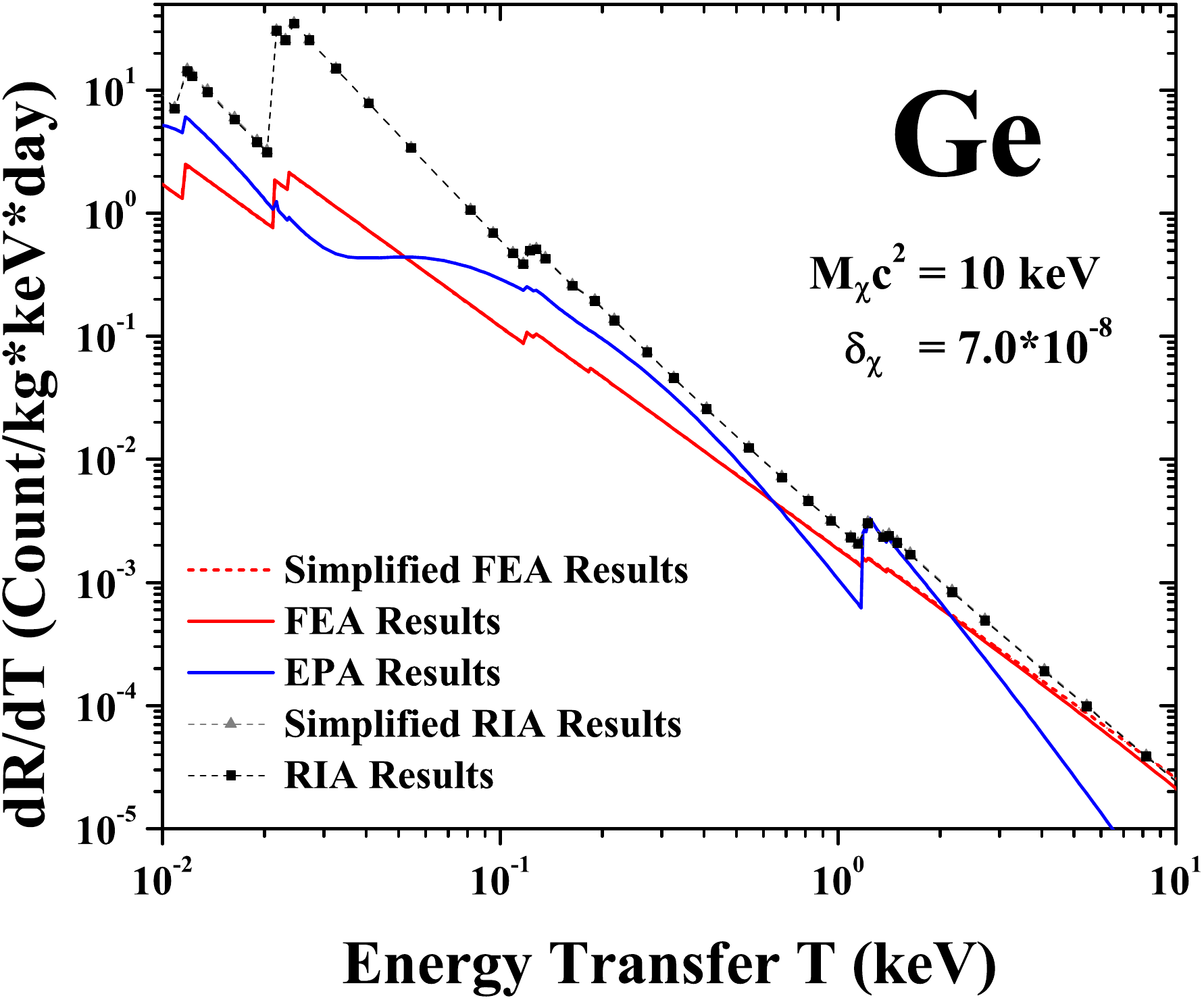}
\includegraphics[scale=0.4]{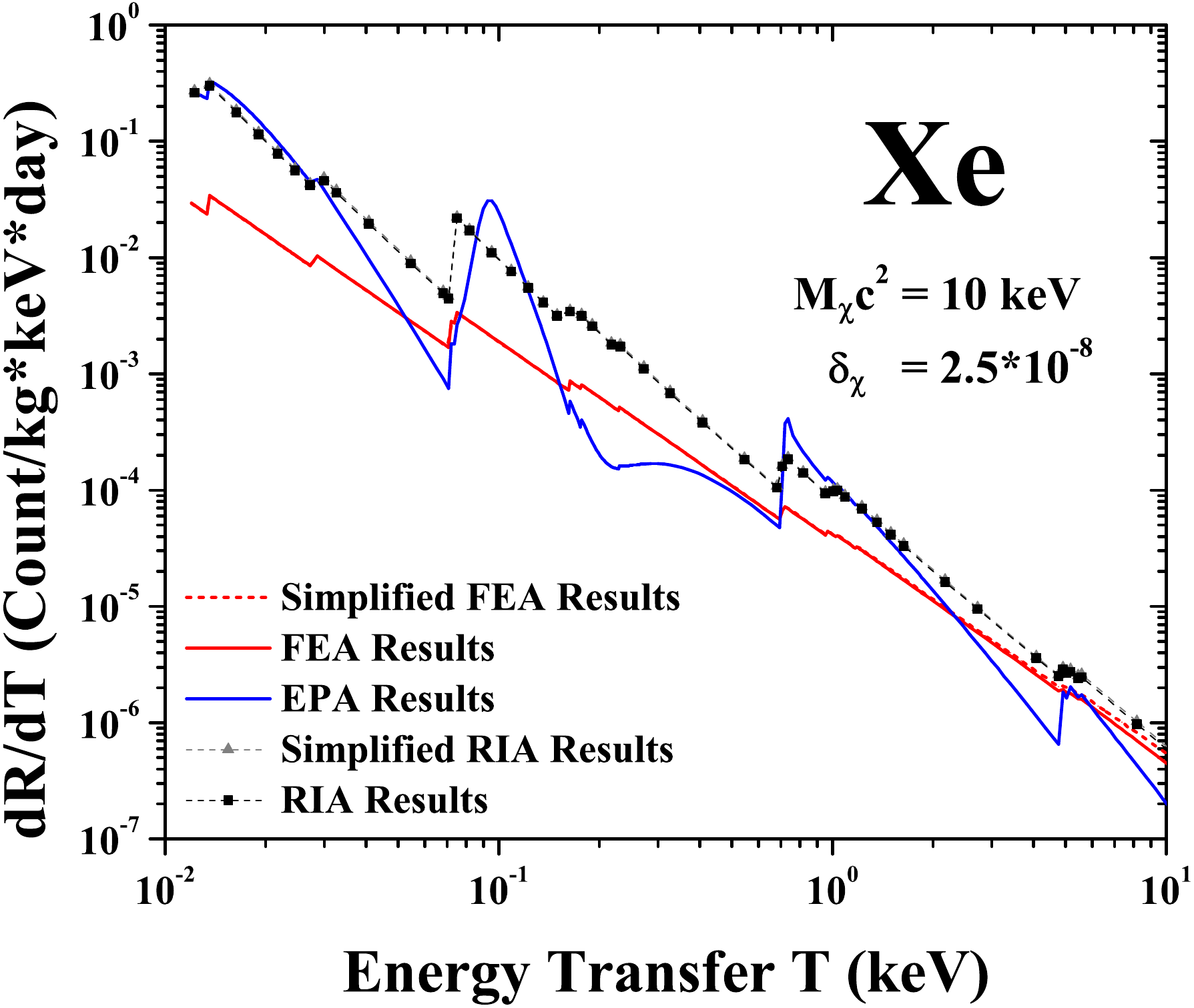}
\includegraphics[scale=0.4]{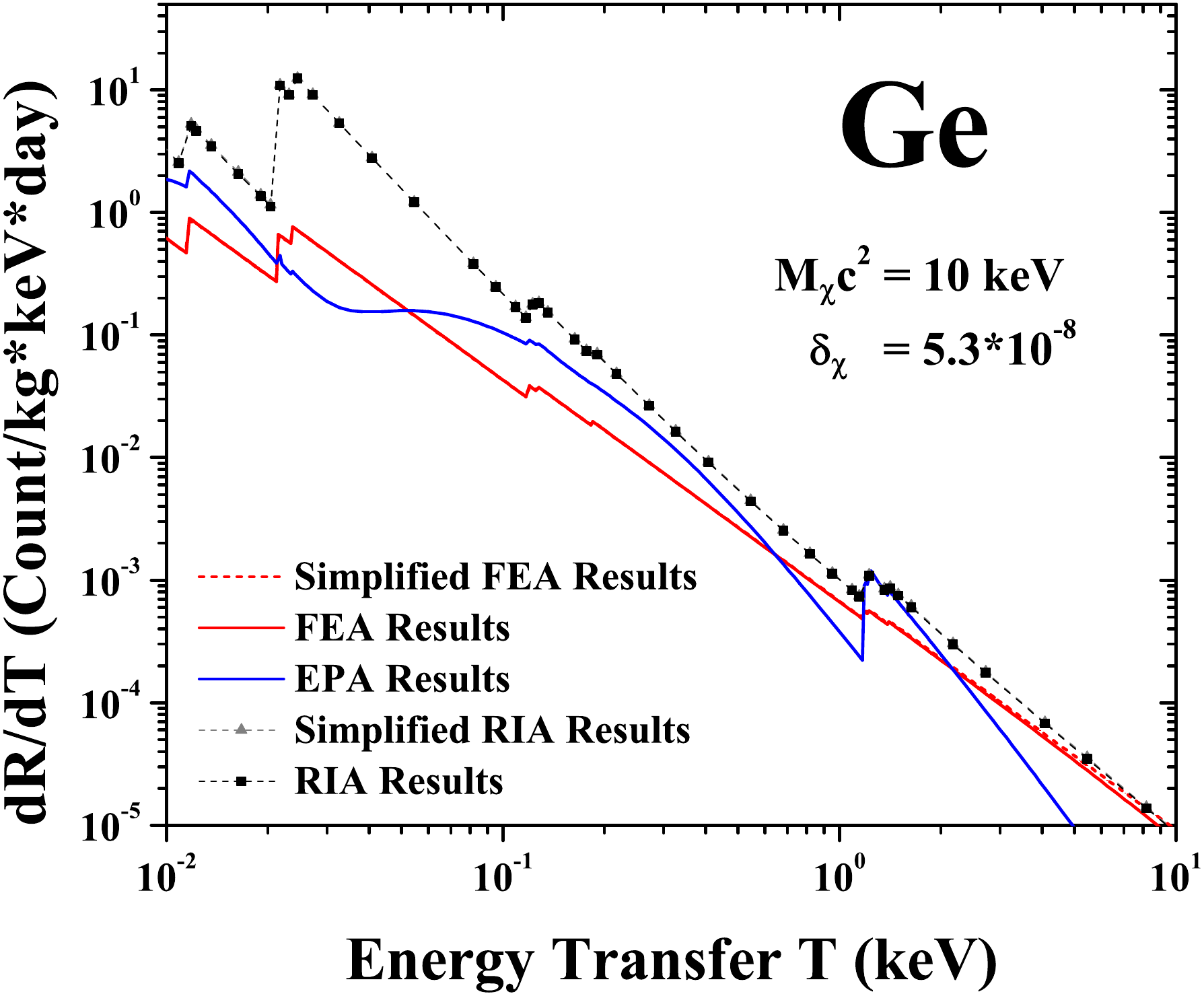}
\includegraphics[scale=0.4]{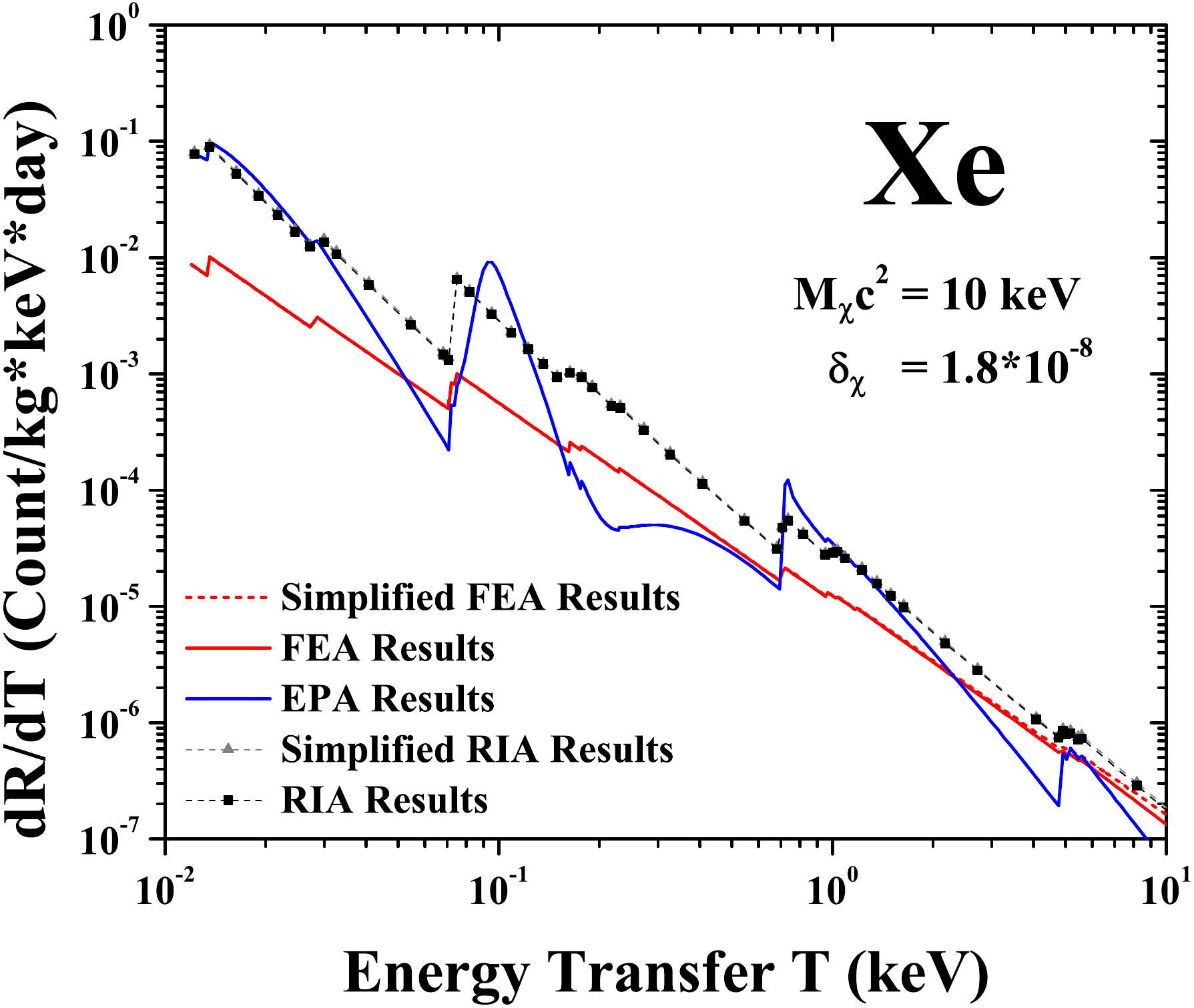}
\includegraphics[scale=0.4]{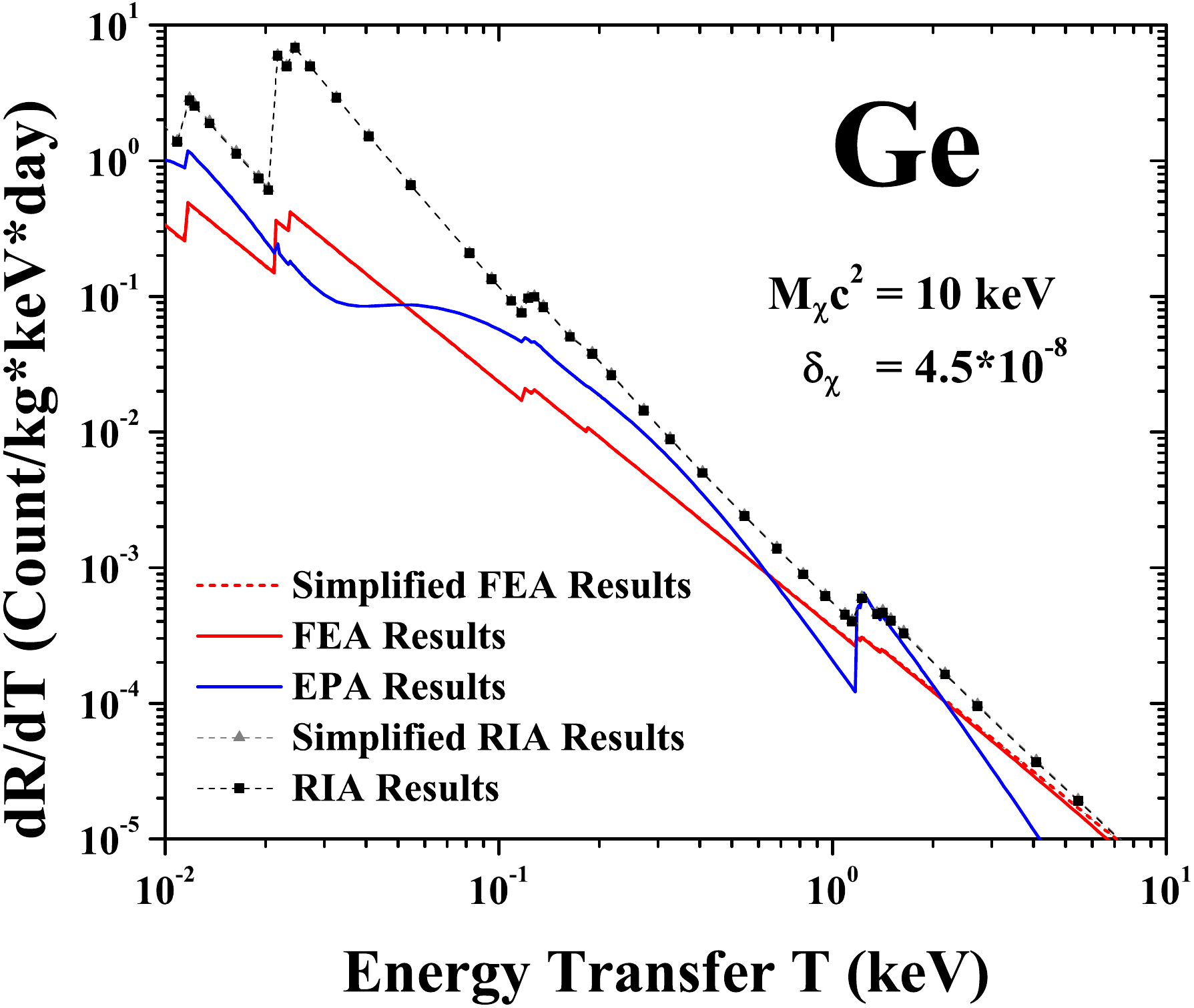}
\includegraphics[scale=0.4]{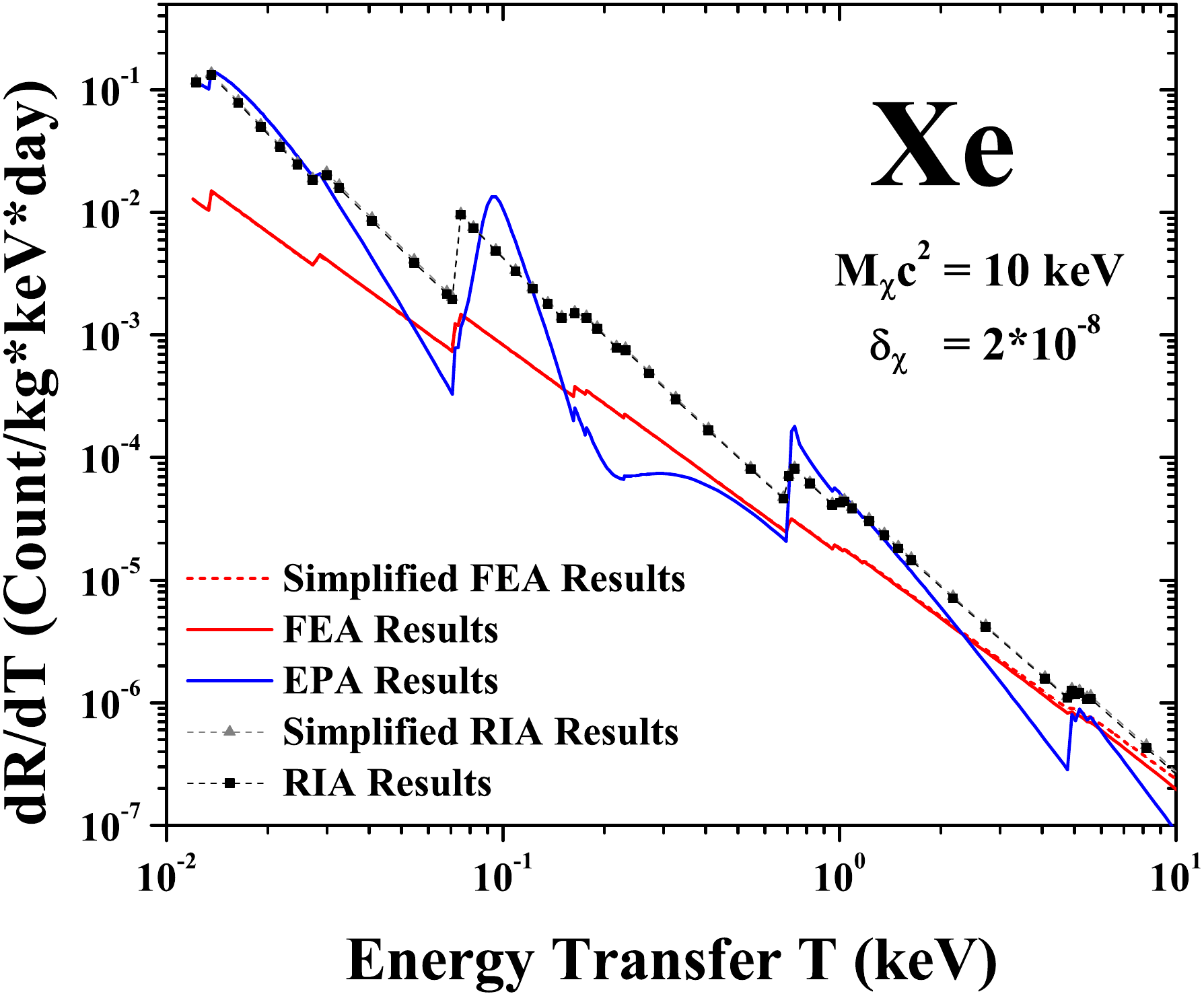}
\caption{The differential reaction event rates for the atomic ionization process induced by millicharged particles in HPGe and LXe detectors. The mass of millicharged dark matter particle is set as $m_{\chi}c^{2}=10$ keV, and its millicharge $\delta_{\chi}$ is chosen such that the reaction event rates in HPGe and LXe detectors match the experimental background levels for next-generation experiments, respectively. In this figure, the horizontal axis represents the energy transfer $T$, and the vertical axis represents the differential event rate $dR/dT$ in unit of cpkkd. The same as in figure \ref{Millicharge count rate figure} and figure \ref{Millicharge count rate figure2}, the red solid lines correspond to the FEA results; red dashed lines represent the simplified FEA results; blue lines stand for the EPA results; black squares display the RIA results; gray triangles show the simplified RIA results.}
\label{Millicharge count rate figure more4}
\end{figure*}

\begin{figure*}
\centering
\includegraphics[scale=0.4]{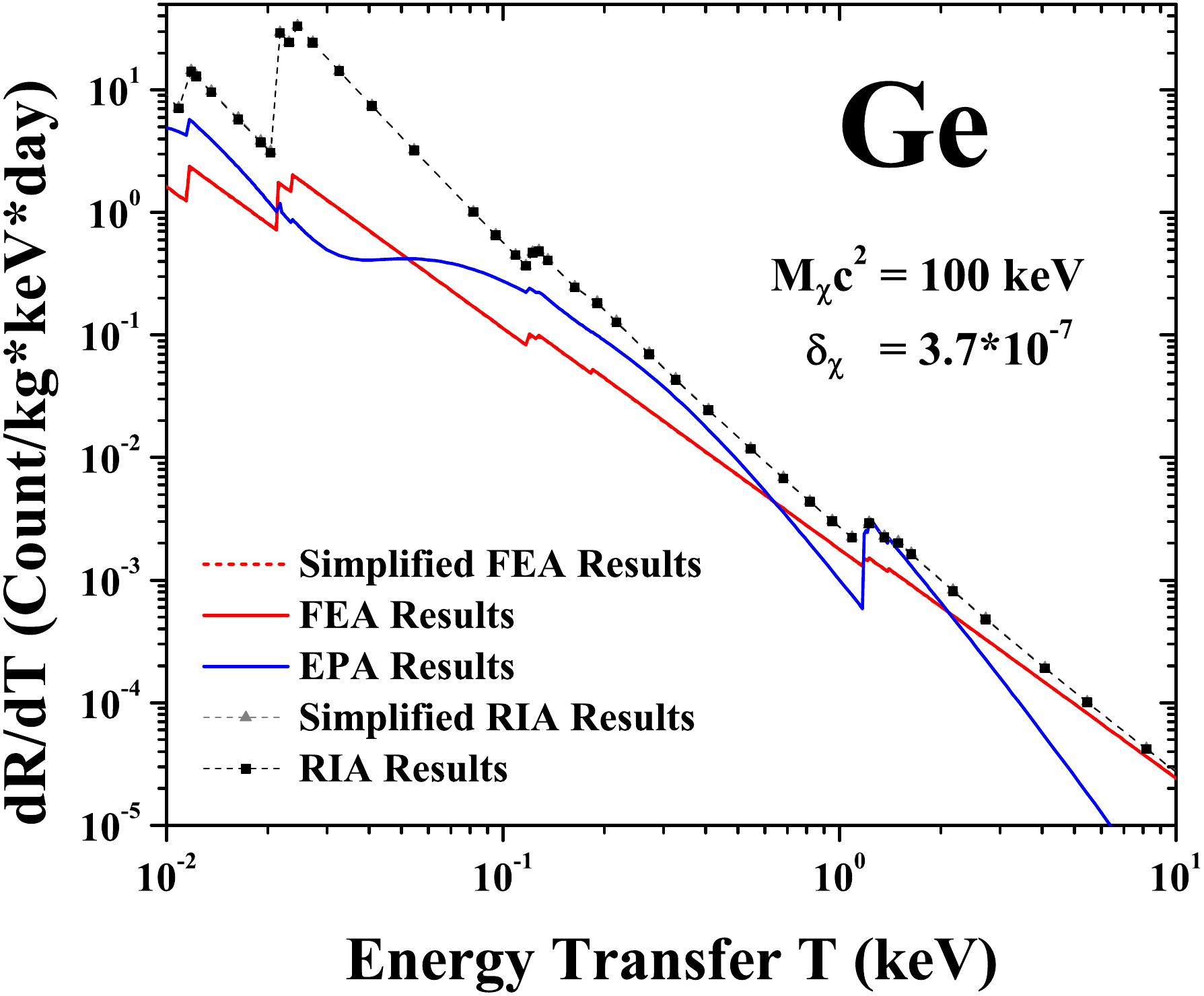}
\includegraphics[scale=0.4]{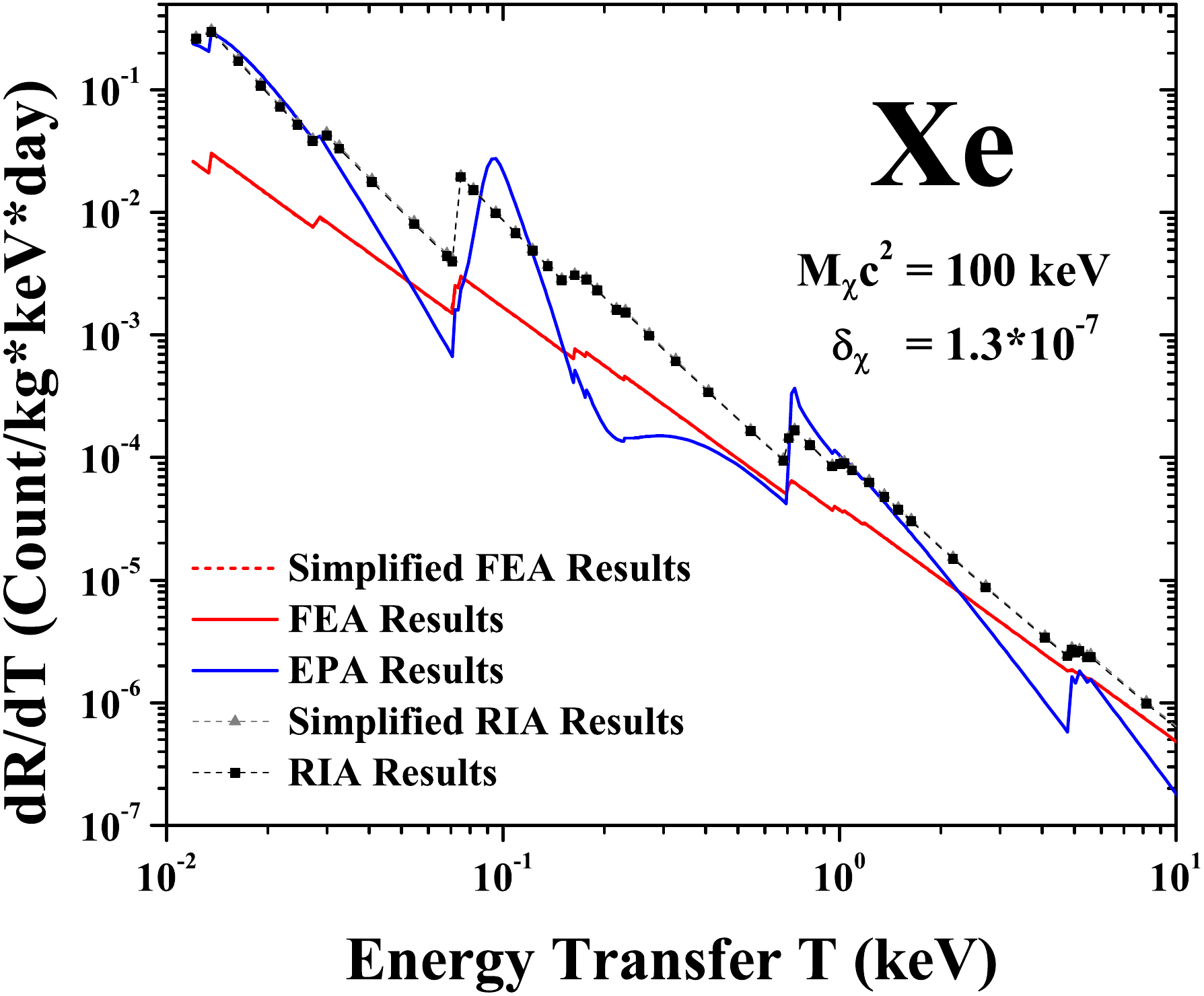}
\includegraphics[scale=0.4]{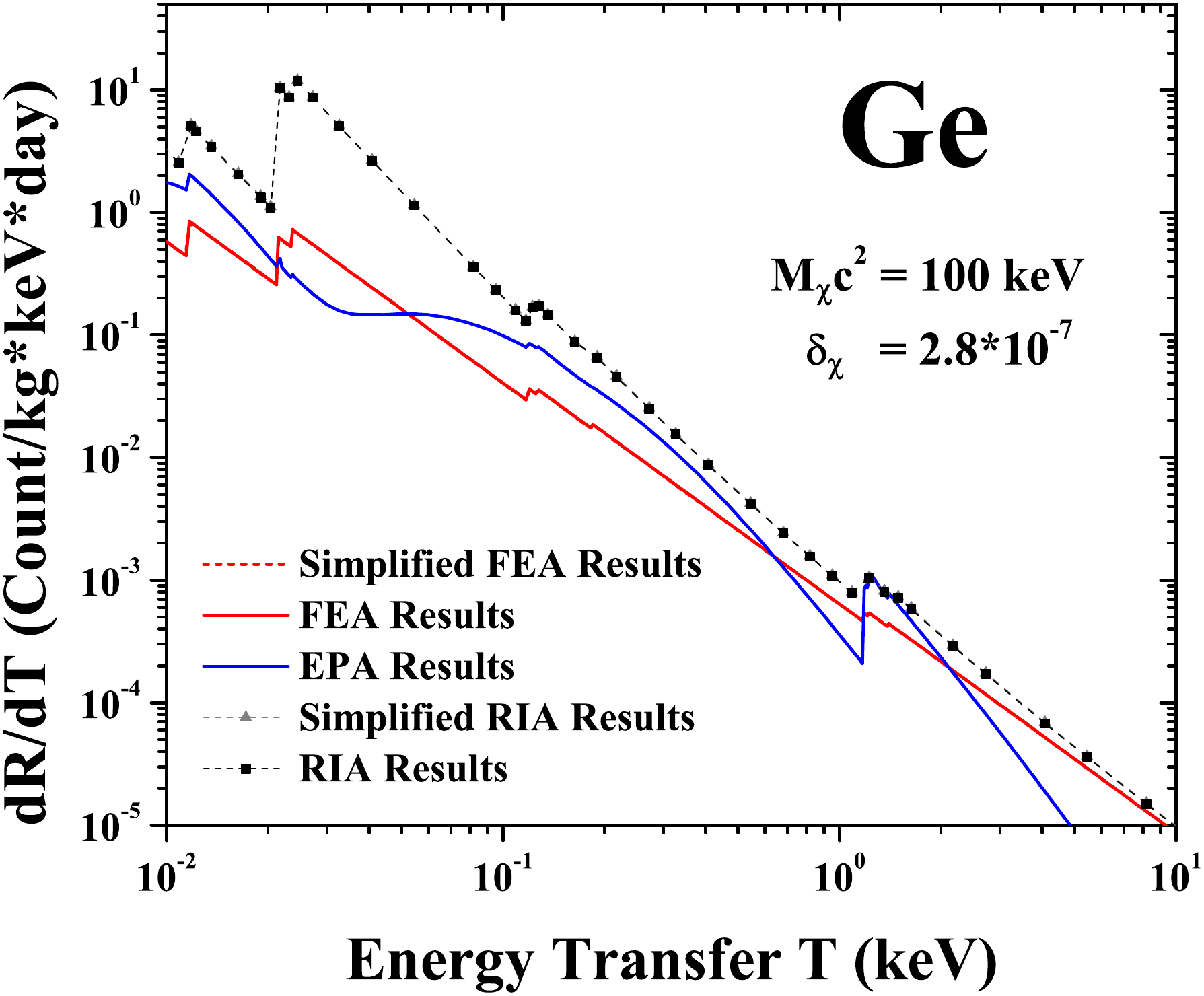}
\includegraphics[scale=0.4]{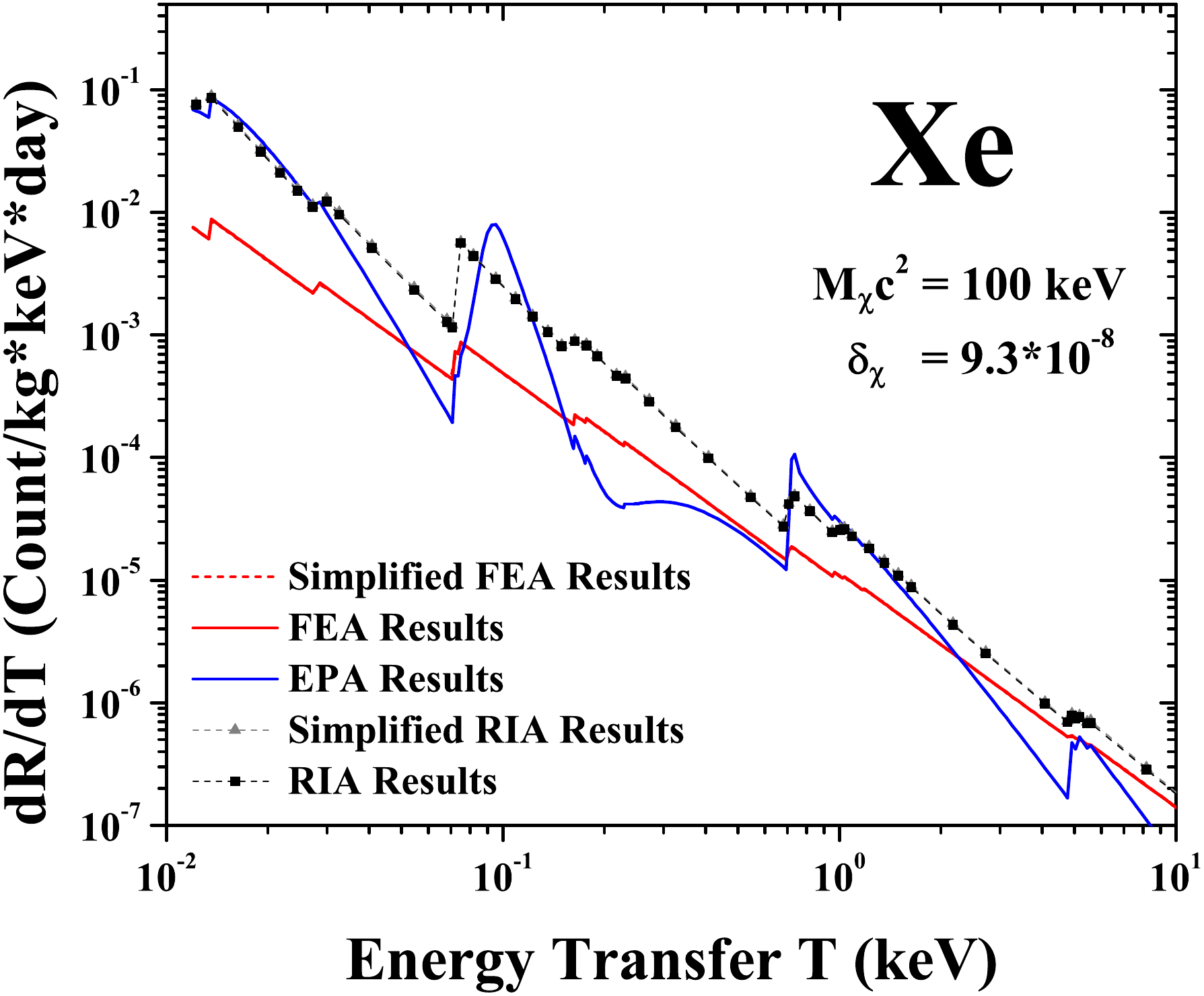}
\includegraphics[scale=0.4]{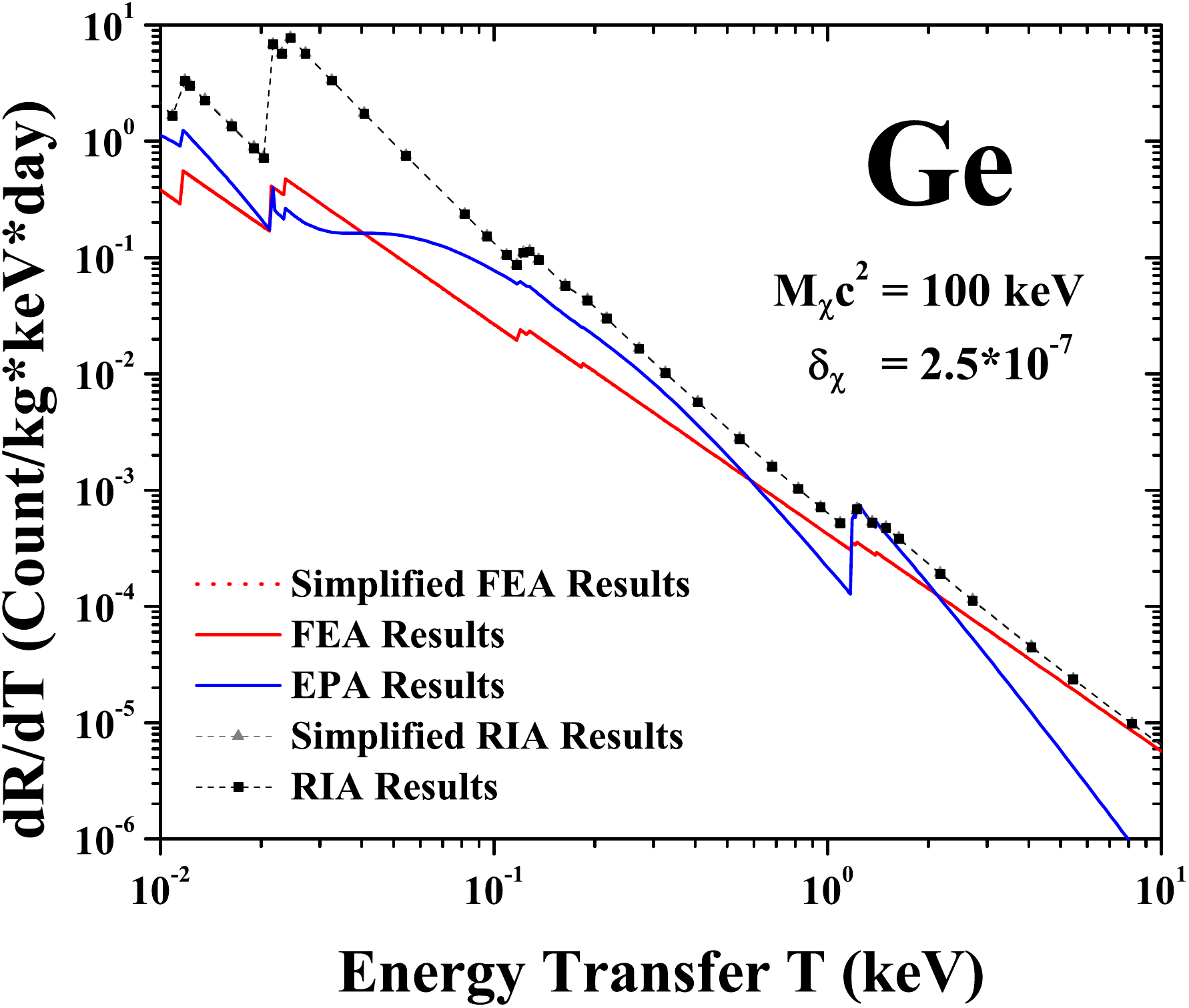}
\includegraphics[scale=0.4]{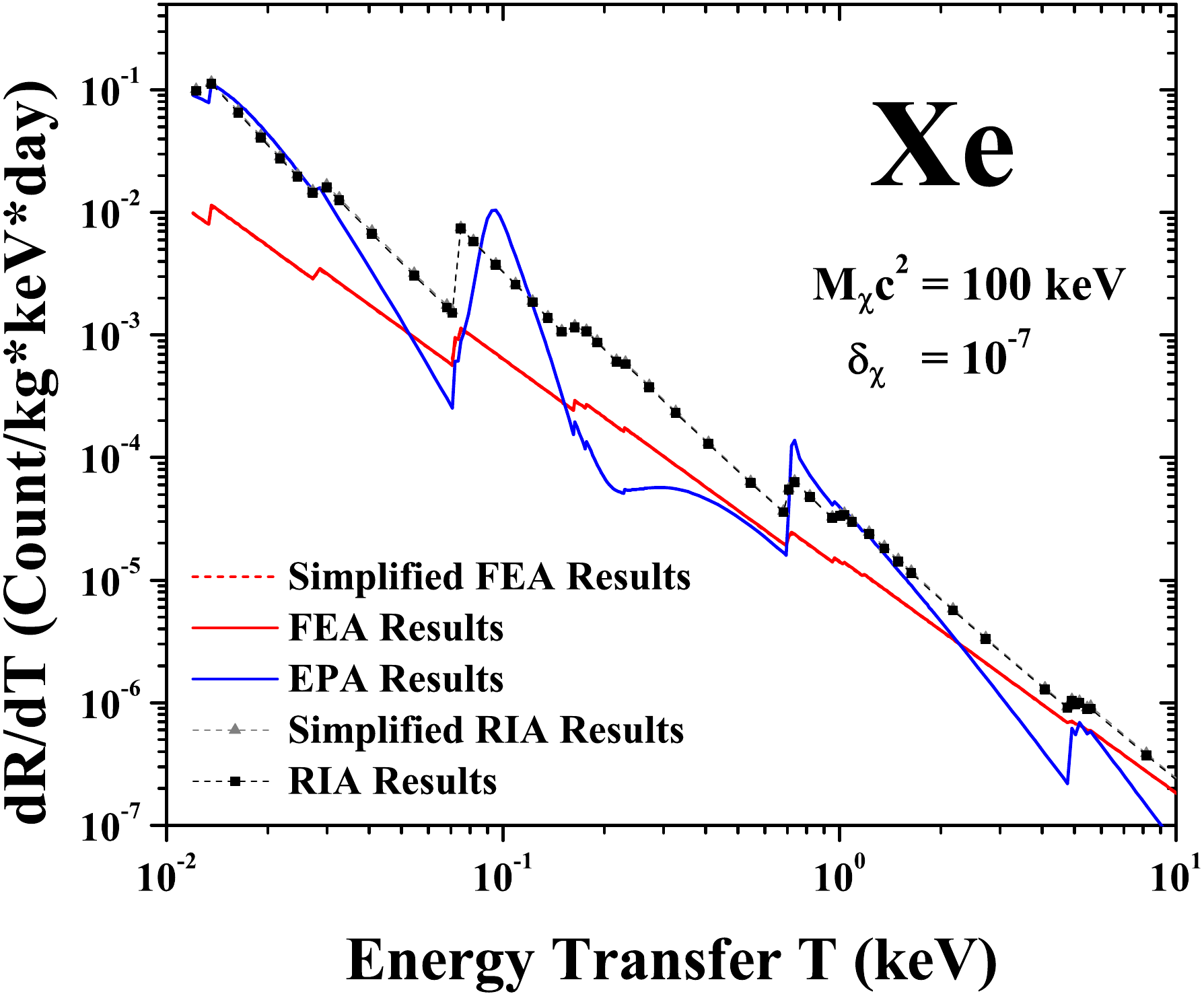}
\caption{The differential reaction event rates for the atomic ionization process induced by millicharged particles in HPGe and LXe detectors. The mass of millicharged dark matter particle is set as $m_{\chi}c^{2}=100$ keV, and its millicharge $\delta_{\chi}$ is chosen such that the reaction event rates in HPGe and LXe detectors match the experimental background levels for next-generation experiments, respectively. In this figure, the horizontal axis represents the energy transfer $T$, and the vertical axis represents the differential event rate $dR/dT$ in unit of cpkkd. The same as in figure \ref{Millicharge count rate figure} and figure \ref{Millicharge count rate figure2}, the red solid lines correspond to the FEA results; red dashed lines represent the simplified FEA results; blue lines stand for the EPA results; black squares display the RIA results; gray triangles show the simplified RIA results.}
\label{Millicharge count rate figure more5}
\end{figure*}

In this appendix, the detailed numerical results on reaction event rates in HPGe and LXe detectors calculated using FEA, EPA and RIA approaches are given for different dark matter particle mass $m_{\chi}$ and their millicharge $\delta_{\chi}$. The estimations of detecting sensitivity on dark matter millicharge $\delta_{\chi}$ for next-generation HPGe and LXe based experiments, which has been summarized in subsection \ref{sec:5c}, are obtained from these results.

From the numerical results in figures \ref{Millicharge count rate figure more1}-\ref{Millicharge count rate figure more5}, it is indicated that our RIA results are approaching to the FEA results when energy transfer $T$ is large. For large dark matter particle mass $m_{\chi}$, i.e. $m_{\chi}c^{2}$ =100 keV, this tendency becomes more apparent. On the other hand, the difference between our RIA results and EPA results becomes tiny in the ultra-low-energy transfer region (namely in the $T\rightarrow0$ limit), especially for LXe detectors. This is consistent with the conclusions in subsection \ref{sec:5b}, indicating the validity of our RIA approach in the entire region of energy transfer $T$.

Furthermore, figures \ref{Millicharge count rate figure more1}-\ref{Millicharge count rate figure more5} also show that, for smaller dark matter particle mass $m_{\chi}$, the differences between FEA and EPA results become larger in the low-energy transfer region. When the dark matter particle mass $m_{\chi}$ reduces, the discrepancies between the FEA results and simplified FEA results become more notable. In the appendix \ref{appendix1}, it is shown that the simplified RIA results reduced to the FEA results only when conditions $m_{\chi} \ll m_{e}$, $T \ll m_{e}c^{2}$, $T \ll E_{\chi}$ are satisfied. For smaller dark matter particle mass $m_{\chi}$, the minimal energy for incoming millicharged dark matter particle, which is chosen as $E_{\chi}^{\text{min}}=10\ m_{\chi}c^{2}$ in our numerical calculations, becomes lower. Therefore, there are large amount of millicharged dark matter particles entering into HPGe and LXe detectors, and the condition $T \ll E_{\chi}$ is harder to satisfy in this case.

\acknowledgments

We acknowledge the helpful discussions with L. Singh and Henry T. Wong. This work was supported by the National Key Basic Research and Develop Program (No. 2017YFA0402203), the National Natural Science Foundation of China (Grants No. 11975159, No. 11975162 and No. 11475117), the Scientific Research Foundation of Chongqing University of Technology (Grant No. 2020ZDZ027), and the Fundamental Research Funds for the Central Universities (Grant No. 20822041C4030). The author Chen-Kai Qiao thanks the comfort and encouragement from Ning Ding, Ran Zhang and Li-Li Gong during the depressed period. Thanks for Kobe Bryant and his Mamba spirit, which has inspired thousands of young people to make desperate efforts to our everyday life. The authors should also thank to the great efforts from all around the world during the pandemic period of Covid-19.

% This is the most common positions for acknowledgments. A macro is available to maintain the same layout and spelling of the heading.

% query inspire, trying to verify the data (this will probably spot

\end{document}